\documentclass[12pt]{iopart}

\usepackage{setstack,amssymb,amsfonts,graphicx,iopams}

\usepackage{color}

\begin{document}
\title[Particle propagator]{%
Particle Propagator of Spin Calogero-Sutherland Model}
\author{
Ryota Nakai$^{1}$ and
Yusuke Kato$^2$
}

\address{
$^1$WPI - Advanced Institute for Materials Research (WPI-AIMR), Tohoku
University, Sendai 980-8577, Japan\\
$^2$Department of Basic Science, The University of Tokyo, Tokyo 153-8902, Japan
}
\ead{rnakai@wpi-aimr.tohoku.ac.jp}
\ead{yusuke@phys.c.u-tokyo.ac.jp}

\begin{abstract}
Explicit-exact expressions for the particle propagator of the spin $1/2$ Calogero-Sutherland model are derived for the system of a finite number of particles and for that in the thermodynamic limit. Derivation of the expression in the thermodynamic limit is also presented in detail. Combining this result with the hole propagator obtained in earlier studies, we calculate the spectral function of the single particle Green's function in the full range of the energy and momentum space. The resultant spectral function exhibits power-law singularity characteristic to correlated particle systems in one dimension.
\end{abstract}

\pacs{02.30.Ik,03.75.Kk,04.20.Jb}
\submitto{\JPA}

\maketitle

\section{Introduction}

The Calogero-Sutherland (CS) model is one of the inverse-square interaction models in one dimension\cite{Su1,Su2,Su3,Su4,KuraKato}. The CS model is different from other models exactly solved by the Bethe ansatz in that the exact eigenfunctions can be written schematically\cite{Sta} and explicit expressions for the integrals of motion are known\cite{Dun}. The CS model with spin internal degrees of freedom\cite{Ha2,Kaw,Min} is also an integrable model.  In this paper, we refer to this model as the spin Calogero-Sutherland model while the CS model for spinless particles is referred to as the scalar CS model. The scalar and the spin CS models are important in the sense that they exhibit simple but nontrivial structures of elementary excitations through exact dynamical correlation functions. In the spin CS model, the SU($N$) spin degrees of freedom give rise to large degeneracy of the energy spectrum and the degenerate eigenstates are decomposed into irreducible representations of Yangian algebra $Y(gl_N)$\cite{Haldane1992,Bernard1993}. An orthogonal basis of degenerate eigenfunctions is called the Yangian Gelfand-Zetlin basis\cite{Naz,Takemura1997}. 

The exact correlation functions of the scalar CS model have been studied since its original discovery of the integrability by Sutherland\cite{Su1,Su2,Su3,Su4,Sim,Min94,For,Hal2}. Exact calculation of dynamical correlation functions has been extended to those with arbitrary integer or rational interaction parameters using the relations of Jack polynomials \cite{Sta,Ha94,Ha95a,Les,Ha95b,zirnbauer95,Ser}. The eigenfunctions of the CS model with the spin degrees of freedom are written in two ways, the Jack polynomials with prescribed symmetry\cite{Dun,Bak} and the Yangian Gelfand-Zetlin basis\cite{Naz,Takemura1997}. With use of the former polynomials and relations of non-symmetric Jack polynomials, the hole propagator\cite{kato97,katyam98} and the density correlation function\cite{KuraKato} have been obtained. On the other hand, the density correlation function, the spin correlation function and the hole propagator have been obtained with use of the Yangian Gelfand-Zetlin basis and the isomorphism to the gl$_2$-Jack polynomials \cite{Uglov1998,Yam1999,hp}.

In our previous paper\cite{hp,hp2}, we have demonstrated the way to calculate the exact dynamical single-particle Green's functions of the spin 1/2 CS model (the spin CS model for particles with the one-half spin). This method, which we call ``the Uglov's method'' naming after the literature \cite{Uglov1998}, uses the Yangian Gelfand-Zetlin basis and an isomorphism between the eigenspace of the spin 1/2 CS model and that of spinless systems. Using this scheme, we have obtained a finite-size representation and the thermodynamic limit of the hole propagator
\begin{eqnarray}
 G^-(x,t)=\frac{\langle {\rm g},N|\psi_{s}^{\dagger}(x,t)
  \psi_{s}(0,0)
  |{\rm g},N\rangle}
 {\langle {\rm g},N|{\rm g},N\rangle}\quad s=\uparrow,\downarrow
\label{eq: hp}
\end{eqnarray}
of the ground state for an arbitrary non-negative integer interaction parameter. Here $|{\rm g}, N\rangle$ is the ground state vector of the spin CS model with $N$ particles.  

The entire set of the single-particle Green's functions of the spin 1/2 CS model are completed by calculating the particle propagator,
\begin{eqnarray}
 G^+(x,t)=\frac{\langle {\rm g},N|\psi_{s}(x,t)
  \psi_{s}^{\dagger}(0,0)
  |{\rm g},N\rangle}
 {\langle {\rm g},N|{\rm g},N\rangle}\quad s=\uparrow,\downarrow.
 \label{pp}
\end{eqnarray}
An explicit-expression for (\ref{pp}) and that for (\ref{eq: hp}) in our previous paper lead to the spectral function of the single-particle Green's function in the full momentum-energy plane. As in the case of exact calculation of the particle propagator of the scalar CS model\cite{Ser}, this calculation is complicated due to complexity of the intermediate states. Some formulae necessary for exact calculation of (\ref{pp}) can be derived from mathematical formulae of the Macdonald symmetric polynomials\cite{Mac}. This gives the reason why we consider the Uglov's method is suited for exact calculations of the single-particle Green's functions, even though the hole propagator of the spin CS model has been already obtained by another method\cite{kato97,katyam98}.

In the present paper, we derive exact expressions for the particle propagator of the spin 1/2 CS model for finite-size systems and in the thermodynamic limit, and we also examine characteristic features of the spectral function. The purpose of this paper is two-fold. One is to present the exact explicit expression for the particle propagator and discuss physical implications of the exact result. The other is to explain derivation of the exact expression in detail. In order for physical aspects of our results to be accessible to readers, we first present the model and the main results in the next section and section~\ref{sec: main result} before explaining derivation of those results. 
In section~\ref{sec: preliminaries}, we briefly summarize fundamental properties of the spin CS model and the method used in this paper. The expression for the particle propagator of finite-size systems is derived in section~\ref{sec: finite-size}. In section~\ref{sec: th}, we derive the expression in the thermodynamic limit. We discuss nontrivial aspects in the derivation in section \ref{sec: discussion} and summarize our conclusion in section~\ref{sec: conclusion}.

\section{Model}
We consider an $N$-particle system in one-dimensional space whose length is $L$. Each particle carries the spin 1/2 as internal degrees of  freedom. The wave function $\Psi(x_1,\sigma_1,\cdots, x_N,\sigma_N)$, which is a function of the spatial coordinate $x_i\in [0,L]$ and the spin coordinate $\sigma_i(=\pm 1/2)$ of the $i$th particle ($i=1,\cdots,N$), satisfies the periodic boundary condition
\begin{equation}
\Psi(x_1,\sigma_1,\cdots, x_i +L,\sigma_i,\cdots, x_N,\sigma_N)
=
\Psi(x_1,\sigma_1,\cdots, x_i ,\sigma_i,\cdots, x_N,\sigma_N).
\label{eq: pbc-Psi}
\end{equation}
The spin 1/2 CS model\cite{Ha2,Kaw,Min} is defined on such a one-dimensional system and the Hamiltonian is given by
\begin{eqnarray}
 \mathcal{H}=-\sum_{i=1}^N\frac{\partial^2}{\partial x_i^2}
 +\frac{2\pi^2}{L^2}\sum_{i<j}\frac{\lambda(\lambda+P_{ij})}
 {\sin^2[(x_i-x_j)\pi/L]}.  \label{eq: spinCS}
\end{eqnarray}
Here $P_{ij}$ is the operator that exchanges the spin coordinates of $i$th and $j$th particles
\begin{equation}
P_{ij}\Psi(\cdots, x_i,\sigma_i,\cdots, x_j,\sigma_j,\cdots)
=
\Psi(\cdots, x_i,\sigma_j,\cdots, x_j,\sigma_i,\cdots).
\label{eq: Pij}
\end{equation}
The symbol $\lambda$ is an interaction parameter that specifies the Hamiltonian. For non-negative real $\lambda$, the eigenenergies and eigenstates of the Hamiltonian can be explicitly written\cite{KuraKato,Bernard1993,Takemura1997,Uglov1998,Kato1995}. In earlier works\cite{kato97,katyam98,hp,hp2}, the hole propagator of the spin 1/2 CS model was derived when $\lambda$ is a non-negative integer. In the following, similarly we take $\lambda$ to be non-negative integer. The statistics of particles are bosonic for odd $\lambda$ and fermionic for even $\lambda$, following the earlier works\cite{kato97,katyam98,hp,hp2} on the hole propagator. 
For convenience, we set the number of particles $N$ to be twice an odd (even) integer for even (odd) $\lambda$ so that the ground state has no degeneracy.

\section{Main results and physical interpretation}
\label{sec: main result}
In this section, main results of the present paper are shown in advance of detailed derivation. At first, the particle propagator of the spin 1/2 CS model in the coordinate space in the thermodynamic limit  (t.d.l.)
\begin{equation}
{\rm (t.d.l.)}: N\rightarrow \infty,\quad L\rightarrow \infty,\quad N/L=d(\mbox{fixed})
\label{eq: tdl}
\end{equation}
is shown with physical interpretation in terms of quasi-particles and quasi-holes. Next, a spectral function, the particle propagator in the energy-momentum space, is shown. The spectral function is drawn numerically, and we reveal its singular behavior at the edge of and the inside of the support.

\subsection{Particle propagator}

Since the non-degenerate ground state of the spin 1/2 CS model is spin-singlet\cite{Ha2,Kaw}, the particle propagator (\ref{pp}) is independent of the spin $s$ of the field operators, and thus we consider $s=\downarrow$ without loss of generality. In the thermodynamic limit, the particle propagator is composed of three parts
\begin{equation}
G^{+}(x,t)=G^{(\rm 0R)}(x,t)+G^{(\rm 0L)}(x,t)+G^{(1)}(x,t).
\label{eq: G+decouple}
\end{equation}
With a parameter $\lambda'=2\lambda+1$, the first two terms in the right-hand side of (\ref{eq: G+decouple}) are given by
\begin{eqnarray}
&G^{(\rm 0R)}(x,t) \nonumber\\
&=\frac{\lambda'd}{4}
  \int_1^{\infty}{\rm d}w_{\rm R}
  F^{(0)}(w_{\rm R})
  \exp\left[-{\rm i}(E_{\rm qp}(w_{\rm R})+\mu)t
 +{\rm i}P_{\rm qp}(w_{\rm R})x\right],\label{eq: G0R-th}
\end{eqnarray}
and
\begin{eqnarray}
&G^{(\rm 0L)}(x,t) \nonumber\\
&=\frac{\lambda'd}{4}
  \int_{-\infty}^{-1}{\rm d}w_{\rm L}
  F^{(0)}(w_{\rm L})
  \exp\left[-{\rm i}(E_{\rm qp}(w_{\rm L})+\mu)t
 +{\rm i}P_{\rm qp}(w_{\rm L})x\right],\label{eq: G0L-th}
\end{eqnarray}
and the last term in the right-hand side of (\ref{eq: G+decouple}) is given by
\begin{eqnarray}
&G^{(1)}(x,t)\nonumber\\
&=C^{(1)}
  \sum_{\tau_{\rm L},\tau_{\rm R},\{\sigma_j\}
  }
 \int_1^{\infty}{\rm d}w_{\rm R}
 \int_{-\infty}^{-1}{\rm d}w_{\rm L}
 \int_{-1}^1{\rm d}u_{\lambda'}
 \int_{-1}^{u_{\lambda'}}{\rm d}u_{\lambda'-1}\cdots
 \int_{-1}^{u_2}{\rm d}u_1\nonumber\\
 &\times
 \delta_{\sum_j\sigma_j+\tau_{\rm L}+\tau_{\rm R},-1/2}
  F(\{u_j\},w_{\rm R},w_{\rm L},\{\sigma_j\},\tau_{\rm{R}},\tau_{\rm{L}})\nonumber\\
 &\times
 \exp\left[-{\rm i}
 \left\{E_{\rm qp}(w_{\rm R})+E_{\rm qp}(w_{\rm L})
 +\sum_{j=1}^{\lambda'}E_{\rm qh}(u_j)
 +\mu\right\}t
 \right]\nonumber\\ 
 &\times
 \exp\left[{\rm i}
 \left\{P_{\rm qp}(w_{\rm R})+P_{\rm qp}(w_{\rm L})
 +\sum_{j=1}^{\lambda'}P_{\rm qh}(u_j)\right\}x
 \right],\label{eq: th-G1}
\end{eqnarray}
where the coefficient $C^{(1)}$ is given by
\begin{equation}
 C^{(1)}
=\frac{d}{2^{2(\lambda+2)}\lambda'^{\lambda}\Gamma(\lambda+1)}
 \prod_{j=1}^{\lambda'}
 \frac{\Gamma\left((\lambda+1)/\lambda'\right)}
 {\Gamma\left(j/\lambda'\right)^2},
\end{equation}
and the chemical potential is $\mu=(\pi\lambda'd/2)^2$, which follows from the ground state energy\cite{Ha2,Kaw,hp} (see for example (22) of \cite{hp}). $\Gamma$ denotes the gamma function.

\begin{figure}
 \begin{center}
  \begin{tabular}{cc}
   \begin{minipage}{0.5\textwidth}
    \vspace{10mm}
    \includegraphics[width=60mm]{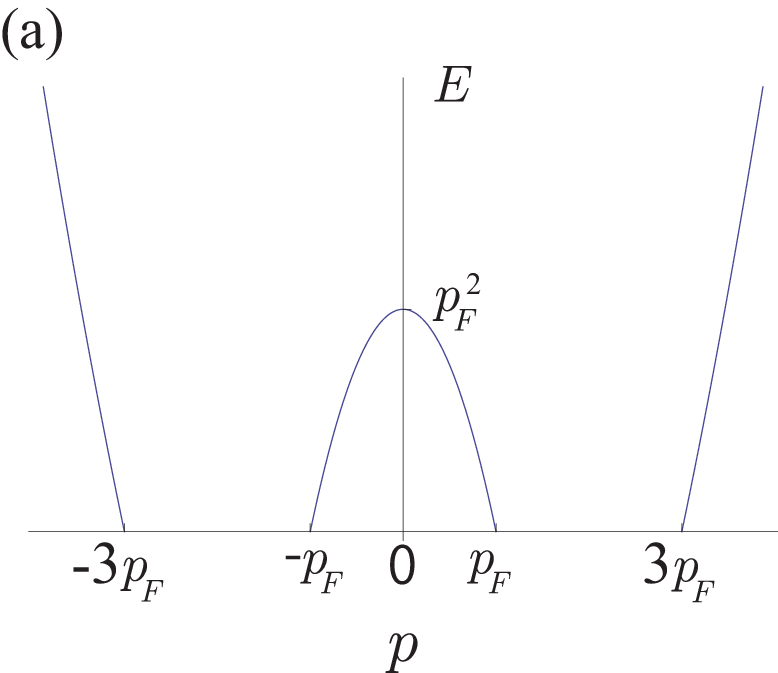}
   \end{minipage}
   & 
   \begin{minipage}{0.5\textwidth}
    \includegraphics[width=60mm]{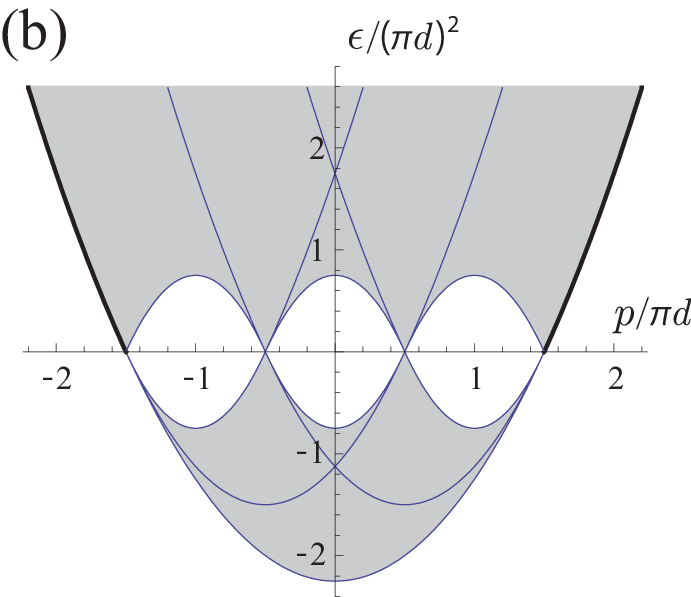} 
   \end{minipage}
  \end{tabular}
 \end{center}
\caption{
 (a) Dispersion relations of a quasi-particle and a quasi-hole for $\lambda=1$, where $p_F=\pi d/2$. (b) The support of the spectral function in energy-momentum space (shaded areas).}\label{fig: dispersion}
\end{figure}

Each part of the particle propagator $G^{(0R)}$, $G^{(0L)}$, and $G^{(1)}$ is written in terms of the spins and the dimensionless momenta of quasi-particles, which are $\tau_{R(L)}(=\pm 1/2)$ and $w_{{\rm R}({\rm L})} (|w_{{\rm R}({\rm L})}|\ge 1)$, and those of quasi-holes, which are $\sigma_j (=\pm 1/2)$ and $u_j$ ($|u_j|\le 1$). The excitation energies and the dimensionful momenta of quasi-particles are, respectively,
\begin{equation}
 E_{\rm qp}(w)
 = (\lambda'p_F)^2\left(w^2 -1\right),
 \quad 
 P_{\rm qp}(w)
 =\lambda'p_F w,
 \label{eq: dispersion-qp}
\end{equation}
and those of quasi-holes are
\begin{equation}
E_{\rm qh}(u)
 =
 \lambda'p_F^2\left(1-u^2\right),
 \quad 
 P_{\rm qh}(u)
 = p_F u,
\label{eq: dispersion-qh}
\end{equation}
where $p_F=\pi d/2$.
The excitation spectra of these two types of excitations are shown in Fig.~\ref{fig: dispersion} (a). 
Note that a quasi-particle with positive (negative) momentum is right-moving (left-moving). The form factors are given by
\begin{eqnarray}
  &F^{(0)}(w)=\left(\frac{|w|-1}{|w|+1}\right)^{\lambda},
\end{eqnarray}
and
\begin{eqnarray}
 &F(\{u_j\},w_{\rm R},w_{\rm L},\{\sigma_j\},\tau_{\rm{R}},\tau_{\rm{L}}) \nonumber\\
 &= 
 \frac{(w_{\rm R}^2-1)^{\lambda}(w_{\rm L}^2-1)^{\lambda}}{(w_{\rm R}-w_{\rm L})^{2\lambda}}
 \prod_{j=1}^{\lambda'}(1-u_j^2)^{-\frac{\lambda}{\lambda'}}
 \prod_{j<k}(u_k-u_j)^{-\frac{2\lambda}{\lambda'}} \nonumber\\
 &\times
 \Bigg(
 \underset{{\rm s.t.}\,
 |Q|=\lambda+1}{\sum_{Q\subset \{1,\cdots, \lambda'\}}}
\varepsilon(Q,\{\sigma_j\})
 \underset{(j,k)\in (Q,Q)\mbox{{\footnotesize or}}(\bar{Q},\bar{Q})}{\prod_{j<k}}
 (u_k-u_j)\,\,\cdot W_{\sigma}^{-1}
 \prod_{j\in Q}\frac{\partial}{\partial u_j}
 W_{\sigma}\Bigg)^2,
\label{eq: def-F}
\end{eqnarray}
where $\varepsilon(Q,\{\sigma_j\})$ denotes a sign factor  
\begin{equation}
\varepsilon(Q,\{\sigma_j\})=(-1)^{|[Q\cap Q_{\sigma}^{(-)}\cap \mbox{{\footnotesize even}} ]\cup
 [\bar{Q}\cap \bar{Q}_{\sigma}^{(-)}\cap\mbox{{\footnotesize odd}}]|}\\
\label{eq: signfactor}
\end{equation}
with $Q_{\sigma}^{(-)}=\{j\in[1,\lambda']|\sigma_j=-1/2\}$, a set of indices of quasi-holes with $-1/2$ spins,
and $W_\sigma$ denotes 
\begin{eqnarray}
&W_{\sigma}
 =\prod_{j<k}(u_k-u_j)^{1-\delta_{\sigma_j\sigma_k}}
 \prod_{j=1}^{\lambda'}
 (w_{\rm R}+u_j)^{-2\lambda(1-\delta_{\sigma_j\tau_{{\rm R}}})}
 (u_j+w_{\rm L})^{-2\lambda(1-\delta_{\sigma_j\tau_{{\rm L}}})}.\nonumber\\
\label{eq: W}
\end{eqnarray}

It is readily seen from the expression of the particle propagator (\ref{eq: G0R-th})-(\ref{eq: th-G1}) that $G^{(\rm 0R)}$ ($G^{(\rm 0L)}$) represents a single excitation of a right- (left-)moving quasi-particle, and $G^{(1)}(x,t)$ represents excitations of a right-moving quasi-particle, a left-moving quasi-particle, and $\lambda'$ excitations of quasi-holes. These three excitation patterns have the same number of particles since a quasi-hole carries a $-1/\lambda'$ charge. The factor $\delta_{\sum_j\sigma_j+\tau_{\rm L}+\tau_{\rm R},-1/2}$ in (\ref{eq: th-G1}) accounts for the conservation law of the $z$-component of the total spin.

\subsection{Spectral Function}

\begin{figure}[t]
 \begin{center}
  \begin{tabular}{cc}
   \begin{minipage}{0.5\textwidth}
    \includegraphics[width=80mm]{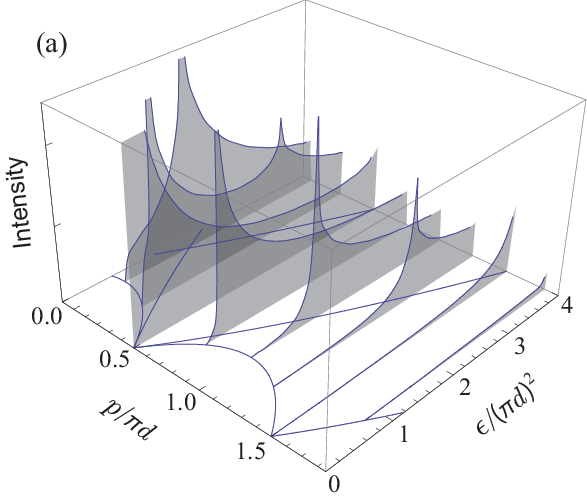}
   \end{minipage}
 & 
   \begin{minipage}{0.5\textwidth}
   \begin{tabular}{cc}
    \begin{minipage}{0.36\textwidth}
     \includegraphics[width=32mm]{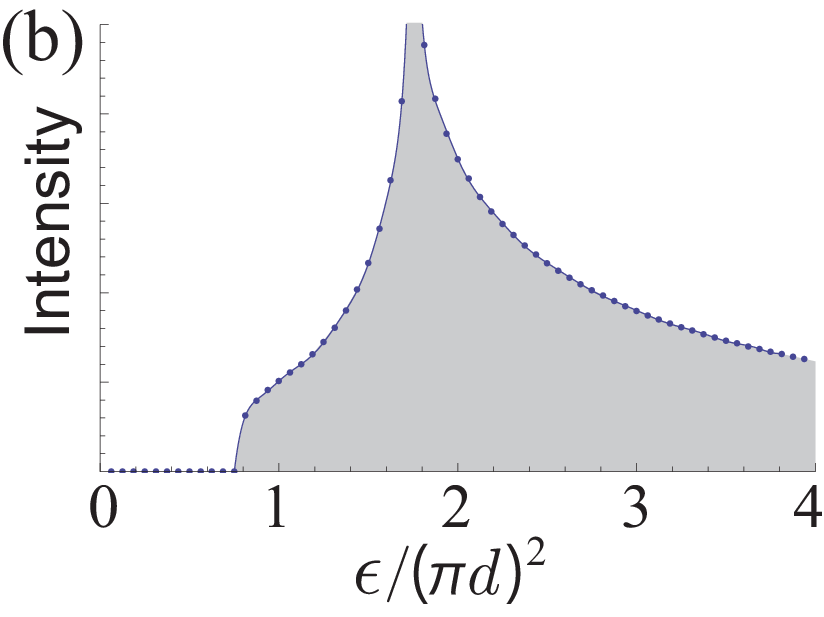}
    \end{minipage}&
    \begin{minipage}{0.36\textwidth}
      \includegraphics[width=32mm]{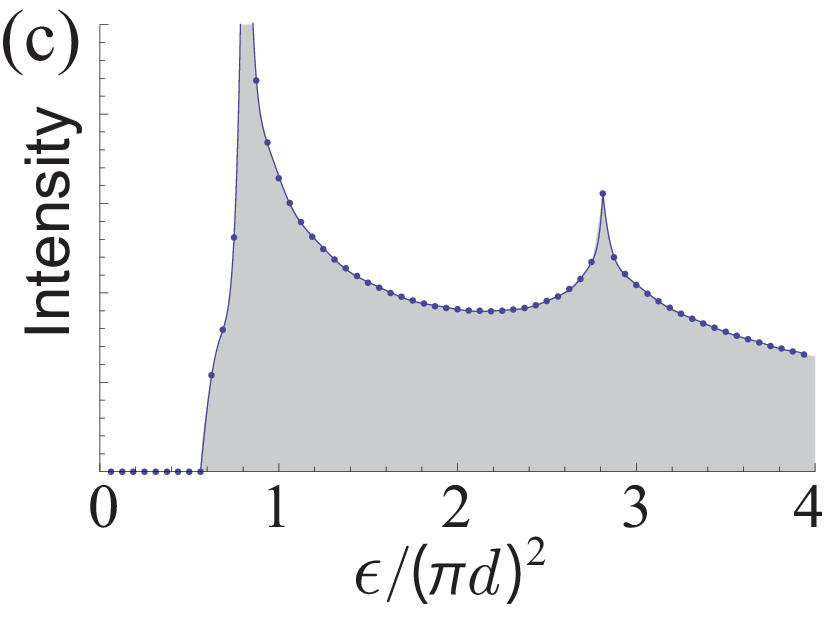}
    \end{minipage}
   \end{tabular}
   \begin{tabular}{cc}
    \begin{minipage}{0.36\textwidth}
     \includegraphics[width=32mm]{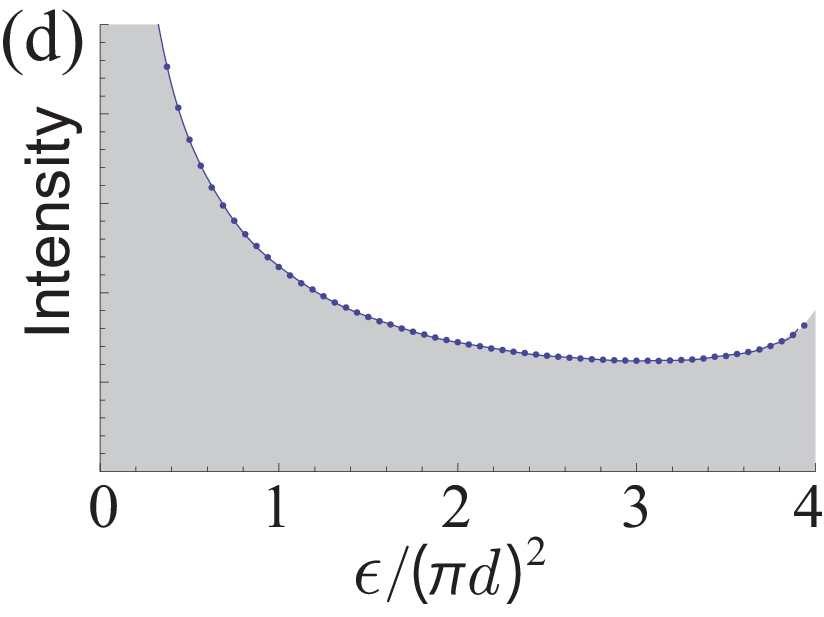}
    \end{minipage}&
    \begin{minipage}{0.36\textwidth}
     \includegraphics[width=32mm]{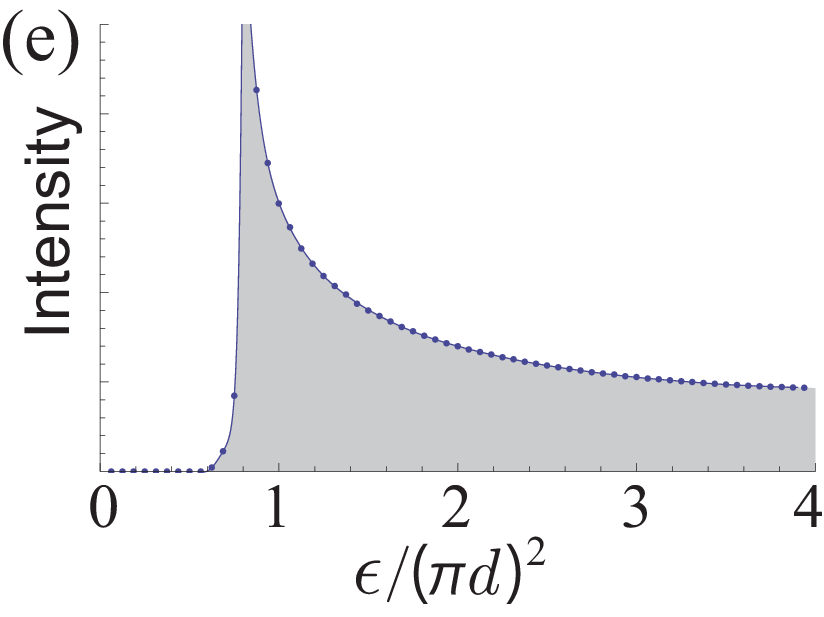}
    \end{minipage}
   \end{tabular}
   \begin{tabular}{cc}
    \begin{minipage}{0.36\textwidth}
     \includegraphics[width=32mm]{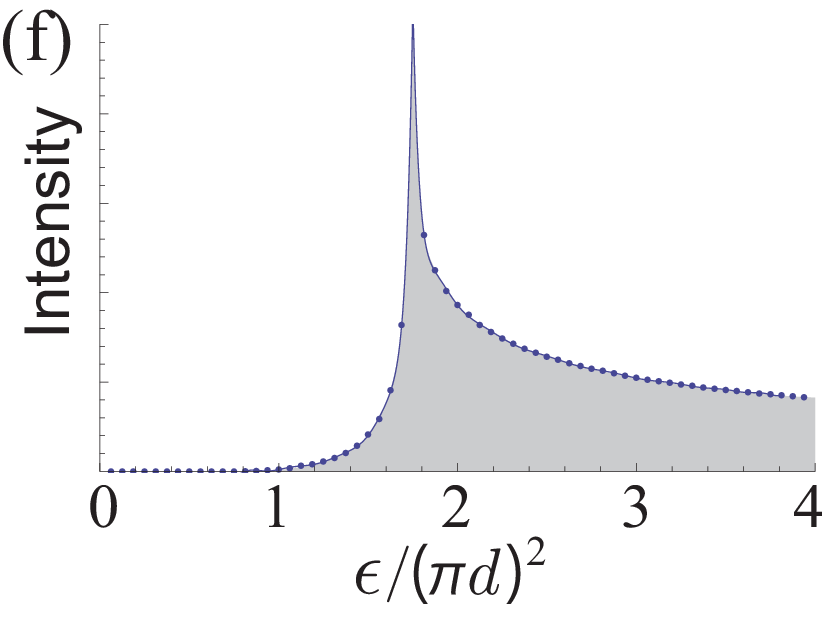}
    \end{minipage}&
    \begin{minipage}{0.36\textwidth}
     \includegraphics[width=32mm]{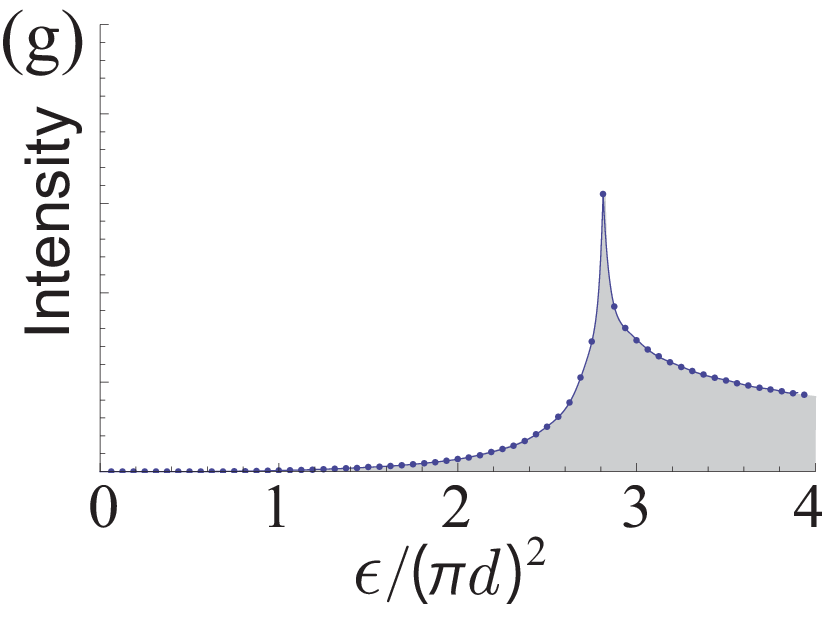}
    \end{minipage}
    \end{tabular}
    \begin{tabular}{cc}
     \begin{minipage}{0.36\textwidth}
      \includegraphics[width=32mm]{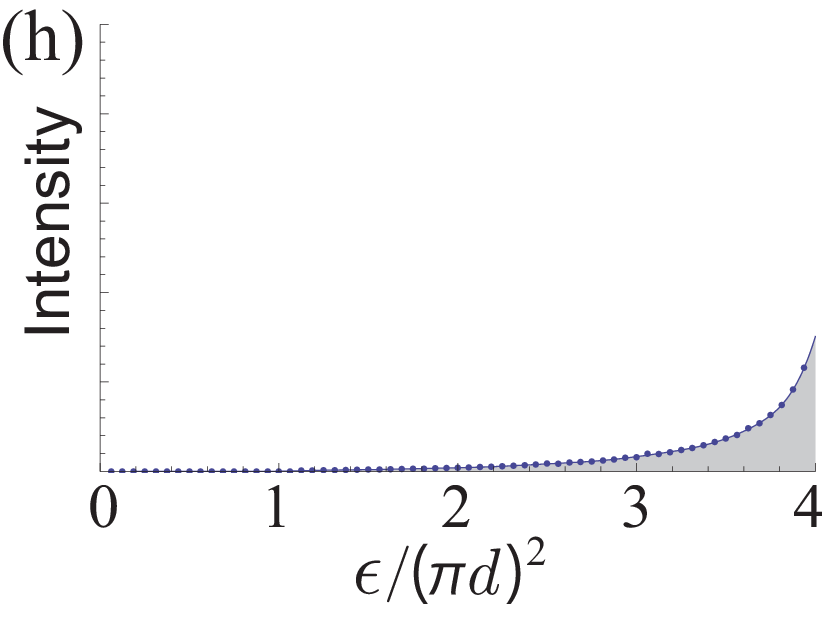}
     \end{minipage}&
     \begin{minipage}{0.36\textwidth}
      \includegraphics[width=32mm]{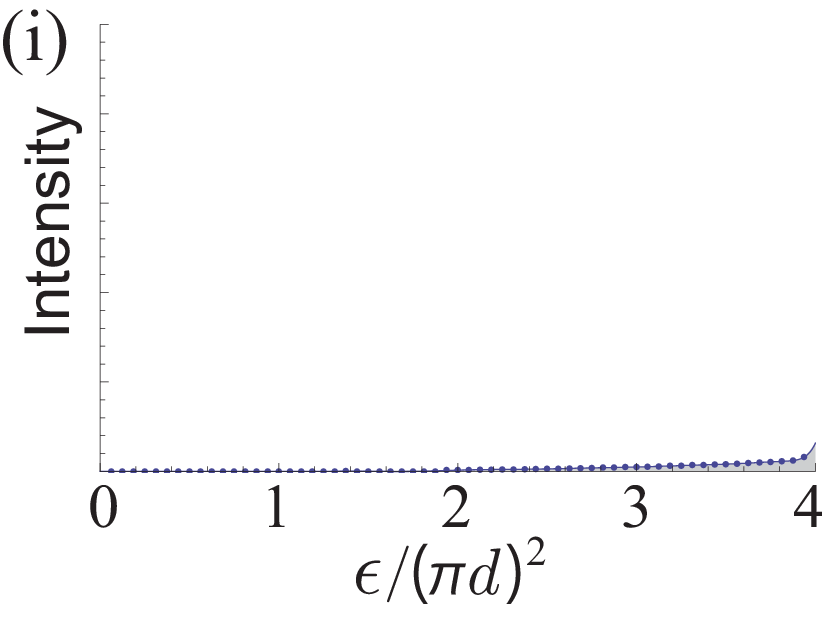}
     \end{minipage}
    \end{tabular}
   \end{minipage}
  \end{tabular}
    \caption{Spectral function of the spin $1/2$ Calogero-Sutherland model for $\lambda=1$. Each graph corresponds to  (a) the spectral function in the energy-momentum space, and a section of the spectral function at $p/\pi d = $ (b) 0, (c) 0.25, (d) 0.5, (e) 0.75, (f) 1, (g) 1.25, (h) 1.5, and (i) 1.75, respectively.}
  \label{sp}
 \end{center}
\end{figure}

In the energy-momentum space, the spectral function is defined by
\begin{eqnarray}
 A^{+}(p,\epsilon)
 &=
 \frac{1}{2\pi}
 \int_{-\infty}^{\infty} {\rm d}x
 \int_{-\infty}^{\infty}{\rm d}t 
 \,
 e^{{\rm i}(\epsilon+\mu)t-{\rm i}px} 
 G^{+}(x,t).\label{eq: A+}
\end{eqnarray}
Sections of the spectral function as a function of the energy $\epsilon$ for some specific values of momenta $p$ are numerically shown in Fig.~\ref{sp} (a)-(i).
A part of the spectral function (\ref{eq: A+}) that results from $G^{\rm (0R)}(x,t)$ and $G^{\rm (0L)}(x,t)$ is given by 
\begin{eqnarray}
 A^{+,0}(p,\epsilon)
 &=
 \left(\frac{|p|-\lambda'p_{\rm F}}{|p|+\lambda'p_{\rm F}}\right)^\lambda
 \delta(\epsilon-p^2+(\lambda'p_{\rm F})^2)
 \theta(|p|-\lambda'p_{\rm F}),
\end{eqnarray}
where $p_{\rm F}=\pi d/2$, and $\theta$ is the step function. $A^{+,0}(p,\epsilon)$ is nonzero only on two semi-infinite lines $\epsilon=p^2-(\lambda'p_{\rm F})^2$ with $|p|\ge \lambda'p_{\rm F}$ (drawn by bold lines at the edge of shaded area in Fig.~\ref{fig: dispersion} (b), but not drawn in Fig.~\ref{sp}), which coincide with the dispersion relation of the quasi-particle (\ref{eq: dispersion-qp}). The remaining part of the spectral function results from $G^{\rm (1)}(x,t)$, which we denote $A^{+,1}(p,\epsilon)$. $A^{+,1}(p,\epsilon)$ is nonzero when $\epsilon\ge p^2-(\lambda'p_{\rm F})^2$ for $|p|\ge \lambda'p_{\rm F}$, and $\epsilon\ge \mbox{Max}\left[\lambda'\left(p_{\rm F}^2-(p-2lp_{\rm F})^2\right)\right]_{l\in [-\lambda,\lambda]}$ for $|p|\le \lambda'p_{\rm F}$. In Fig.~\ref{fig: dispersion} (b), the region on which $A^{+,1}(p,\epsilon)$ is nonzero is shown by a shaded area (we call this region  ``support'') in the upper-half plane of the energy-momentum space. The lower edge of the support for $|p|\ge \lambda'p_{\rm F}$ coincides with the dispersion of the quasi-particle, while that for $|p|\le \lambda'p_{\rm F}$ coincides with shifted dispersions of the quasi-hole.  As a reference, the support of the spectral function of the hole propagator 
\begin{eqnarray}
 A^{-}(p,\epsilon)&=\frac{1}{2\pi}
 \int_{-\infty}^{\infty} {\rm d}x\int_{-\infty}^{\infty}{\rm d}t 
 \,e^{{\rm i}(\epsilon-\mu)t-{\rm i}px} 
 \frac{\langle{\rm g},N|\psi_{\downarrow}^{\dagger}(0,0)
 \psi_{\downarrow}(x,t)|{\rm g},N\rangle} 
 {\langle{\rm g},N|{\rm g},N\rangle}
\end{eqnarray}
is shown by the shaded area in the lower-half plane in Fig.~\ref{fig: dispersion} (b), which results from the hole propagator with reversed position and time. The particle propagator together with the hole propagator leads to the spectral function of the single-particle Green's function in the whole range of the energy-momentum space.

The fact that the form factor $F(\{u_j\},w_{\rm R},w_{\rm L},\{\sigma_j\},\tau_{\rm{R}},\tau_{\rm{L}})$ becomes singular when 
\begin{equation}
\begin{array}{ll}
w_{\rm R}\sim 1,\quad w_{\rm L}\sim -1,\quad u_j\sim \pm -1,&\\ 
u_i\sim u_j\mbox{ when }\sigma_i =-\sigma_j, &\\
w_{\rm R}+u_j\sim 0\mbox{ when }\tau_{\rm R}=-\sigma_j, \\
w_{\rm L}+u_j\sim 0\mbox{ when }\tau_{\rm L}=-\sigma_j. 
\end{array}
\end{equation}
is reflected in singular behavior of $A^{+,1}(\epsilon,p)$ at the boundary of the support and the internal curves shown in Fig.~\ref{fig: dispersion} (b). Power-law singularity near the lower edge is a property characteristic to the Tomonaga-Luttinger liquid. Note that the spectral function in the upper-half plane also has delta-function type divergences at the boundaries of the support, which result from $G^{(\rm 0L)}$ and $G^{(\rm 0R)}$. All the lines in the upper-half plane on which the intensity diverges are continuously connected to those in the lower-half lines at $p=\pm \pi d/2,\pm 3\pi d/2$ (see Fig. 5 in \cite{hp}).

\section{Preliminaries for derivation of the particle propagator in a finite size system}\label{sec: preliminaries}
In this section, fundamental properties of the spin CS model and related mathematical tools are reviewed together with the notations which are used throughout this paper. Most of this section is covered in our previous paper\cite{hp}.

\subsection{Eigenstates}
Introducing the complex variables $z_i=\exp[2\pi{\rm i}x_i/L]$ in place of the spatial coordinate $x_i\in[0,L]$, the exact eigenfunction of the spin CS model (\ref{eq: spinCS}) is written by the product $\Phi_{\kappa,\alpha}\Psi_{0,N}$ of a Jastrow-type ground state wave function
\begin{eqnarray}
 \Psi_{0,N}(z)=\prod_{i=1}^N z_i^{-\lambda(N-1)/2}
  \prod_{i<j}(z_i-z_j)^{\lambda}, 
\end{eqnarray} 
and the Yangian Gelfand-Zetlin basis $\Phi_{\kappa,\alpha}(z,\sigma)$. Here $N$ is the number of particles, $z=(z_1,\cdots,z_N)$ is a set of the coordinate variables, and $\sigma=(\sigma_1,\cdots,\sigma_N)$ is a set of the spin variables\cite{Takemura1997,Uglov1998,hp}. The subscripts $\kappa$ and $\alpha$ in $\Phi_{\kappa,\alpha}(z,\sigma)$ are indices of a momentum state and a spin state of spin 1/2 particles, respectively. The subscript $\kappa$ belongs to the set
\begin{eqnarray}
 &{\cal L}_{N,2}=\left\{ 
 \kappa\in {\cal L}_N|
 \,\forall s\in {\bf{Z}},\sharp\{\kappa_i \: | \: \kappa_i =  s\}  
 \leq 2 \right\}
 \quad &\lambda\mbox{ even or }N\mbox{ odd}\\
 &{\cal L}'_{N,2}=\left\{ \kappa\,
 |\,\kappa+1/2\in 
 {\cal L}_{N,2}\right\} 
 \quad &\lambda\mbox{ odd and }N\mbox{ even},
\end{eqnarray}
where
\begin{equation}
 {\cal L}_{N}=
 \left\{ 
 \kappa\in  {\bf Z}^N|
 \,\kappa_i\ge \kappa_{i+1}\mbox{ for }i\in [1,N-1]\right\},
\end{equation}
and the subscript $\alpha=(\alpha_1,\cdots,\alpha_N)\in W_{\kappa}$ belongs to the set
\begin{equation}
 W_\kappa=\left\{\alpha=(\alpha_1,\cdots,\alpha_N)\in [1,2]^N|\,\alpha_i 
 <\alpha_{i+1} \mbox{  if  }\kappa_i=\kappa_{i+1}\right\}.
\end{equation}
$\alpha_i=1(2)$ corresponds to the spin $1/2(-1/2)$ of the $i$th particle. Thus the eigenvalue of $S^{z}_{\rm tot}$ of a state $\Phi_{\kappa,\alpha}\Psi_{0,N}$ is given by
\begin{equation}
S^{z}_{\rm tot}=\sum_{i=1}^N (3/2-\alpha_i).
\label{eq: Sztot}
\end{equation}

As prerequisites, we introduce the Slater determinant,
\begin{equation}
u_{\kappa,\alpha}={\rm Asym}\left[\,\prod_{i=1}^N z_i^{\kappa_i} \varphi_{\alpha_i}(\sigma_i)\right]
\label{eq:ukappaalpha}
\end{equation}
for a set of momenta $\kappa\in {\cal L}_{N,2}\mbox{ or }{\cal L}'_{N,2}$ and a spin configuration $\alpha\in W_\kappa$. Here a one-particle spin function $\varphi_{\alpha_i}(\sigma_i)$ is given by $\delta_{3/2-\alpha_i,\sigma_i}$. The symbol Asym$[\cdots]$ denotes anti-symmetrization of the function of $z$ and $\sigma$
\begin{eqnarray}
&{\rm Asym}\ f(z_1,\sigma_1,\cdots, z_N,\sigma_N) \nonumber\\
=&\sum_{P\in S_N}\left(-1\right)^{P}f(z_{P(1)},\sigma_{P(1)},\cdots, 
z_{P(N)},\sigma_{P(N)}),
\end{eqnarray}
where $\left(-1\right)^{P}$ is the sign of a permutation $P$ in the symmetric group $S_N$.

In order to define the ordering between the basis functions, 
we introduce dominance partial order\cite{Mac}
\begin{eqnarray}
 \nu\,\ge\,\mu \quad\Leftrightarrow\quad |\nu|=|\mu|
 \,\,\,\,\, \mbox{and}\,\,\,\,\, 
 \forall r>0 \,\,\,,\,\, \sum_{i=1}^r \nu_i\,\ge\,
 \sum_{i=1}^r \mu_i
\end{eqnarray}
between $\nu,\mu\in {\cal L}_{N,2}$ or $\nu,\mu\in {\cal L}'_{N,2}$. 
Further we define the order for spin configurations by
\begin{eqnarray}
 \alpha>\alpha'\,\,\,\Leftrightarrow\,\,\,
 &\sum_{i=1}^N\alpha_i=\sum_{i=1}^N\alpha'_i, \,\, \nonumber\\
 &\mbox{and }\mbox{nonzero }
 \alpha'_i-\alpha_i\,
 \mbox{ at the least $i$ is positive}.
\end{eqnarray}
The order of $(\kappa,\alpha)$ is then defined by
\begin{eqnarray}
 (\kappa,\alpha)>(\kappa',\alpha')\quad\Leftrightarrow\quad
 \,\kappa>\kappa'\, ,
 \,\,\,\,\mbox{or}\quad 
 \kappa=\kappa' \,\,\,\mbox{and}\,\,\, \alpha>\alpha'.
\end{eqnarray} 

Now we are ready to introduce $\Phi_{\kappa,\alpha}(z,\sigma)$, which is uniquely defined by the following two conditions:  
\begin{enumerate}
 \item[(i)] triangularity. 
       $\Phi_{\kappa,\alpha}(z,\sigma)$ is expanded by $u_{\kappa',\alpha'}$
       satisfying $(\kappa',\alpha')\le(\kappa,\alpha)$
\begin{eqnarray}
 \Phi_{\kappa,\alpha}(z,\sigma)=u_{\kappa,\alpha}
 +\sum_{(\kappa',\alpha')(<(\kappa,\alpha))}
 a_{(\kappa,\alpha)(\kappa',\alpha')}u_{\kappa',\alpha'}. \label{YGZcon1}
\end{eqnarray}
 \item[(ii)] orthogonality. Orthogonal with respect to the scalar product $\langle\cdots\rangle_{N,\lambda}$ 
\begin{eqnarray}
 \langle\Phi_{\kappa',\alpha'},
 \Phi_{\kappa,\alpha}\rangle_{N,\lambda}=0
 \quad \mbox{for} \,\,(\kappa',\alpha')\neq(\kappa,\alpha), \label{YGZcon2}
\end{eqnarray}
\end{enumerate}
where the scalar product $\langle\cdots\rangle_{N,\lambda}$ is defined
by a weighted integral 
\begin{eqnarray}
 &\langle \Phi',\Phi \rangle_{N,\lambda}
 = \nonumber\\
 &\,\,\frac{1}{N!} 
 \left[\prod_{i=1}^N
 \oint\frac{{\rm d}z_i}{2\pi \rmi z_i}\,\sum_{\sigma_i}\right]
 \prod_{i\neq j}\left(1-\frac{z_i}{z_j}\right)^{\lambda}
 \overline{\Phi'(z,\sigma)}\,\Phi(z,\sigma) \label{ynorm}
\end{eqnarray}
($\overline{\Phi(z,\sigma)}$
means the complex conjugate of $\Phi(z,\sigma)$).
The scalar product (\ref{ynorm}) is directly related to the usual inner product of wave functions of spin 1/2 particles
\begin{eqnarray}
&\langle\Psi'|\Psi\rangle\nonumber\\
&=
\sum_{\sigma_1=\pm 1/2}\cdots\sum_{\sigma_N=\pm 1/2}\int_0^L{\rm d}x_1\cdots 
\int_0^L{\rm d} x_N 
\overline{\Psi'(\{x_i\},\{\sigma_i\})}
          \Psi(\{x_i\},\{\sigma_i\})\label{eq:physical-norm}
\end{eqnarray}
through the relation $\langle\Psi'|\Psi\rangle=N! L^N\langle\Phi',\Phi\rangle_{N,\lambda}$, where $\Psi=\Phi\Psi_{0,N}$ and $\Psi'=\Phi'\Psi_{0,N}$. The eigenenergy of the Hamiltonian (\ref{eq: spinCS}) which corresponds to the eigenfunction $\Phi_{\kappa,\alpha}\Psi_{0,N}$ is given by
\begin{equation}
E_{N}(\kappa)=(\pi /L)^2\sum_i^{N} 
 \left(2\kappa_i+\lambda (N+1-2i)\right)^2
 \label{eq:ENkappa}.
\end{equation}
The ground state is specified by 
\begin{equation}
\kappa=\kappa^0\equiv\left(\frac{N-2}{4},\frac{N-2}{4},\frac{N-6}{4},\frac{N-6}{4},\cdots,\frac{-N+2}{4},\frac{-N+2}{4}\right) \label{eq:gs-partition}
\end{equation}
and $\alpha=\alpha^0\equiv(1,2,1,2,\cdots,1,2,1,2)$. The total spin of the ground state $S_z(\alpha=\alpha^0)$ is zero.

\subsection{gl$_2$ Jack polynomials and Macdonald polynomials}
An index that specifies a symmetric polynomial, e.g. a wave function of spinless free bosons, is given by a partition, which is a set of non-negative integers arranged in the non-increasing order. The set of partitions with length equal to or shorter than $N$ is denoted by 
\begin{eqnarray}
 \Lambda_N=&\{
 \nu=(\nu_1,\nu_2,\cdots,\nu_N)\in{\bf Z}^N 
 \,| \nu_1\ge\nu_2\ge\cdots\ge\nu_N\ge 0\,\}.
\end{eqnarray}
The monomial symmetric polynomial $m_{\nu}$ ($\nu\in \Lambda_N$) is defined by symmetrization of a monomial $z^{\nu}=z_1^{\nu_1}z_2^{\nu_2}\cdots z_N^{\nu_N}$ as 
\begin{eqnarray}
 m_{\nu}=
 \sum_{\sigma\in S_N}z_1^{\nu_{\sigma(1)}}z_2^{\nu_{\sigma(2)}}
 \cdots z_N^{\nu_{\sigma(N)}},
\end{eqnarray}
where the sum is taken over all permutations of the elements of $\nu$.
The Macdonald polynomial $P_{\nu}(z;q,t)$ for $\nu\in \Lambda_N$ 
is uniquely defined 
by the following two conditions\cite{Mac}:
\begin{enumerate}
 \item[(i)] triangularity. $P_{\nu}(z;q,t)$ is expanded by $m_{\mu}$
       satisfying $\mu\le \nu$
\begin{eqnarray}
 P_{\nu}(z;q,t)=m_{\nu}+\sum_{\mu(<\nu)}v_{\nu\mu}m_{\mu}. \label{maccon1}
\end{eqnarray}
 \item[(ii)] orthogonality. Orthogonal with respect to the scalar product $\langle \cdots \rangle_{N,q,t}$
\begin{eqnarray}
 \langle P_{\mu}(z;q,t),P_{\nu}(z;q,t) \rangle_{N,q,t} = 0
 \quad \mbox{for} \,\,\mu\neq\nu, \label{maccon2}
\end{eqnarray}
\end{enumerate}
where the scalar product in (\ref{maccon2}) is defined by a weighted integral
using the function $(x;q)_{\infty}=\prod_{r=0}^{\infty}(1-xq^r)$, as
\begin{eqnarray}
 \langle f,g \rangle_{N,q,t} =
 \frac{1}{N!} 
 \left[\prod_{i=1}^N
 \oint\frac{{\rm d}z_i}{2\pi iz_i}\right]
 \prod_{i\neq j}
 \frac{\displaystyle (z_i/z_j;q)_{\infty}}
 {\displaystyle (tz_i/z_j;q)_{\infty}}
 \overline{f(z)}\,g(z). \label{mnorm}
\end{eqnarray}
The gl$_2$-Jack polynomials\cite{Uglov1998,Yam1999} are defined from the Macdonald polynomials as  
\begin{equation}
P_{\nu}^{(\lambda')}(z)=
\lim_{q=-p,t=-p^{\lambda'},p\rightarrow 1} P_{\nu}(z;q,t), 
\label{eq:gl2-def}
\end{equation}
with $\lambda'\equiv 2\lambda+1$. 
From (\ref{maccon1}), (\ref{maccon2}) and (\ref{eq:gl2-def}), it follows that 
\begin{enumerate}
 \item[(i)] triangularity. $P_{\nu}^{(\lambda')}(z)$ is expanded by $m_{\mu}$
       satisfying $\mu\le \nu$
\begin{eqnarray}
 P_{\nu}^{(\lambda')}(z)=m_{\nu}+\sum_{\mu(<\nu)}c_{\nu\mu}m_{\mu}.
 \label{gl2con1}
\end{eqnarray}
 \item[(ii)] orthogonality. Orthogonal with respect to the scalar product $\{ \cdots\}_{N,\lambda}$
\begin{eqnarray}
 \{ P_{\mu}^{(\lambda')}
 , P_{\nu}^{(\lambda')}\}_{N,\lambda} = 0
 \quad \mbox{for} \,\,\mu\neq\nu. \label{gl2con2}
\end{eqnarray}
\end{enumerate}
The scalar product in (\ref{gl2con2}) is given by
\begin{eqnarray}
 &\left\{f,g\right\}_{N,\lambda}
 = \nonumber\\
 &\,\,\frac{1}{N!} 
 \left[\prod_{i=1}^N
 \oint\frac{{\rm d}z_i}{2\pi iz_i}\right]
 \prod_{i\neq j}\left(1-\frac{z_i}{z_j}\right)^{\lambda+1}\!\!
 \left(1+\frac{z_i}{z_j}\right)^{\lambda}
 \overline{f(z)}g(z).\label{gnorm}
\end{eqnarray}
The definition of the Macdonald polynomials $P_\nu$ and the gl$_2$-Jack polynomials $P^{(\lambda')}_\nu$ with $\nu\in\Lambda_N$ are extended to those with $\nu\in$ ${\cal L}_N$ (or ${\cal L}_N'$) as follows. When $\nu\in {\cal L}_N$ is written as 
$$\nu=\mu-J=(\mu_1 -J,\cdots,\mu_N-J)$$ 
with an integer (half integer) $J$ and a partition $\mu\in \Lambda_N$, $P_\nu$ and $P^{(\lambda')}_\nu$ is defined as
\begin{equation}
P_\nu(z)\equiv (z_1 \cdots z_N)^{-J}P_\mu(z),\quad
P^{(\lambda')}_\nu(z)\equiv (z_1 \cdots z_N)^{-J}P^{(\lambda')}_\mu(z),
\end{equation}
respectively. 

\subsection{Uglov's mapping of eigenfunctions and the field annihilation operator}
A linear mapping $\Omega$ between the set of functions spanned by $u_{\kappa,\alpha}$ with $(\kappa,\alpha)\in ({\cal L}_{N,2},W_\kappa)$ or $({\cal L}'_{N,2},W_\kappa)$ and the set of symmetric functions is defined as $\Omega(u_{\kappa,\alpha})=s_{\nu}$, where $ s_{\nu}(z)$ denotes the Schur symmetric function with $\nu\in \Lambda_N$, 
\begin{equation}
 s_{\nu}(z)=\frac{\mbox{Asym}\left[z_1^{\nu_1+N-1}z_2^{\nu_2+N-2}
 \cdots z_N^{\nu_N}\right]}
 {\mbox{Asym}\left[z_1^{N-1}z_2^{N-2}\cdots z_N^{0}\right]}.
\end{equation}
The relation between $(\kappa,\alpha)$ and $\nu$ is given by
\begin{eqnarray}
 \nu_i=\alpha_{N+1-i}-2\kappa_{N+1-i}-N+i+K \label{parttrans}
\end{eqnarray}
with an even integer $K$ and this relation for $K=0$ is schematically illustrated in Fig.~1 of \cite{hp}.

The mapping $\Omega$ has the following properties:
\begin{enumerate}
\item[(i)] Isometry. The scalar product is preserved under the mapping $\Omega$. For functions $\Phi'(z,\sigma)$ and $\Phi(z,\sigma)$, the relation 
\begin{equation}
\langle \Phi',\Phi\rangle_{N,\lambda}=
\left\{\Omega(\Phi'),\Omega(\Phi)\right\}_{N,\lambda}
\label{eq:isometry}
\end{equation}
holds.
\item[(ii)]
The correspondence between the Yangian Gelfand-Zetlin basis and the gl$_2$-Jack polynomials 
\begin{equation}
\Omega(\Phi_{\kappa,\alpha})=P^{(\lambda')}_\nu.
\end{equation}
\end{enumerate}
The mapping of the field annihilation operator under $\Omega$ has been given in our previous paper\cite{hp},  and shown briefly in the following. The field annihilation operator of spinless particles and that of spin $1/2$ particles act on wave functions $f(\{x\})$ and $g(\{x\},\{\sigma\})$, respectively, as
\begin{eqnarray}
 &\psi(x)f(x_1,\cdots,x_N)
  =\sqrt{N}\xi^{N-1} f(x_1,\cdots,x_{N-1},x), \\
 &\psi_{s}(x)g(x_1,\sigma_1,\cdots,x_N,\sigma_N)
  =\sqrt{N}\xi^{N-1} g(x_1,\sigma_1,\cdots,x_{N-1},\sigma_{N-1},x,s),
\end{eqnarray}
where $\xi=1(-1)$ for bosons (fermions), $\psi(x)=\psi(x,t=0)$ and $\psi_{s}(x)=\psi_{s}(x,t=0)$. Here we identify $s=\uparrow(\downarrow)$ with $s=1/2(-1/2)$ for a notational convenience. We introduce the similarity-transformed field operators $\tilde{\psi}_s(0,0)$ and $\tilde{\psi}(0,0)$ as
\begin{eqnarray}
 &\tilde{\psi}(0,0)f\equiv
 (\tilde{\Psi}_{0,N-1})^{-1}\,\,\psi(0,0)\,\,f\tilde{\Psi}_{0,N}\label{tildepsi}\\
 &\tilde{\psi}_{s}(0,0)g\equiv
 (\Psi_{0,N-1})^{-1}\,\,\psi_{s}(0,0)\,\,g\Psi_{0,N},
 \quad s=\uparrow\mbox{or}\downarrow \label{tildepsi-s},
\end{eqnarray}
with
\begin{eqnarray}
 \tilde{\Psi}_{0,N}(z)=
 \prod_{i=1}^N z_i^{-\lambda'(N-1)/2}
 \prod_{i<j}(z_i-z_j)^{\lambda+1}(z_i+z_j)^{\lambda}.\nonumber\\
\end{eqnarray}
It follows that
\begin{eqnarray}
 &\!\!\!\!\!
 \tilde{\psi}(0,0)f(z)\nonumber\\
 &=\sqrt{N}\xi^{N-1}f(z_1,\cdots,z_{N-1},z_N=1) 
 \prod_{i=1}^{N-1} z_i^{-\lambda-1/2}
 (z_i-1)^{\lambda+1}(z_i+1)^{\lambda}.\label{action-tilde-psi}
\end{eqnarray}
In \cite{hp}, we showed that 
\begin{eqnarray}
 \Omega\left((\tilde{\psi}_{\uparrow}(0,0)+\tilde{\psi}_{\downarrow}(0,0))
 \Phi
 \right)
 &=(-1)^{(N-1)\lambda}\left(\prod_{i=1}^{N-1}z_i^{1/2}\right)
 \tilde{\psi}(0,0)\Omega(\Phi).\label{eq:tilde-psi-trasf}
\end{eqnarray}

\subsection{Notations of combinatorial quantities}
One of the advantages in using the Uglov's mapping lies in the fact that several useful formulae are available in the theory of the Macdonald polynomials. Those formulae are expressed in terms of the combinatorial quantities related to the Young diagram\cite{Mac}. A partition $\nu\in \Lambda_{N}$ can be graphically expressed by a Young diagram. The Young diagram corresponding to a partition $\nu$ is denoted by $D(\nu)$, in which the number of squares in the $i$th row is equal to the $i$th element $\nu_i$ of the partition $\nu$. Each square in a Young diagram is specified by two-dimensional coordinate with setting the upper-left square $s=(1,1)$ (See e.g. Fig.~2 of \cite{hp}.) The first (second) coordinate of a square $s=(i,j)$ represents the vertical (horizontal) axis and increases from top to bottom (from left to right). Let $\nu'_j$ be the length of the $j$th column. Four functions that measure the lengths between a square $s=(i,j)$ and  edges of the Young diagram are introduced as
\begin{eqnarray*}
&& a(s)=\nu_i-j, \quad l(s)=\nu_j'-i\\
&& a'(s)=j-1, \quad    l'(s)=i-1. 
\end{eqnarray*}
These functions are used to represent the formulae specified by partitions\cite{Mac}. In this paper, we sometimes use the generalized Young diagram in order to express $\nu\in {\cal L}_N$ (e.g. Fig.~\ref{4intermediate} (ii), (iii) and Fig.~\ref{2qp3qh}) when nobody would misunderstand.  

\subsection{Formulae of the gl$_2$ Jack polynomials}\label{sec: formulae}
In this subsection, we summarize two important formulae in the following calculations. The first formula is a kind of `` the binomial formula'' given by
\begin{eqnarray}
 \prod_{i=1}^{N}\, (1-z_i)^{\lambda+1}&(1+z_i)^{\lambda}=
 \!\!\!\!
 \sum_{
  \mu\in\Lambda_{N}\atop
  {{\rm s.t.}\,|C_2(\mu)|+|H_2(\mu)|=|\mu|}}
 \!\!\!\!
b_\mu P_{\mu}^{(\lambda')}(z)  \label{eq: binomial}.
\end{eqnarray}
Here $C_2(\mu)$ and $H_2(\mu)$ are the subsets of $D(\mu)$ defined as
\begin{eqnarray}
 &C_2(\mu)=\{s\in D(\mu)|\,a'(s)+l'(s)\equiv 0\,\,\,\,{\rm mod}\,2\}, \\
 &H_2(\mu)=\{s\in D(\mu)|\,a(s)+l(s)+1\equiv 0\,\,\,\, {\rm mod}\,2\}.
\end{eqnarray}   
An illustration of $C_2(\mu)$ and $H_2(\mu)$ is given in Fig.~3 of \cite{hp}. The coefficient $b_\mu$ is given by
\begin{eqnarray}
b_\mu=(-1)^{|\mu|+\sum l'(s)} 
\displaystyle
 \frac{\displaystyle \prod_{s\in D(\mu)\setminus
 C_2(\mu)}(a'(s)-\lambda'(l'(s)+1))}
 {\displaystyle \prod_{s\in H_2(\mu)}(a(s)+1+\lambda'l(s))} 
 \label{left_exp},
\end{eqnarray}
where the set $A\setminus B$ denotes the complementary set of $B$ in $A$.
The expression (\ref{left_exp}) can be obtained from the formula in \cite{hp} by replacing $N-1$ by $N$.

The second formula is given in the form
\begin{equation}
 P_{\nu}^{(\lambda')}(z_1,\cdots,z_N,1)=\sum_{\mu\in \Lambda_N}\psi_{\nu\mu}^{(\lambda')}
 P_{\mu}^{(\lambda')}(z_1,\cdots,z_N)\label{eq: expansiongl2}
\end{equation}
for $\nu\in \Lambda_{N+1}$. 
The expression for $\psi_{\nu\mu}^{(\lambda')}$ is given by
\begin{eqnarray}
\psi_{\nu\mu}^{(\lambda')}&=&\underset{{\rm s.t}\,s\in {\rm C}_{\nu/\mu}\setminus{\rm R}_{\nu/\mu}}
 {\prod_{s\in H_2(\nu)}}\!\!\!\!
 \left(\frac{a(s)+1+\lambda'l(s)}{a(s)+\lambda'(l(s)+1)}\right)_{\nu}\nonumber\\&\times&
 \underset{{\rm s.t.}\,s\in {\rm C}_{\nu/\mu}\setminus{\rm R}_{\nu/\mu}}
 {\prod_{s\in H_2(\mu)}}\!\!\!\!
 \left(\frac{a(s)+\lambda'(l(s)+1)}{a(s)+1+\lambda'l(s)}\right)_{\mu},\label{eq: psi_numu_CR}
\end{eqnarray}
when $\mu\in \Lambda_N$ and $\nu/\mu$ is a horizontal strip, which means that all the columns of $\nu$ and $\mu$ satisfy $\nu_j'-\mu_j'=0\mbox{ or }1$\cite{Mac}. When $\nu/\mu$ is not a horizontal strip, $\psi_{\nu\mu}^{(\lambda')}$ vanishes. Here ${\rm C}_{\nu/\mu}$ is the set of columns satisfying $\nu_j'-\mu_j'=0$ in $j\in [\nu_{N+1},\lambda']$. The symbol ${\rm R}_{\nu/\mu}$ denotes the set of rows satisfying $\nu_i-\mu_i=0$. The notation $s\in {\rm C}_{\nu/\mu}\setminus{\rm R}_{\nu/\mu}$ means the element $s=(i,j)$ with $j\in {\rm C}_{\nu/\mu}$ and $i\notin {\rm R}_{\nu/\mu}$. The subscript $\mu$ ($\nu$) of the large parenthesis in the right-hand side of (\ref{eq: psi_numu_CR}) means that $a(s)$ and $l(s)$ are evaluated in $D(\mu)$ ($D(\nu)$). 

When $\nu\in \mathcal{L}_{N+1}$ and $\mu\in \mathcal{L}_{N}$, $\psi_{\nu\mu}^{(\lambda')}$ is given by
\begin{equation}
\psi_{\nu,\mu}^{(\lambda')}=\psi_{\nu_+,\mu_+}^{(\lambda')},
\end{equation}
with 
\begin{equation}
\left(\begin{array}{l}
\nu_+=(\nu_1-\nu_{N+1},\nu_2-\nu_{N+1},\cdots,\nu_{N}-\nu_{N+1},0)\in
 \Lambda_{N+1} \\
\mu_+=(\mu_1-\nu_{N+1},\mu_2-\nu_{N+1},\cdots,\mu_{N}-\nu_{N+1})\in
\Lambda_{N}
\end{array}\right. .\end{equation}
The formula (\ref{eq: expansiongl2}) with (\ref{eq: psi_numu_CR}) is derived from the corresponding formula that relates the Macdonald polynomial of $N+1$ variables $(z_1,\cdots,z_N,y)$ to that of $N$ variables $(z_1,\cdots,z_N)$
\begin{eqnarray}
 P_{\nu}(z_1,\cdots,z_N,y;q,t)=\underset{{\rm s.t.}\,\nu/\mu:{\rm h.s.}}
 {\sum_{\mu\in \Lambda_N}}
 &\psi_{\nu\mu}(q,t)y^{|\nu|-|\mu|}P_{\mu}(z_1,\cdots,z_N;q,t)
 \label{macppre},
\end{eqnarray}
for $\nu\in \Lambda_{N+1}$ (see chap.~VI.~6 in \cite{Mac}). The notation ``$\nu/\mu:{\rm h.s.}$'' implies that $\mu$ is taken into account in the summation only when $\nu/\mu$ is a horizontal strip. The expansion coefficient $\psi_{\nu\mu}(q,t)$ is given by
\begin{eqnarray}
 \psi_{\nu\mu}(q,t)&=\sum_{i<j}
 \frac{f(q^{\mu_i-\mu_j}t^{j-i})}{f(q^{\nu_i-\mu_j}t^{j-i})}
 \frac{f(q^{\nu_i-\nu_{j+1}}t^{j-i})}{f(q^{\mu_i-\nu_{j+1}}t^{j-i})}, \\
 &\left(
 \begin{array}{l}
  \displaystyle f(u)=\frac{(tu;q)_{\infty}}{(qu;q)_{\infty}}\\
  \displaystyle (a;q)_{\infty}=\prod_{r=0}^{\infty}(1-aq^r).
 \end{array}
 \right.
\end{eqnarray}
The coefficient $\psi_{\nu\mu}^{(\lambda')}$ is obtained by substituting $y=1$, and taking the limit $q=-p,t=-p^{\lambda'},p\to1$ on both sides of (\ref{macppre}), and it is given by
\begin{eqnarray}
\psi_{\nu\mu}^{(\lambda')}=\underset{{\rm s.t}\,j\in {\rm C}_{\nu/\mu}}
 {\prod_{s=(i,j)\in H_2(\nu)}}\!\!\!\!
 \left(\frac{a(s)+1+\lambda'l(s)}{a(s)+\lambda'(l(s)+1)}\right)_{\nu}
 \underset{{\rm s.t.}\,j\in {\rm C}_{\nu/\mu}}
 {\prod_{s=(i,j)\in H_2(\mu)}}\!\!\!\!
 \left(\frac{a(s)+\lambda'(l(s)+1)}{a(s)+1+\lambda'l(s)}\right)_{\mu},\nonumber\\\label{eq: psi_numu}
\end{eqnarray}
which can be further reduced to (\ref{eq: psi_numu_CR}).

\section{Particle propagator in a finite size system}
\label{sec: finite-size}
In this section, we derive an expression for the particle propagator in a finite-size system with use of the formulae summarized in the previous section. 

\subsection{Mapping of the particle propagator}
In terms of the Yangian Gelfand-Zetlin basis, the particle propagator (\ref{pp}) is rewritten by
\begin{eqnarray}
&G^{+}(x,t)\nonumber\\
&=
 \sum_{\kappa\in \mathcal{L}_{N+1,2}}
 \sum_{\alpha\in W_\kappa}
 \frac{|\langle{\rm g},N|
 \psi_{\downarrow}(0,0)|(\kappa,\alpha),N+1\rangle|^2
 e^{-{\rm i}\omega_{\kappa,\alpha}t+{\rmi}P_{\kappa,\alpha}x}}
 {\langle(\kappa,\alpha),N+1|(\kappa,\alpha),N+1\rangle
 \cdot\langle{\rm g},N|{\rm g},N\rangle}, 
 \label{pp-YGZ}
\end{eqnarray}
where $|(\kappa,\alpha),N+1\rangle$ is the state vector whose wave function is $\Phi_{\kappa,\alpha}(z,\sigma)\Psi_{0,N+1}$, and the complete set of the state vectors with $N+1$ particles is inserted between two operators in the numerator of the right-hand side of (\ref{pp}). Note that only excited states with $S_z^{\rm tot}(\alpha)=-1/2$ contribute to (\ref{pp-YGZ}), since $S_z^{\rm tot}(\alpha^0)=0$ in the ground state. The excitation energy $\omega_{\kappa,\alpha}$ is difference of the energy of an excited state of $N+1$ particles from that of the ground state of $N$ particles given by
\begin{equation}
\omega_{\kappa,\alpha}=E_{N+1}(\kappa)-E_{N}({\rm g}),
\end{equation}
where the eigenenergy is given in (\ref{eq:ENkappa}), and the character g denotes the partition of the ground state given by (\ref{eq:gs-partition}). The total momentum of the excited state is given by
\begin{equation}
P_{\kappa,\alpha}=(2\pi/L)\sum_{i=1}^{N+1} \kappa_i.
\label{Pkappaalpha}
\end{equation}

The matrix element is  transformed by the Uglov's mapping in the same manner as in the preceding paper\cite{hp}. By the isometry of the mapping (\ref{eq:isometry}) and the transformation formula of the field operator (\ref{eq:tilde-psi-trasf}), the particle propagator is described in terms of the gl$_2$-Jack polynomials as
\begin{eqnarray}
 &G^{+}(x,t)=\frac{1}{L(N+1)}{\sum_{\nu\in\mathcal{L}_{N+1},S_{{\rm tot}}^z=-1/2}}\!\!
 \,\,e^{-{\rm i}\tilde{\omega}_{\nu}t+{\rm i}\tilde{P}_{\nu}x} 
 \frac{|\{P_{{\rm g}-1/2}^{(\lambda')},
 \tilde{\psi}(0,0)P_{\nu}^{(\lambda')}\}_{N,\lambda}|^2}
 {\{1,1\}_{N,\lambda}
 \{P_{\nu}^{(\lambda')},P_{\nu}^{(\lambda')}\}_{N+1,\lambda}
 },\nonumber\\
 \label{ppgl2}
\end{eqnarray}
where the Uglov's mapping (\ref{parttrans}) with $K=N/2+\lambda-1$ is used. Substituting the ground state indices $\kappa^{0}=(N/4-1/2,\cdots,-N/4+1/2)$ and $\alpha^{0}=(1,2,\cdots,1,2)$ into (\ref{parttrans}), the gl$_2$-Jack polynomial corresponding to the ground state is $P_{{\rm g}}^{(\lambda')}(z)=\prod_{i=1}^N z_i^{\lambda+1}$. Here, $P_{{\rm g}-1/2}^{(\lambda')}$ stands for $ P_{{\rm g}}^{(\lambda')}(z)\prod_{i=1}^N z_i^{-1/2}=\prod_{i=1}^N z_i^{\lambda'/2}$. As for a state of $N+1$ particles, the set of indices $(\kappa,\alpha)$ is transformed to $\nu$ by (\ref{parttrans}) replacing $N$ by $N+1$ as
\begin{equation} 
 \nu_i=\alpha_{N+2-i}-2\kappa_{N+2-i}-(N+1)+i+(N/2+\lambda-1). \label{nui-def}
\end{equation}
The excitation energy and the total momentum in terms of $\nu$ and the spin of the $i$th particle $\sigma^{\rm p}_i=3/2-\alpha_{N+2-i}$ are, respectively, 
\begin{equation}
\omega_{\kappa,\alpha}=\tilde{\omega}_{\nu}=(\pi/L)^2\sum_{i=1}^{N+1}
 (\nu_i+\lambda'(N+1-2i)/2+\sigma^{\rm p}_i)^2 -E_N({\rm g})
\label{tildeomeganu}
\end{equation}
and
\begin{equation}
P_{\kappa,\alpha}=\tilde{P}_{\nu}=-(\pi/L)\sum_{i=1}^{N+1}
 (\nu_i+\lambda'(N+1-2i)/2+\sigma^{\rm p}_i).
\end{equation}
Note that, from (\ref{nui-def}) and the definition of $\sigma^{\rm p}$, $\sigma^{\rm p}_i=1/2(-1/2)$ when $\nu_i-i$ is even (odd). The total spin is
\begin{equation}
 S^z_{\rm tot}=\sum_{i=1}^N \sigma^{\rm p}_i. \label{eq:total_spin}
\label{eq: sztotal}
\end{equation}
We denote the spin of the $i$th particle $\sigma^{\rm P}_i$ (instead of $\sigma_i$); we will reserve the notation $\sigma_j$ for the spin variables of quasi-holes.

\subsection{Combinatorial description of the particle propagator}

In this subsection, we reduce (\ref{ppgl2}) to a combinatorial expression. 
$\tilde{\psi}(0,0)P_{\nu}^{(\lambda')}$ in the numerator of (\ref{ppgl2})
is decomposed into the product of two factors; 
one originates from a gl$_2$-Jack polynomial $P_{\nu}^{(\lambda')}(z,1)$
 of $N+1$ particles with one of the variables fixed,
and the other from the ground state wave function of $N+1$ particles 
with one of the variables fixed. 
The numerator of the summand in (\ref{ppgl2}) is thus expressed as
\begin{eqnarray}
 &\{P_{{\rm g}-1/2}^{(\lambda')},
 \tilde{\psi}(0,0)P_{\nu}^{(\lambda')}\}_{N,\lambda}\nonumber\\
 &=\sqrt{N+1}\xi^N
 \left\{
 \prod_{i=1}^N z_i^{\lambda'/2},
 \prod_{i=1}^Nz_i^{-\lambda'/2}(z_i-1)^{\lambda+1}
 (z_i+1)^{\lambda}
 P_{\nu}^{(\lambda')}(z_1,\cdots,z_N,1)\right\}_{N,\lambda}\nonumber\\
 &=\sqrt{N+1}\xi^N
 \left\{
 \prod_{i=1}^N (1-z_i)^{\lambda+1}(1+z_i)^{\lambda}
 ,
 P_{\nu}^{(\lambda')}(z_1,\cdots,z_N,1)\right\}_{N,\lambda}. \label{ppnum}
\end{eqnarray}
With use of the formulae in section \ref{sec: formulae}, the particle propagator is rewritten as 
\begin{eqnarray}
 &G^+(x,t)=\frac{1}{L}\!\!\underset{{\rm s.t.}\,S_{{\rm tot}}^z=-1/2}
 {\sum_{\nu\in\mathcal{L}_{N+1}}}
 e^{-{\rm i}\tilde{\omega}_{\nu}t+{\rm i}\tilde{P}_{\nu}x} \cdot
 \frac{\left(\sum'_{\mu\in \Lambda_{N}}b_\mu \psi^{(\lambda')}_{\nu\mu}\{P^{(\lambda')}_\mu,P^{(\lambda')}_\mu\}_{N,\lambda}\right)^2}{\{1,1\}_{N,\lambda}\{P^{(\lambda')}_\nu,P^{(\lambda')}_\nu\}_{N+1,\lambda}}.
\label{ppbpshipmupmu}
\end{eqnarray}
The sum $\sum'_{\mu\in \Lambda_{N}}$with respect to $\mu$ is taken over the partitions satisfying $|C_2(\mu)|+|H_2(\mu)|=|\mu|$. The scalar products in (\ref{ppbpshipmupmu}) are $\{1,1\}_{N,\lambda}=c^{(\lambda',2)}_N$,
\begin{equation}
\{P^{(\lambda')}_\mu,P^{(\lambda')}_\mu\}_{N,\lambda}=c^{(\lambda',2)}_N\frac{Y_\mu(1/2)Z_\mu(1/(2\lambda'))}{Y_\mu(1/(2\lambda'))Z_\mu(1/2)},
\end{equation}
and 
\begin{equation}
\{P^{(\lambda')}_\nu,P^{(\lambda')}_\nu\}_{N+1,\lambda}=c^{(\lambda',2)}_{N+1} \frac{Y'_\nu(1)Z_\nu(1/(2\lambda'))}{Y'_\nu((\lambda'+1)/(2\lambda'))Z_\nu(1/2)},\label{norm-nu-def}
\end{equation}
where the definitions of $c^{(\lambda',2)}_N$ and $c^{(\lambda',2)}_{N+1}$ are given in (74) in \cite{hp}. The symbols $Y_{\mu}(r)$, $Y'_{\nu}(r)$ and $Z_{\nu}(r)$ are defined as
\begin{eqnarray}
 &Y_\nu(r) \equiv \prod_{s\in C_2(\nu)}
  \left(\frac{a'(s)}{2\lambda'}+r +\frac{N-1-l'(s)}{2}\right),\label{eq:Y-def}\\
 &Y'_\nu(r) \equiv \prod_{s\in D(\nu)\setminus C_2(\nu)}
  \left(\frac{a'(s)}{2\lambda'}+r +\frac{N-1-l'(s)}{2}\right),\label{eq:Yp-def}\\
 &Z_\nu(r) \equiv \prod_{s\in H_2(\nu)}\left(\frac{a(s)}{2\lambda'}+r+\frac{l(s)}{2}\right),
 \label{eq:Z-def}
\end{eqnarray}
respectively. $Y'_{\nu}(r)$ is used instead of $Y_{\nu}(r)$ when $2\lambda'\times r$ is even.

\subsection{Particle propagator in terms of rapidities and spins of elementary excitations}\label{sec: pp-ee}
The matrix element in (\ref{ppbpshipmupmu}) is nonzero only for a certain class of excited states $\nu$. Taking account of the selection rule, the expression for the particle propagator can be reduced so that character of underlying elementary excitations becomes manifest. 

\subsubsection{Selection rule}

Since the formula (\ref{eq: binomial}) contains the expansion over partitions, we only consider $\mu\in\Lambda_{N}\subset\mathcal{L}_N$. Furthermore in (\ref{eq: binomial}) the sum over $\mu$ is restricted to the partitions that do not have the square $s=(1,\lambda'+1)$, that is, $\mu$ has at most $\lambda'$ columns.

Another restriction on $\mu$ comes from the relation $|C_2(\mu)|+|H_2(\mu)|=|\mu|$, which can be described as the condition that the number of columns such that $\mu_j'-j$ is even is $\lambda$ or $\lambda+1$\cite{hp} (we refer to the set of these columns as $Q_{\mu}$ and refer to the number of them as $n_{\mu}$, adhering to \cite{hp}. )
 
\begin{figure}[t]
 \begin{center}
  \input{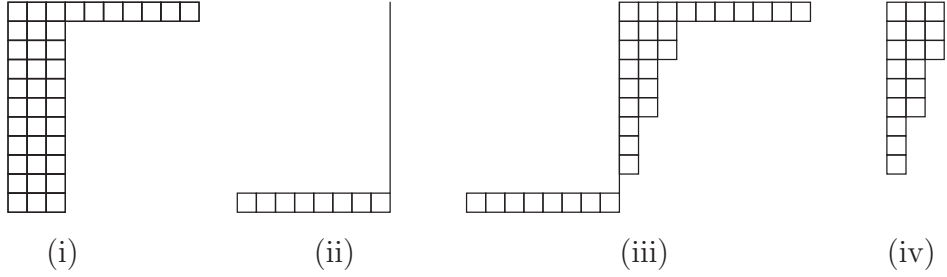} 
 \end{center}
 \caption{Generalized Young diagrams for typical intermediate states
 $\nu$ for $\lambda=1$.}
 \label{4intermediate}
\end{figure}

The condition $\psi_{\nu\mu}^{(\lambda')}\ne 0$ imposes a restriction on $\nu=(\nu_1,\cdots,\nu_N,\nu_{N+1})$ that  $\nu'_j-\mu'_j=0\mbox{ or }1$. Therefore $\nu$ satisfies
\begin{eqnarray}
 &\nu_2\le\nu_1\le\infty, \nonumber\\
 &0\le\nu_N\le\nu_{N-1}\le\cdots\le\nu_3\le\nu_2\le \lambda', \label{nu-restriction}\\
 &-\infty\le\nu_{N+1}\le\nu_{N}. \nonumber
\end{eqnarray}
We classify excited states $\nu$ satisfying (\ref{nu-restriction})
into four types of states (Fig.~\ref{4intermediate}).

\begin{enumerate}
\item One left-moving quasi-particle (0L) states specified by $\nu$ with
\begin{equation}
 \nu_1>\nu_2=\cdots=\nu_{N+1}=\lambda',
\end{equation}
\item one right-moving quasi-particle (0R) states specified by $\nu$ with
\begin{equation}
 \nu_2=\cdots=\nu_{N}=0>\nu_{N+1},
\end{equation}
\item states of one right-moving quasi-particle, one left-moving quasi-particle and $\lambda'$ quasi-holes specified by $\nu$ satisfying
\begin{equation}
\nu_1>\lambda'\ge\nu_2\ge \cdots\ge \nu_{N}\ge 0>\nu_{N+1},\label{eq: 0L0Rlambdahole}
\end{equation}
\item and the other states with $\nu$ satisfying 
\begin{equation}
 \nu_{N+1}\in [0,\lambda'),\quad\mbox{ or }\nu_1\in (0,\lambda'].
\end{equation}
\end{enumerate} 
The states (iv) do not contribute to the thermodynamic limit and we do not consider them. The contributions from (i), (ii) and (iii) to the particle propagator are denoted, respectively, by $G^{\rm (0L)}$, $G^{\rm (0R)}$ and $G^{(1)}$, expressions of which are derived in the following subsections.

\subsubsection{Derivation of $G^{(0{\rm L})}$}
First, we derive $G^{(0{\rm L})}$, which corresponds to a contribution from the one-left-moving quasi-particle (0L) states
\begin{equation}
\nu=(\rho_{\rm L}+\lambda',\overbrace{\lambda',\cdots,\lambda'}^{N}),\quad \rho_{\rm L}>0. 
\label{eq: nu_s}
\end{equation}
From  (\ref{tildeomeganu}), the eigenenergy of the state (\ref{eq: nu_s}) is 
\begin{eqnarray}
E_{N+1}&=&\left(\frac{\pi}{L}\right)^2\left(\rho_{\rm L}+\lambda'(N+1)/2+\sigma^{\rm p}_1\right)^2\nonumber\\&+&\left(\frac{\pi}{L}\right)^2\sum_{i=2}^{N+1}\left(\lambda'(N+3-2i)/2+\sigma^{\rm p}_i\right)^2\label{eq: EN+1-0L},
\end{eqnarray}
where the spins of the 0L states are, by definition of $\sigma^{\rm p}_{i}$ (see the sentence above (\ref{eq:total_spin})), $\sigma^{\rm p}_i=-1/2$ for even $i \in [2,N]$, and $\sigma^{\rm p}_i=+1/2$ for odd $i \in [3,N+1]$.
The second term of the right-hand side of (\ref{eq: EN+1-0L}) coincides with the ground state energy of $N$ particles. Now we introduce the spin and the momentum of the left-moving quasi-particle as
\begin{equation}
\tau_{\rm L}=\sigma_1^{\rm p},\quad  \tilde{\rho}_{\rm L}=\rho_{\rm L}+\lambda'(N+1)/2.
\label{eq: tauLrhoL}
\end{equation}
Then we obtain
\begin{equation}
E_{N+1}-E_N({\rm g})=\left(\frac{\pi}{L}\right)^2\left(\tilde{\rho}_{\rm L}+\tau_{\rm L}\right)^2.
\end{equation}
 Similarly, the momentum for the 0L states is given by
\begin{equation}
P=-\left(\frac{\pi}{L}\right)\left(\tilde{\rho}_{\rm L}+\tau_{\rm L}\right). \end{equation}
Since the 0L states relevant to (\ref{ppbpshipmupmu}) have the total spin $S^{\rm tot}_z=-1/2$, the spin of the quasi-particle is $\tau_{\rm L}=-1/2$ and $\rho_{\rm L}$ is an odd integer.

In the following part, we show
\begin{eqnarray}
&&G^{(\rm 0L)}(x,t)\nonumber\\
&&=\frac{1}{L}
 \sum_{\rho_{\rm L}=1,3,5,\cdots}
 \exp\left[-{\rm i}\left(\frac{\pi}{L}\right)^2\left(\tilde{\rho}_{\rm L}-\frac12\right)^2 t-{\rm i}\left(\frac{\pi}{L}\right)\left(\tilde{\rho}_{\rm L}-\frac12\right)x\right] F^{(\rm 0L)}\nonumber\\
\label{eq: G0L-finite}
 \end{eqnarray}
with
\begin{eqnarray}
F^{(\rm 0L)}&=&\frac{\Gamma\left((\rho_{\rm L}+1+\lambda' N)/2)\right)\Gamma\left((\rho_{\rm L}+\lambda' )/2)\right)}{\Gamma\left(\rho_{\rm L}+\lambda'(N+1))/2\right)\Gamma\left((\rho_{\rm L}+1)/2\right)}\nonumber\\
&& \nonumber\\
&=&
\frac{\Gamma\left(\left(\tilde{\rho}_{\rm L}-\tilde{\rho}_{{\rm R},0} +1\right)/2)\right)\Gamma\left(\left(\tilde{\rho}_{\rm L}-\tilde{\rho}_{{\rm L},0}\right)/2\right)}{\Gamma\left(\left(\tilde{\rho}_{\rm L}-\tilde{\rho}_{{\rm R},0} +\lambda'\right)/2\right)\Gamma\left(\left(\tilde{\rho}_{\rm L}-\tilde{\rho}_{{\rm L},0}+1-\lambda'\right)/2\right)}\label{eq: F0L}.
\end{eqnarray}
Here we have introduced the notations $\tilde{\rho}_{{\rm  L},0}=-\tilde{\rho}_{{\rm  R},0}=\lambda'(N-1)/2$ for convenience. 

First we show that for $\nu$ representing a (0L) state, the expression (\ref{ppbpshipmupmu}) reduces to $G^{(\rm 0L)}(x,t)$ with 
\begin{eqnarray}
 &G^{(\rm 0L)}(x,t)\equiv\frac{1}{L}\frac{c_N^{(\lambda',2)}}{c_{N+1}^{(\lambda',2)}}
 \!\!\underset{{\rm s.t.}\,\tau_{\rm L}=-1/2}
 {\sum_{\nu\in\mbox{\scriptsize{0L}}}}
 e^{-{\rm i}\tilde{\omega}_{\nu}t+{\rm i}\tilde{P}_{\nu}x} \cdot
 \frac{Y'_{\nu}(1/2+1/(2\lambda'))}{Y'_{\nu}(1)}
 \frac{Z_{\nu}(1/2)}{Z_{\nu}(1/(2\lambda'))},\nonumber\\
 \label{ppVXYZ-0L}
\end{eqnarray}
where we have used 0L in (\ref{ppVXYZ-0L}) as the set of the 0L states.
For $\nu=(\rho_{\rm L}+\lambda',\overbrace{\lambda',\cdots,\lambda'}^{N})$, 
\begin{equation}
\mu=(\overbrace{\lambda',\cdots,\lambda'}^{N})\label{mu0R} 
\end{equation}
is the only partition such that $\nu_/\mu$ is a horizontal strip and the relation $|C_2(\mu)|+|H_2(\mu)|=|\mu|$ holds. 
For those $\nu$ and $\mu$, ${\rm C}_{\nu/\mu}=\emptyset$ and hence $\psi_{\nu\mu}^{(\lambda')}=1$. The expansion coefficient $b_\mu$ in (\ref{left_exp}) for (\ref{mu0R}) is $(-1)^{N(\lambda+1)}=1$ because the coefficient of $\prod_{i=1}^N z_i^{\lambda'}$ in the left-hand side of (\ref{eq: binomial}) is $(-1)^{N(\lambda+1)}$ and the monomial $\prod_{i=1}^N z_i^{\lambda'}$ in the right-hand side appears only in $P^{(\lambda')}_{\mu}$ with $\mu$ in (\ref{mu0R}). Furthermore, (\ref{mu0R}) is a Galilean shifted partition of $(0,\cdots,0)$ and thus $\{P^{(\lambda')}_{\mu},P^{(\lambda')}_{\mu}\}_{N,\lambda}=c^{(\lambda',2)}_N$. 
From the above consideration, we arrive at (\ref{ppVXYZ-0L}). The remaining task is to evaluate the factor $\frac{Y'_{\nu}(1/2+1/(2\lambda'))}{Y'_{\nu}(1)}\frac{Z_{\nu}(1/2)}{Z_{\nu}(1/(2\lambda'))}$, which results from the scalar product $\{P^{(\lambda')}_{\nu},P^{(\lambda')}_{\nu}\}_{N+1,\lambda}$ of the (0L) states. This factor can be evaluated through $\frac{Y'_{\nu_{-}}(1/2+1/(2\lambda'))}{Y'_{\nu_-}(1)}\frac{Z_{\nu_-}(1/2)}{Z_{\nu_-}(1/(2\lambda'))}$ with  $\nu_-=(\rho_{\rm L},0,\cdots,0)$ since the scalar product is invariant under the Galilean shift. $l'(s)=0$ in $\nu_-$ and hence $s=(1,j)\in D(\nu_-)\setminus C_2(\nu_-)$ when $j$ is even. Maximum value of $j$ is $\rho_{\rm L}-1$ since $\rho_{\rm L}$ is odd.  $Y'_{\nu_-}(r)$ is expressed as
 \begin{eqnarray}
 Y'_{\nu_-}(r)&=&\prod_{j\in \{2,4,\cdots\}}^{\rho_{\rm L}-1}\left(\frac{j-1}{2\lambda'}+r+\frac{N-1}{2}\right)\nonumber\\
 &=&(\lambda')^{-(\rho_{\rm L}-1)/2}\frac{\Gamma\left(\rho_{\rm L}/2+\lambda'(r+(N-1)/2)\right)}{\Gamma\left(1/2+\lambda'(r+(N-1)/2)\right)}.\label{eq: Ynus-}
 \end{eqnarray}
Since the squares $s\in H_{2}(\nu)\cup D(\nu_{-})$ are parameterized as $s=(1,j)$ with $j=2,4,\cdots,\rho_{\rm L} -1$, we obtain
\begin{equation}
Z_{\nu_{-}}(r)=(\lambda')^{-(\rho_{\rm L}-1)/2}\frac{\Gamma(\rho_{\rm L}/2+\lambda' r)}{\Gamma(1/2+\lambda'r)}.
\label{eq: Znus-} 
\end{equation}
From (\ref{eq: Ynus-}) and (\ref{eq: Znus-}), it follows that
\begin{eqnarray}
\frac{Y'_{\nu}(1/2+1/(2\lambda'))}{Y'_{\nu}(1)}
 \frac{Z_{\nu}(1/2)}{Z_{\nu}(1/(2\lambda'))} 
&=&\frac{\Gamma((1+(N+1)\lambda')/2)}{\Gamma(1+N\lambda'/2)\Gamma((1+\lambda')/2)}\nonumber\\
&\times&\frac{\Gamma\left((\rho_{\rm L}+1+\lambda' N)/2)\right)\Gamma\left((\rho_{\rm L}+\lambda' )/2)\right)}{\Gamma\left(\rho_{\rm L}+\lambda'(N+1))/2\right)\Gamma\left(\rho_{\rm L}+1)/2\right)}.\nonumber\\
\end{eqnarray}
From this and (\ref{ppVXYZ-0L}), we obtain (\ref{eq: G0L-finite}). To evaluate the overall factor, we have used the relation
\begin{equation}
\frac{c_N^{(\lambda',2)}}{c_{N+1}^{(\lambda',2)}}\frac{\Gamma((1+(N+1)\lambda')/2)}{\Gamma(1+N\lambda'/2)\Gamma((1+\lambda')/2)}=1.\label{eq: K0L}
\end{equation}

\subsubsection{Derivation of $G^{(0{\rm R})}$}
A contribution from the one-right-moving quasi-particle (0R) states can be derived in the same manner as in the previous subsection. The (0R) states are specified by 
\begin{equation}
\nu=(\overbrace{0,\cdots,0}^{N},-\rho_{\rm R}),\quad \rho_{\rm R}>0. \label{eq: nu_t}
\end{equation}
The spin of each particle is $\sigma^{\rm p}_i=-1/2$ for odd $i \in [1,N-1]$ and  $\sigma^{\rm p}_i=+1/2$ for even $i \in [2,N]$. The energy and the momentum corresponding to (\ref{eq: nu_t}) are given by
\begin{eqnarray}
&&\tilde{\omega}_\nu=E_{N+1}-E_N({\rm g})=\left(\frac{\pi}{L}\right)^2 \left(\tilde{\rho}_{\rm R}+\tau_{\rm R}\right)^2\label{eq: 0R-energy}\\
&&\tilde{P}_\nu=-\frac{\pi}{L}\left(\tilde{\rho}_{\rm R}+\tau_{\rm R}\right)\label{eq: 0R-momentum}
\end{eqnarray}
with 
\begin{equation}
\tilde{\rho}_{\rm R}=\nu_{N+1}-\frac{\lambda'(N+1)}{2}=-\left(\rho_{\rm R}+\frac{\lambda'(N+1)}{2}\right),
\quad \tau_{\rm R}=\sigma_{N+1}^{\rm p}
. \label{eq: rhoRtauR}
\end{equation}
From the condition $S^{\rm tot}_z=-1/2$,  the spin of the right-moving quasi-particle is fixed to be $\tau_{\rm R}=-1/2$ and hence $\rho_{\rm R}$ is even.  $G^{(0{\rm R})}$ is described in terms of $\rho_{\rm R}$ or $\tilde{\rho}_{\rm R}$, as
\begin{eqnarray}
&&G^{(\rm 0R)}(x,t)\nonumber\\
&&=\frac1L
 \sum_{\rho_{\rm R}=2,4,6,\cdots}
 \exp\left[-{\rm i}\left(\frac{\pi}{L}\right)^2\left(\tilde{\rho}_{\rm R}-\frac12\right)^2 t-{\rm i}\left(\frac{\pi}{L}\right)\left(\tilde{\rho}_{\rm R}-\frac12\right)x\right] F^{(\rm 0R)}\nonumber\\ 
&&\label{ppVXYZ-0R-finite-final}
 \end{eqnarray}
with
\begin{eqnarray}
F^{(\rm 0R)}&=&
\frac{\Gamma\left(\left(\rho_{\rm R}+2+\lambda' N\right)/2\right)
\Gamma\left(\left(\rho_{\rm R}+1+\lambda' \right)/2\right)}{
\Gamma\left(\left(\rho_{\rm R}+1+\lambda'(N+1)\right)/2\right)
\Gamma\left(\left(\rho_{\rm R}+2\right)/2\right)}.\nonumber\\
&=&
\frac{
\Gamma\left(\left(\tilde{\rho}_{{\rm L},0}-\tilde{\rho}_{\rm R}+2\right)/2\right)
\Gamma\left(\left(\tilde{\rho}_{{\rm R},0}-\tilde{\rho}_{\rm R}+1\right)/2\right)}{
\Gamma\left(\left(\tilde{\rho}_{{\rm L},0}-\tilde{\rho}_{\rm R} +\lambda'+1\right)/2\right)
\Gamma\left(\left(\tilde{\rho}_{{\rm R},0}-\tilde{\rho}_{\rm R}+2-\lambda'\right)/2\right)}\label{eq: F0R}.
\end{eqnarray}

We consider $\psi_{\nu+\rho_{\rm R},\mu+\rho_{\rm R}}^{(\lambda')}$ instead of $\psi_{\nu\mu}^{(\lambda')}$ because $\nu$ is not a partition. $\mu=0^N=(\overbrace{0,\cdots,0}^{N})$ is the only partition such that $0\le \mu_{i}\le \lambda'$ for $i\in[1,N]$ and $(\nu+\rho_{\rm R})/(\mu+\rho_{\rm R})$ is a horizontal strip. This $\mu$ obviously satisfies $|C_2(\mu)|+|H_2(\mu)|=|\mu|$. ${\rm C}_{(\nu+\rho_{\rm R})/(\mu+\rho_{\rm R})}$ is thus $\emptyset$ and $\psi^{(\lambda')}_{(\nu+\rho_{\rm R})/(\mu+\rho_{\rm R})}=1$. Consequently, for $\nu$ representing the (0R) states,  (\ref{ppbpshipmupmu}) reduces to 
\begin{eqnarray}
 &G^{(\rm 0R)}(x,t)\equiv\frac{1}{L}\frac{c_N^{(\lambda',2)}}{c_{N+1}^{(\lambda',2)}}
 \!\!\underset{{\rm s.t.}\,\tau_{\rm R}=-1/2}
 {\sum_{\nu\in\mbox{\scriptsize{0R}}}}
 e^{-{\rm i}\tilde{\omega}_{\nu}t+{\rm i}\tilde{P}_{\nu}x} \cdot
 \frac{Y'_\nu(1/2+1/(2\lambda'))}{Y'_\nu(1)}
 \frac{Z_\nu(1/2)}{Z_\nu(1/(2\lambda'))}.\nonumber\\
 \label{ppVXYZ-0R}
\end{eqnarray}
For $\nu+\rho_{\rm R}=(\overbrace{\rho_{\rm R},\cdots,\rho_{\rm R}}^{N},0)$, $Y'_\nu(r)$ and $Z'_\nu(r)$ are written, respectively, by
\begin{eqnarray}
Y'_\nu(r)&=&\prod_{j=1,3,\cdots}^{\rho_{\rm R}-1}\left(\prod_{i=2,4,\cdots}^{N}\left(\frac{j-1}{2\lambda'}+r+\frac{N-i}{2}\right)\right)\nonumber\\
         &&\times\prod_{j=2,4,\cdots}^{\rho_{\rm R}}\left(\prod_{i=1,3,\cdots}^{N-1}\left(\frac{j-1}{2\lambda'}+r+\frac{N-i}{2}\right)\right)\label{eq: Ynur-0R}
\end{eqnarray}
and 
\begin{eqnarray}
Z_\nu(r)&=&\prod_{j=1,3,\cdots}^{\rho_{\rm R}-1}\left(\prod_{i=2,4,\cdots}^{N}
\left(\frac{\rho_{\rm R}-j}{2\lambda'}+r+\frac{N-i}{2}\right)\right)\nonumber\\
         &&\times
         \prod_{j=2,4,\cdots}^{\rho_{\rm R}}\left(\prod_{i=1,3,\cdots}^{N-1}   \left(\frac{\rho_{\rm R}-j}{2\lambda'}+r+\frac{N-i}{2}\right)\right)\nonumber\\
&=&
\prod_{j'=1,3,\cdots}^{\rho_{\rm R}-1}\left(\prod_{i=1,3,\cdots}^{N-1}\left(\frac{j'-1}{2\lambda'}+r+\frac{N-i}{2}\right)\right)\nonumber\\
         &&\times
         \prod_{j'=2,4,\cdots}^{\rho_{\rm R}}\left(\prod_{i=2,4,\cdots}^{N}   \left(\frac{j'-1}{2\lambda'}+r+\frac{N-i}{2}\right)\right).
         \label{eq: Znu-0R}         
\end{eqnarray}
In the second equality of (\ref{eq: Znu-0R}), we have changed a dummy variables from $j$ to $j'=\rho_{\rm R}+1-j$. From (\ref{eq: Ynur-0R}), we obtain 
\begin{eqnarray}
Y'_\nu(r+1/2)&=&
\prod_{j'=1,3,\cdots}^{\rho_{\rm R}-1}\left(\prod_{i=1,3,\cdots}^{N-1}\left(\frac{j'-1}{2\lambda'}+r+\frac{N-i}{2}\right)\right)\nonumber\\
         &&\times
         \prod_{j'=2,4,\cdots}^{\rho_{\rm R}}\left(\prod_{i=0,2,4,\cdots}^{N-2}   \left(\frac{j'-1}{2\lambda'}+r+\frac{N-i}{2}\right)\right).
         \label{eq: Ynur12-0R}         
\end{eqnarray}
Dividing (\ref{eq: Ynur12-0R}) by (\ref{eq: Znu-0R}), we obtain 
\begin{eqnarray}
\frac{Y'_\nu(r+1/2)}{Z_\nu(r)}&=&\prod_{j'=2,4,\cdots}^{\rho_{\rm R}}
\frac{\frac{j-1}{2\lambda'}+r+\frac{N}{2}}{\frac{j-1}{2\lambda'}+r}\nonumber\\
&=&\frac{\Gamma\left((\rho_{\rm R}+1+\lambda' N)/2+\lambda' r \right)\Gamma\left(1/2+\lambda'r \right)}{\Gamma\left((\rho_{\rm R}+1)/2+\lambda'r\right)\Gamma\left((1+\lambda'N)/2+\lambda'r\right)},
\end{eqnarray}
from which 
\begin{eqnarray}
&&\frac{Y'_\nu(1/2+1/(2\lambda'))}{Y'_\nu(1)}
 \frac{Z_{\nu}(1/2)}{Z_{\nu}(1/(2\lambda'))}\nonumber\\ 
&&=\frac{\Gamma((1+(N+1)\lambda')/2)}{\Gamma(1+N\lambda'/2)\Gamma((1+\lambda')/2)}
\nonumber\\
&&\times
\frac{
\Gamma\left(\left(\rho_{\rm R}+2+\lambda' N\right)/2\right)
\Gamma\left(\left(\rho_{\rm R}+1+\lambda' \right)/2\right)}{
\Gamma\left(\left(\rho_{\rm R}+1+\lambda'(N+1)\right)/2\right)
\Gamma\left(\left(\rho_{\rm R}+2\right)/2\right)}\nonumber\\
\end{eqnarray}
follows.
From this, (\ref{eq: 0R-energy}), (\ref{eq: 0R-momentum}), (\ref{ppVXYZ-0R}) and (\ref{eq: K0L}), we arrive at (\ref{ppVXYZ-0R-finite-final}) with (\ref{eq: F0R}).

\subsubsection{Derivation of $G^{(1)}$}
\begin{figure}[t]
 \begin{center}
  \vspace{-0pt}
  \input{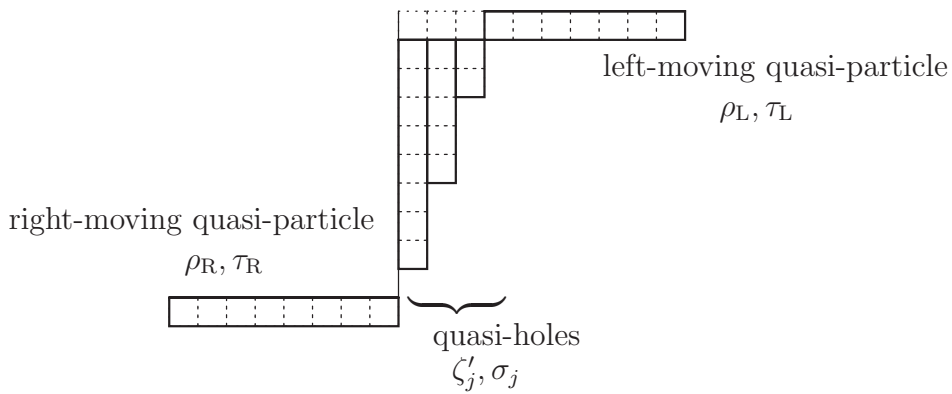} 
 \end{center}
 \caption{Quasi-hole and quasi-particle description of 
 intermediate states. Quasi-particles are characterized by the 
 momenta and the spins $\rho_{{\rm L/R}},\tau_{{\rm L/R}}$,
 and quasi-holes by $\zeta'_{j},\sigma_{j}$.
 Here nonzero $\rho_{{\rm L}}$ and $\rho_{{\rm R}}$ prevent
 the momenta of quasi-holes being $\nu'_j=0$ or $N+1$,
 and it is natural to take the new variables $\zeta'_{j}=\nu'_j-1$ 
 as the momenta of quasi-holes in this case.}
 \label{2qp3qh}
\end{figure}

When $\nu$ satisfies (\ref{eq: 0L0Rlambdahole}), 
$\nu$ is expressed by a generalized Young diagram shown in Fig.~\ref{2qp3qh}.
The notations $\rho_{\rm L}=\nu_1-\lambda'$, $\rho_{\rm R}=-\nu_{N+1}$, 
\begin{eqnarray}
 \tau_{\rm L}=
 \left\{\begin{array}{ll}
  +1/2 &(\rho_{\rm L}:{\rm even})\\
  -1/2 &(\rho_{\rm L}:{\rm odd})
 \end{array}\right., \,\,
 \tau_{\rm R}=
 \left\{\begin{array}{ll}
  +1/2 &(\rho_{\rm R}:{\rm odd})\\
  -1/2 &(\rho_{\rm R}:{\rm even})
 \end{array}\right. \label{eq: tauLtauR}
\end{eqnarray} 
have been introduced in the previous two subsections. The quantities $\zeta_j'$ and $\sigma_j$ will be introduced in the following calculations. 

We decompose the excitation energy $\tilde{\omega}_\nu$ (\ref{tildeomeganu}) into terms with $i=1,N+1$ and the others. The former two terms can be treated as in the previous subsections. With use of the notations (\ref{eq: tauLrhoL}) and (\ref{eq: rhoRtauR}), $\tilde{\omega}_\nu$ is rewritten as
\begin{equation}
\tilde{\omega}_\nu=(\pi/L)^2\{(\tilde{\rho}_{\rm L}+\tau_{\rm L})^2 +
(\tilde{\rho}_{\rm R}+\tau_{\rm R})^2\}+\tilde{\omega}'_\nu
\end{equation}
with 
\begin{equation}
\tilde{\omega}'_\nu=\left(\frac{\pi}{L}\right)^2\sum_{i=2}^{N}\left(\nu_i+\frac{\lambda'(N+1-2i)}{2}+\sigma_i^{\rm p}\right)^2 -E_N({\rm g}).
\label{Eprimenu}
\end{equation}
Introducing a partition $\zeta=(\zeta_1,\cdots,\zeta_{N-1})\in \Lambda_{N-1}$ by
\begin{equation}
\zeta_i=\nu_{i+1},\quad i\in [1,N-1],
\end{equation}
the energy (\ref{Eprimenu}) becomes
\begin{equation}
\tilde{\omega}'_\nu=\left(\frac{\pi}{L}\right)^2\sum_{i=1}^{N-1}\left(\zeta_i+\frac{\lambda'(N-1-2i)}{2}+\sigma_{i+1}^{\rm p}\right)^2-E_N({\rm g}).
\label{Eprimenu-zeta}
\end{equation}
The first term in the right-hand side coincides with (B.2) in \cite{hp} when we replace $\zeta_i$ by $\mu_i$ and $\sigma_{i+1}^{\rm p}$ by $3/2-\alpha_{N-i}$. Further, the relation
$$
\sigma_{i+1}^{\rm p}=\left\{
\begin{array}{rl}
-1/2&\quad (i,\zeta_i)\in C_2(\zeta)\\
1/2&\quad (i,\zeta_i)\in D(\zeta)\setminus C_2(\zeta)\\
\end{array}
\right.
$$
coincides with (B.4) in \cite{hp} under the same replacement. 
Therefore we can rewrite (\ref{Eprimenu-zeta}) following the argument of Appendix B in \cite{hp}. 
For $j\in [1,\lambda']$, let $\zeta'_j$  be the length of $j$th column in $D(\zeta)$ and $\sigma_j$ be ``the spin variable''
 defined by
\begin{equation}
\sigma_j=\left\{
\begin{array}{rl}
1/2,& \zeta'_j-j\mbox{ is odd}\\
-1/2,& \zeta'_j-j\mbox{ is even}.\\
\end{array}
\right.
\label{eq: sigma-j}
\end{equation}
Furthermore, we introduce the rapidity
\begin{equation}
\check{\zeta}_j=\zeta'_j -\frac{N-1}{2}+\frac{\lambda+1-j}{\lambda'}.
\label{eq: mutildej}
\end{equation}
In terms of the spin (\ref{eq: sigma-j}) and the rapidity(\ref{eq: mutildej}), $\tilde{\omega}'_\nu$ is rewritten as
\begin{eqnarray}
 &\tilde{\omega}'_{\nu}=
  \left(\frac{\pi}{L}\right)^2\left[-\lambda'
  \sum_{j=1}^{\lambda'}\left(\check{\zeta}_j+\sigma_j\right)^2
  +\frac{(\lambda')^2-2}{3}\right], 
   \end{eqnarray} 
from which 
\begin{eqnarray}
 &\tilde{\omega}_{\nu}=
  \left(\frac{\pi}{L}\right)^2
 \left[ \sum_{i={\rm L,R}}(\tilde{\rho}_i+\tau_i)^2
     -\lambda'\sum_{j=1}^{\lambda'}\left(\check{\zeta}_j+\sigma_j\right)^2
  +\frac{(\lambda')^2-2}{3}\right] 
   \label{finite-energy}
   \end{eqnarray} 
follows. 
Similarly, the total momentum is rewritten as
\begin{eqnarray}
 &\tilde{P}_{\nu}=-\frac{\pi}{L}
  \left[\sum_{i={\rm L,R}}(\tilde{\rho}_i+\tau_i)
              +\sum_{j=1}^{\lambda'}\left(\check{\zeta}_j+\sigma_j\right)
  \right]. \label{finite-momentum}
\end{eqnarray} 
Following the argument of Appendix A in \cite{hp}, $z$ component of the total spin (\ref{eq: sztotal}) is 
\begin{eqnarray}
 S_{{\rm tot}}^z&=\tau_{\rm L}+\tau_{\rm R}+\sum_{j=1}^{\lambda'}\sigma_j, 
\label{eq: sztotal2}
\end{eqnarray}
From (\ref{finite-energy}), (\ref{finite-momentum}) and (\ref{eq: sztotal2}), $\tilde\rho_{\rm L}$, $\tilde\rho_{\rm R}$ and $\check{\zeta}_j$ can be identified as the rapidities of the left-moving quasi-particle, the right-moving quasi-particle and the quasi-holes, respectively. Similarly, $\tau_{\rm L}$, $\tau_{\rm R}$ and $\sigma_j$ can be identified as the spins of the left-moving quasi-particle, the right-moving quasi-particle and the quasi-holes, respectively. Figure~\ref{2qp3qh} shows the excited state specified by $\nu$. The excitation content of this state consists of a left-moving quasi-particle, a right-moving quasi-particle and $\lambda'$ quasi-holes. 

We rewrite the expression in the parenthesis in the numerator in the right-hand side of (\ref{ppbpshipmupmu}) as
\begin{eqnarray}
&\sum_{\mu\in \Lambda_{N}}\delta_{|C_2(\mu)|+|H_2(\mu)|,|\mu|} b_\mu \psi^{(\lambda')}_{\nu\mu}\{P^{(\lambda')}_\mu,P^{(\lambda')}_\mu\}_{N,\lambda}\nonumber\\
&=c_{N}^{(\lambda',2)}\sum_{\mu\in \Lambda_{N}}(-1)^{|\mu|+\sum_s l'(s)}\delta_{|C_2(\mu)|+|H_2(\mu)|,|\mu|}\frac{X_\mu Y_\mu(1/2)}{Z_\mu(1/2)Y_\mu(1/(2\lambda'))}\psi_{\nu\mu}^{(\lambda')},\label{bpmupmupsi} 
\end{eqnarray}
where $X_\mu=(-1)^{|\mu|-|C_2(\mu)|}Y'_\mu (-N/2)$.
In a way similar to that used in \cite{hp}, the expressions $X_\mu/ Z_\mu(1/2)$ and $Y_\mu(1/2)/ Y_\mu(1/(2\lambda'))$ can be written in terms of $\mu'_j$ and $\sigma^\mu_j$. The quantity $\sigma_j^\mu$ is defined as $+1/2$ $(-1/2)$ when $\mu'_j-j$ is even (odd). 

Let the columns with $\zeta_j'-\mu_j'=0\,\,(j\in[1,\lambda'])$ be denoted by $I$ and those with $\zeta_j'-\mu_j'=-1\,\,(j\in[1,\lambda'])$ be denoted by $J$. Then,  $X_\mu/ Z_\mu(1/2)$ and $Y_\mu(1/2)/ Y_\mu(1/(2\lambda'))$ can be written in terms of $\check{\zeta}_j$, $\sigma_j$. As shown in \ref{psinumu}, $\psi^{(\lambda')}_{\nu\mu}$ can be written in terms of  $\check{\zeta}_j$, $\sigma_j$, $\tilde{\rho}_{\rm L}$, $\tau_{\rm L}$, $\tilde{\rho}_{\rm R}$ and $\tau_{\rm R}$.
Then (\ref{bpmupmupsi}) is rewritten as $F_\nu M_\nu$ with
\begin{eqnarray}
F_\nu&=&\frac{c_N^{(\lambda',2)}}{(\lambda')^{\lambda}\prod_{j=1}^{\lambda'}\Gamma[j/\lambda']}
\frac{\Gamma[(\tilde{\rho}_{\rm L}-\tilde{\rho}_{\rm R}+2-\lambda'-\delta_{\tau_{\rm L},\tau_{\rm R}})/2]}
     {\Gamma[(\tilde{\rho}_{\rm L}-\tilde{\rho}_{\rm R}+1-\delta_{\tau_{\rm L},\tau_{\rm R}})/2]}\nonumber\\
&\times&
\frac{\Gamma[(\tilde{\rho}_{\rm L,0}-\tilde{\rho}_{\rm R}+\lambda'+\delta_{\tau_{\rm R},\downarrow})/2]}
     {\Gamma[(\tilde{\rho}_{\rm L,0}-\tilde{\rho}_{\rm R}+1+\delta_{\tau_{\rm R},\downarrow})/2]}
\frac{\Gamma[(\tilde{\rho}_{\rm L}-\tilde{\rho}_{\rm R,0}+\lambda'+\delta_{\tau_{\rm L},\uparrow})/2]}
     {\Gamma[(\tilde{\rho}_{\rm L}-\tilde{\rho}_{\rm R,0}+1+\delta_{\tau_{\rm L},\uparrow})/2]}\nonumber\\
&\times&
\prod_{i=0,\lambda'+1}\prod_{j=1}^{\lambda'}\frac{
\Gamma[(|\check{\zeta}_i-\check{\zeta}_j|+1-\delta_{\sigma_i,\sigma_j})/2]}
{\Gamma[(|\check{\zeta}_i-\check{\zeta}_j|+2-1/\lambda'-\delta_{\sigma_i,\sigma_j})/2]}\nonumber\\
&\times&\prod_{1\le j<k\le \lambda'}\frac{
\Gamma[(\check{\zeta}_j-\check{\zeta}_k+1+\delta_{\sigma_j,\sigma_k})/2]}
{\Gamma[(\check{\zeta}_j-\check{\zeta}_k+2-1/\lambda'-\delta_{\sigma_j,\sigma_k})/2]}\label{Fnu}
\end{eqnarray}
and 
\begin{eqnarray}
 &M_\nu=\underset{n_{\mu}=\lambda,\lambda+1,\,\,\mu_1\le\lambda'}
  {\underset{{\rm s.t.}\nu/\mu:{\rm h.s.}}
  {\sum_{\mu \in\Lambda_N}}}\!\!\!\!
  (-1)^{|{\rm odd}\cap I|}
  \prod_{j\in J}
\left(\frac{\tilde{\rho}_{\rm L}+\lambda'\check{\zeta}_j+(1-\lambda')/2}
           {\tilde{\rho}_{\rm L}+\lambda'\check{\zeta}_j-(1-\lambda')/2}
\right)
  ^{1-\delta_{\sigma_j\tau_{\rm L}}}\nonumber\\
  &\times\prod_{j\in I}
  \left(\frac{-\tilde{\rho}_{\rm R}-\lambda'\check{\zeta}_j+(1-\lambda')/2}
             {-\tilde{\rho}_{\rm R}-\lambda'\check{\zeta}_j-(1-\lambda')/2}
  \right)
  ^{1-\delta_{\sigma_j\tau_{\rm R}}} \nonumber\\
  &\times\underset{j\in J,k\in I}{\prod_{j<k}}
  \left(\frac{1}{2\lambda'}+\frac{\check{\zeta}_j-\check{\zeta}_k}{2}\right)
  ^{1-\delta_{\sigma_j\sigma_k}}
  \left(\frac{\check{\zeta}_j-\check{\zeta}_k}{2}\right)
  ^{-\delta_{\sigma_j\sigma_k}}\nonumber\\
  &\times\underset{j\in I,k\in J}{\prod_{j<k}}
  \left(\frac{-1}{2\lambda'}+\frac{\check{\zeta}_j-\check{\zeta}_k}{2}\right)
  ^{1-\delta_{\sigma_j\sigma_k}}
  \left(\frac{\check{\zeta}_j-\check{\zeta}_k}{2}\right)
  ^{-\delta_{\sigma_j\sigma_k}},\label{Mnufinitefinal}
\end{eqnarray}
where $n_{\mu}$ is the number of $j\in[1,\lambda']$ for which $\mu'_j-j$ is even. 
Here, we have introduced auxiliary quantities $\zeta'_0=N-1$, $\zeta'_{\lambda'+1}=0$, $\sigma_0=1/2$, and $\sigma_{\lambda'+1}=-1/2$, and the definition of the spin (\ref{eq: sigma-j}) and the rapidity (\ref{eq: mutildej}) is extended to $j\in [0,\lambda'+1]$.

The factor $\{P^{(\lambda')}_\nu,P^{(\lambda')}_\nu\}_{N+1,\lambda}$ in the denominator of (\ref{ppbpshipmupmu}) can also be written in terms of the rapidities and the spins of elementary excitations. The explicit expression is given in (\ref{Pnunormfinal}), (\ref{mathcalK}), (\ref{mathcalA}) and (\ref{mathcalB}) and those expressions will be derived in \ref{subsec: normofnu}. 

From (\ref{Fnu}), (\ref{Mnufinitefinal}) and (\ref{Pnunormfinal}), 
 $G^{(1)}(x,t)$ is given as 
\begin{eqnarray}
 &G^{(1)}(x,t)=  \frac{K^{(1)}}{L}
\sum_{\tau_{\rm L},\tau_{\rm R},\{\sigma_j\}}
\sum_{\rho_{\rm L},\rho_{\rm R},\{\check{\zeta}_j\}}
  \delta_{\sum\sigma_j+\tau_{\rm L}+\tau_{\rm R},-1/2}
  L_{\nu}M_{\nu}^2
  e^{-{\rm i}\tilde{\omega}_{\nu}t+{\rm i}\tilde{P}_{\nu}x}
\label{eq: G1}
\end{eqnarray}
with
\begin{equation}
K^{(1)}=\frac{\left(\Gamma\left[(\lambda'+1)/(2\lambda')\right]\right)^{\lambda'}}{(\lambda')^{\lambda}\Gamma((\lambda'+1)/2)}
 \prod_{j=1}^{\lambda'}\Gamma\left[j/\lambda'\right]^{-2}
\label{eq: K1}
\end{equation}
and
\begin{equation}
L_\nu=L^{(\rm I)}_\nu L^{(\rm II)}_\nu L^{(\rm III)}_\nu L^{(\rm IV)}_\nu L^{(\rm V)}_\nu 
\end{equation}
with 

\begin{eqnarray}
 &L^{(\rm I)}_\nu=
  \prod_{j=1}^{\lambda'}
  \left(\frac{\tilde{\rho}_{\rm L}+\lambda'\check{\zeta}_j+(\lambda'-1)/2}
             {\tilde{\rho}_{\rm L}+\lambda'\check{\zeta}_j-(\lambda'-1)/2}
\right)
   ^{1-\delta_{\sigma_j\tau_{\rm L}}}\left(
\frac
{\tilde{\rho}_{\rm R}+\lambda'\check{\zeta}_j-(\lambda'-1)/2}
             {\tilde{\rho}_{\rm R}+\lambda'\check{\zeta}_j+(\lambda'-1)/2}
  \right)
   ^{1-\delta_{\sigma_j\tau_{\rm R}}} \nonumber\\
 &L^{(\rm II)}_\nu=
   \frac{\Gamma
\left[(\tilde{\rho}_{\rm L}-\tilde{\rho}_{{\rm L},0}+\delta_{\tau_{\rm L},\uparrow})/2\right]}
   {\Gamma
\left[(\tilde{\rho}_{\rm L}-\tilde{\rho}_{{\rm L},0}+1-\lambda'+\delta_{\tau_{\rm L},\uparrow})/2\right]
}\frac{\Gamma\left[\tilde{\rho}_{\rm L}-\tilde{\rho}_{{\rm R},0}+\lambda'+\delta_{\tau_{\rm L},\uparrow})/2\right]}
{\Gamma\left[\tilde{\rho}_{\rm L}-\tilde{\rho}_{{\rm R},0}+1+\delta_{\tau_{\rm L},\uparrow})/2\right]}
\nonumber\\
 &\quad\quad\times
  \frac{\Gamma\left[(\tilde{\rho}_{{\rm L},0}-\tilde{\rho}_{\rm R}+\lambda'+\delta_{\tau_{\rm R},\downarrow})/2\right]}
{\Gamma\left[(\tilde{\rho}_{{\rm L},0}-\tilde{\rho}_{\rm R}+1+\delta_{\tau_{\rm R},\downarrow})/2\right]}
   \frac{\Gamma\left[(\tilde{\rho}_{{\rm R},0}-\tilde{\rho}_{\rm R}+
    \delta_{\tau_{{\rm R}},\downarrow})/2\right]}
{\Gamma\left[(\tilde{\rho}_{{\rm R},0}-\tilde{\rho}_{\rm R}+1-\lambda'+
    \delta_{\tau_{{\rm R}},\downarrow})/2\right]}
\nonumber\\
 &L^{(\rm III)}_\nu=
  \frac{\Gamma\left[(\tilde{\rho}_{\rm L}-\tilde{\rho}_{\rm R}+2-\lambda'-
    \delta_{\tau_{\rm L},\tau_{\rm R}})/2\right]}
{\Gamma\left[(\tilde{\rho}_{\rm L}-\tilde{\rho}_{\rm R}+1-
    \delta_{\tau_{\rm L},\tau_{\rm R}})/2\right]}
  \frac{\Gamma
\left[(\tilde{\rho}_{\rm L}-\tilde{\rho}_{\rm R}+1+
    \delta_{\tau_{\rm L},\tau_{\rm R}})/2\right]
}
{\Gamma\left[(\tilde{\rho}_{\rm L}-\tilde{\rho}_{\rm R}+\lambda'+
    \delta_{\tau_{\rm L},\tau_{\rm R}})/2\right]},\nonumber\\
&L^{(\rm IV)}_\nu=
\prod_{i=0,\lambda'+1}\prod_{j=1}^{\lambda'}\frac{
\Gamma[(|\check{\zeta}_i-\check{\zeta}_j|+1-\delta_{\sigma_i,\sigma_j})/2]}
{\Gamma[(|\check{\zeta}_i-\check{\zeta}_j|+2-1/\lambda'-\delta_{\sigma_i,\sigma_j})/2]}\nonumber\\
&L^{(\rm V)}_\nu=
\prod_{1\le j<k\le \lambda'}\frac{
\Gamma[(\check{\zeta}_j-\check{\zeta}_k+1+\delta_{\sigma_j,\sigma_k})/2]
\Gamma[(\check{\zeta}_j-\check{\zeta}_k+1/\lambda'+\delta_{\sigma_j,\sigma_k})/2]}
{\Gamma[(\check{\zeta}_j-\check{\zeta}_k+2-1/\lambda'-\delta_{\sigma_j,\sigma_k})/2]
 \Gamma[(\check{\zeta}_j-\check{\zeta}_k+1-\delta_{\sigma_j,\sigma_k})/2]}.\nonumber\\
\end{eqnarray}

\section{Thermodynamic Limit}\label{sec: th}

In this section, we derive an expression for the particle propagator in the thermodynamic limit, from the results in section \ref{sec: pp-ee}. We derive (\ref{eq: G0L-th}) for $G^{(\rm 0L)}$ in section \ref{subsec: th-G0LR} and (\ref{eq: th-G1}) for $G^{(\rm 1)}$ in section \ref{subsec: th-G1}. 
\subsection{$G^{(\rm 0L)}$}\label{subsec: th-G0LR}
We introduce ``the reduced rapidity'' of the left-moving quasi-particle as
\begin{equation}
w_{\rm L}=-\frac{2}{N\lambda'}\left(\tilde{\rho}_{\rm L}-\frac12\right)
\label{eq: def-w}
\end{equation}
and describe each part of $G^{(\rm 0L)}(x,t)$ in terms of $w_{\rm L}$. The excitation energy and the momentum in (\ref{eq: G0L-finite}) are written as $(\pi d\lambda'w_{\rm L})^2/4$ and $\pi d\lambda'w_{\rm L}/2$, respectively. 

In the summation in (\ref{eq: G0L-finite}), the increment $\Delta\rho_{\rm L}$ of $\rho_{\rm L}$ is two. From (\ref{eq: def-w}) and (\ref{eq: tauLrhoL}), we see that $\Delta\rho_{\rm L}=2$ corresponds to $\Delta w_{\rm L}=-4/(N\lambda')$. We also note that $w_{\rm L}=-1+{\cal O}(1/N)$ when $\rho_{\rm L}=1$ and $w_{\rm L}\rightarrow -\infty$ when $\rho_{\rm L}\rightarrow \infty$. It thus follows that 
\begin{equation}
\frac{1}{L}\sum_{\rho_{\rm L}=1,3,5,\cdots}=-\frac{\lambda' d\Delta w_{\rm L}}{4}\sum_{\rho_{\rm L}=1,3,5,\cdots}\rightarrow \frac{\lambda' d}{4}\int_{-\infty}^{-1}\rmd w_{\rm L}
\end{equation} 
in the thermodynamic limit. 
The form factor $F^{(\rm 0L)}$ (\ref{eq: F0L}) reduces to
\begin{equation}
F^{(\rm 0L)}\sim \left(\frac{\tilde{\rho}_{\rm L}-\tilde{\rho}_{\rm R,0}}{2}\right)^{-\lambda}\left(\frac{\tilde{\rho}_{\rm L}-\tilde{\rho}_{\rm L,0}}{2}\right)^{\lambda}\sim \left(\frac{|w_{\rm L}|-1}{|w_{\rm L}|+1}\right)^\lambda,
\end{equation}
which results from the relation 
\begin{equation}
\Gamma(n+a)/\Gamma(n+b)\sim n^{a-b}$ for $n\gg a,b={\cal O}(1).
\label{eq: GammaGamma}
\end{equation}
Combining the above results, we obtain (\ref{eq: G0L-th}). In the same way, we can evaluate $\lim_{\rm t.d.l.}G^{(\rm 0R)}(x,t)$, which reduces to (\ref{eq: G0R-th}). 

\subsection{$G^{(1)}$}\label{subsec: th-G1}
Introducing the reduced rapidities 
\begin{equation}
w_{\rm R}=-\frac{2(\tilde{\rho}_{\rm R}+\tau_{\rm R})}{N\lambda'},\quad
w_{\rm L}=-\frac{2(\tilde{\rho}_{\rm L}+\tau_{\rm L})}{N\lambda'}
\label{eq: def-w12}
\end{equation}
of quasi-particles and 
\begin{equation}
u_j=-\frac{2(\check{\zeta}_j+\sigma_j)}{N}
\label{eq: def-uj}
\end{equation}
of quasi-holes,
the excitation energy $\tilde{\omega}_\nu$ (\ref{finite-energy}) and the momentum $\tilde{P}_\nu$ (\ref{finite-momentum}) appearing
in (\ref{eq: G1}) are rewritten, respectively, as
\begin{equation}
\lim_{\rm t.d.l}\tilde{\omega}_\nu=
\left(\frac{\pi \lambda' d}{2}\right)^2\left(w_{\rm R}^2+w_{\rm L}^2\right)-
\frac{\pi^2 \lambda' d^2}{4}\sum_{j=1}^{\lambda'}u_j^2
\label{eq: omega-th}
\end{equation}
and
\begin{equation}
\lim_{\rm t.d.l}\tilde{P}_\nu=
\frac{\pi \lambda' d\left(w_{\rm R}+w_{\rm L}\right)}{2}+
\frac{\pi d}{2}\sum_{j=1}^{\lambda'}u_j
.\label{eq: P-th}
\end{equation}
In the summation in (\ref{eq: G1}) with respect to $\rho_{\rm L}$, $\rho_{\rm R}$ and $\{\zeta_j\}$ under a set of fixed values of $\tau_{\rm L}$, $\tau_{\rm R}$ and $\{\sigma_j\}$, we see from (\ref{eq: tauLtauR}) and (\ref{eq: sigma-j}) that 
\begin{equation}
\Delta\rho_{\rm L}=\Delta\rho_{\rm R}=\Delta\zeta_j=2
\label{eq: Deltarho}
\end{equation}
for the increments $\Delta\rho_{\rm L}$, $\Delta\rho_{\rm R}$ and $\Delta\zeta_j$ in the summation. From (\ref{eq: def-w12}), (\ref{eq: def-uj}), and (\ref{eq: Deltarho}), it follows that $\Delta w_{\rm R}=\Delta w_{\rm L}=-4/(N\lambda')$ and $\Delta u_j=-4/N$, and then the summation over rapidities becomes
\begin{eqnarray}
&\frac{1}{L}
\sum_{\tau_{\rm L},\tau_{\rm R},\{\sigma_j\}}
\sum_{\rho_{\rm L},\rho_{\rm R},\{\check{\zeta}_j\}}\nonumber\\
&\rightarrow
 \frac{\lambda'^2}{L}\left(\frac{N}{4}\right)^{2\lambda+3}
 \sum_{\tau_{\rm L},\tau_{\rm R},\{\sigma_j\}}
 \int_1^{\infty}{\rm d}w_{\rm R}
 \int_{-\infty}^{-1}{\rm d}w_{\rm L}\nonumber\\
&\times\int_{-1}^1{\rm d}u_{\lambda'}\int_{-1}^{u_{\lambda'}}{\rm d}u_{\lambda'-1}\cdots\int_{-1}^{u_2}{\rm d}u_1.
\label{eq: sum-th}
\end{eqnarray}

When $\tilde{\rho}_{\rm L},\tilde{\rho}_{\rm R},\{\check{\zeta}_j\}$ are of the order of $N$, the expression of each part of $L_\nu$ in the thermodynamic limit is derived with use of 
(\ref{eq: GammaGamma}) as 
\begin{eqnarray}
&&L^{(\rm I)}_\nu\sim 1, \nonumber\\
&&L^{(\rm II)}_\nu\sim \left(\frac{N\lambda'}{4}\right)^{4\lambda}\left(w_{\rm R}^2-1\right)^\lambda\left(w_{\rm L}^2-1\right)^\lambda\nonumber\\
&&L^{(\rm III)}_\nu\sim \left(\frac{N\lambda'}{4}\right)^{-2\lambda}\left(w_{\rm R}-w_{\rm L}\right)^{-2\lambda}\nonumber\\
&&L^{(\rm IV)}_\nu\sim \left(\frac{N}{4}\right)^{-2\lambda}\prod_{j=1}^{\lambda'}\left(1-u_j^2\right)^{-\lambda/\lambda'}\nonumber\\
&&L^{(\rm V)}_\nu\sim \prod_{1\le j<k\le \lambda'}\left(\frac{N}{4}\right)^{-2\lambda/\lambda' +2\delta_{\sigma_j\sigma_k}}
\prod_{1\le j<k\le \lambda'}\left(u_k-u_j\right)^{-2\lambda/\lambda' +2\delta_{\sigma_j\sigma_k}}.\label{eq: th-L}
\end{eqnarray}
It is not straightforward to derive the expression for $M_\nu$ in the thermodynamic limit, because cancellation occurs in the summation with respect to $\mu$, owing to the sign factor $(-1)^{|{\rm odd}\cap I|}$. In order to single out leading contribution in the thermodynamic limit, we introduce variables $e_j=1(j\in J),-1(j\in I)$ in a way similar to the calculation in the scalar Calogero-Sutherland model\cite{Ser}, and rewrite $M_{\nu}$ as 
\begin{equation}
M_\nu=M^{({\rm I})}_\nu M^{({\rm II})}_\nu M^{({\rm III})}_\nu
\end{equation}
with 
\begin{equation}
M^{({\rm I})}_\nu=\left[{\prod_{j<k}}\left(\frac{\check{\zeta}_j-\check{\zeta}_k}{2}\right)
  ^{-\delta_{\sigma_j\sigma_k}}\right],
\end{equation}
\begin{eqnarray}
M^{({\rm II})}_\nu=
 \prod_{j}
  \left(\frac{\frac{\tilde{\rho}_{\rm L}}{\lambda'}+\check\zeta_j}{\frac{\tilde{\rho}_{\rm L}}{\lambda'}+\check\zeta_j+\frac{\lambda}{\lambda'}}\right)^{1-\delta_{\sigma_j\tau_{\rm L}}}
\left(\frac{\frac{\tilde{\rho}_{\rm R}}{\lambda'}+\check\zeta_j+\frac{2\lambda}{\lambda'}}{\frac{\tilde{\rho}_{\rm R}}{\lambda'}+\check\zeta_j+\frac{\lambda}{\lambda'}}\right)^{1-\delta_{\sigma_j\tau_{\rm R}}}\nonumber\\
\end{eqnarray}
and
\begin{eqnarray}
 M^{({\rm III})}_\nu&=
  \underset{n_{\mu}=\lambda,\lambda+1,\,\,\mu_1\le\lambda'}
  {\underset{{\rm s.t.}\nu/\mu:{\rm h.s.}}
  {\sum_{\mu \in\Lambda_N}}}
  \left[\prod_{j<k}\left(\frac{\check{\zeta}_j-\check{\zeta}_k}{2}\right)^{\delta_{\sigma_j^\mu\sigma_k^\mu}}\right]\left[\prod_{j:{\rm odd}}^{\lambda'}e_j\right]\nonumber\\
&\times
\prod_{j}
  \left(1-\frac{\frac{\lambda}{\lambda'}e_j}{\frac{\tilde{\rho}_{\rm L}}{\lambda'}+\check\zeta_j}\right)^{1-\delta_{\sigma_j\tau_{\rm L}}}
 \left(1-\frac{\frac{\lambda}{\lambda'}e_j}{\frac{\tilde{\rho}_{\rm R}}{\lambda'}+\check\zeta_j+\frac{2\lambda}{\lambda'}}\right)^{1-\delta_{\sigma_j\tau_{\rm R}}}\nonumber\\
 &\times
  \prod_{j<k}
  \left(1+\frac{\displaystyle 
   e_j-e_k}
  {\displaystyle
  2\lambda'(\check{\zeta}_j-\check{\zeta}_k)}\right)
  ^{1-\delta_{\sigma_j\sigma_k}}. 
\end{eqnarray}
Here $\sigma_j^{\mu}$ is 1/2 (-1/2) when $\mu'_j-j$ is even (odd) (see the sentence below (\ref{bpmupmupsi})). 
When $\tilde{\rho}_{\rm L}$, $\tilde{\rho}_{\rm R}$ and $\{\check{\zeta}_j\}$ are of the order of $N$, the expressions for $M^{(\rm I)}_\nu$, $M^{(\rm II)}_\nu$ and $M^{(\rm III)}_\nu$ reduce to 
\begin{eqnarray}
&&M^{(\rm I)}_{\nu}\sim \prod_{j<k}\left(\frac{N}{4}\right)^{-\delta_{\sigma_j\sigma_k}}\prod_{j<k}\left(u_k-u_j\right)^{-\delta_{\sigma_j\sigma_k}}\label{eq: th-MI}\\
&&M^{(\rm II)}_{\nu}\sim 1,\label{eq: th-MII}\\
&&M^{(\rm III)}_{\nu}\sim \left(\frac{N}{4}\right)^{\lambda^2}
 \underset{n_{\mu}=\lambda,\lambda+1,\,\,\mu_1\le\lambda'}
  {\underset{{\rm s.t.}\nu/\mu:{\rm h.s.}}
  {\sum_{\mu \in\Lambda_N}}}
 \left[\prod_{j<k}(u_k-u_j)^{\delta_{\sigma_j^{\mu}\sigma_k^{\mu}}}\right]
 \left[\prod_{j:{\rm odd}}e_j\right]
 {\cal M}(\{e_j\}),\nonumber\\
\end{eqnarray}
with 
\begin{eqnarray}
{\cal M}(\{e_j\})&=&\prod_{j}
  \left(1+\frac{2\lambda\tilde{e}_j}{w_{\rm R}+u_j}\right)^{1-\delta_{\sigma_j\tau_{\rm R}}}
  \left(1+\frac{2\lambda\tilde{e}_j}{w_{\rm L}+u_j}\right)^{1-\delta_{\sigma_j\tau_{\rm L}}}\nonumber\\
&&\times  \prod_{j<k}
  \left(1+\frac{\displaystyle 
   \tilde{e}_j-\tilde{e}_k}
  {\displaystyle
  u_k-u_j}\right)
  ^{1-\delta_{\sigma_j\sigma_k}}\label{finiteM}
\end{eqnarray}
and $\tilde{e}_j=e_j/(N\lambda')$. 
We see that the expansion with respect to $N^{-1}$ in ${\cal M}(\{e_j\})$ is equivalent to that with $\tilde{e}_j$. The lowest order terms with respect to $\tilde{e}_j$ which give non-vanishing contributions to $M^{(\rm III)}_\nu$ are given as follows.

Let $f(\{e_j\})$ be a polynomial of $\{e_j\}$. In Appendix B, we show the following (i), (ii) and (iii): 
\begin{enumerate}
\item
\begin{eqnarray}
 \underset{n_{\mu}=\lambda,\lambda+1,\,\,\mu_1\le\lambda'}
  {\underset{{\rm s.t.}\nu/\mu:{\rm h.s.}}
  {\sum_{\mu \in\Lambda_N}}}
 \left[\prod_{j<k}(u_k-u_j)^{\delta_{\sigma_j^{\mu}\sigma_k^{\mu}}}\right]
 \left[\prod_{j:{\rm odd}}e_j\right]f(\{e_j\}). \label{Msumex}
\end{eqnarray}
becomes zero when the degree of $f(\{e_j\})$ is lower than $\lambda+1$.
\item
When the degree of $f$ is $\lambda+1$, non-vanishing contributions to 
(\ref{finiteM}) result only from monomials $\prod_{j\in Q}e_j$ with $Q\subset\{1,2,\cdots,\lambda'\}$ satisfying
\begin{equation}
|Q\cap\mbox{odd}|=|\bar{Q}\cap\mbox{even}|,
\label{eq: Q-condition}
\end{equation}
from which $|Q|=\lambda+1$ follows. 
\item For $Q$ satisfying (\ref{eq: Q-condition}), the relation
\begin{eqnarray}
 &&\underset{n_{\mu}=\lambda,\lambda+1,\,\,\mu_1\le\lambda'}
  {\underset{{\rm s.t.}\nu/\mu:{\rm h.s.}}
  {\sum_{\mu \in\Lambda_N}}}
 \left[\prod_{j<k}(u_k-u_j)^{\delta_{\sigma_j^{\mu}\sigma_k^{\mu}}}\right]
 \left[\prod_{j:{\rm odd}}e_j\right]\left[\prod_{j\in Q}e_j\right]\nonumber\\
&&=2^{\lambda+1}
\varepsilon(Q,\{\sigma_j\})\prod_{(j<k)\in (Q,Q)\cup(\bar{Q},\bar{Q})}(u_k-u_j)\end{eqnarray}
holds. The symbol $\varepsilon(Q,\{\sigma_j\})$ denotes the sign factor defined in (\ref{eq: signfactor}). 
\item
Following the similar calculation in the scalar Calogero-Sutherland model (in \cite{Ser} or Sec.2.7.2 in \cite{KuraKato}), we can show that when ${\cal M}(\{e_j\})$ is written in the form
\begin{equation}
{\cal M}(\{e_j\})=\sum_{Q\mbox{ \scriptsize s.t.(\ref{eq: Q-condition})}}A_Q\prod_{j\in Q}e_j+\mbox{ other terms},
\end{equation}
the coefficient $A_Q$ can be written as
\begin{equation}
A_Q=\left(\frac{-1}{N\lambda'}\right)^{\lambda+1}W_\sigma(w_{\rm R},w_{\rm L},\{u_j\})^{-1}\left[\prod_{j\in Q}\frac{\partial}{\partial u_j}\right]W_\sigma(w_{\rm R},w_{\rm L},\{u_j\})
\end{equation}
with $W_\sigma$ defined in (\ref{eq: W}). 
\end{enumerate}
From (i), (ii), (iii) and (iv), leading terms in $M_\nu^{(\rm III)}$ in the thermodynamic limit are obtained. Combining the resultant expression for $M_\nu^{(\rm III)}$ with (\ref{eq: th-L}), (\ref{eq: th-MI}) and (\ref{eq: th-MII}), we obtain 
\begin{equation}
L_\nu M_\nu^2\sim \frac{1}{\lambda'^\lambda}\left(\frac{2}{N}\right)^{2(\lambda+1)}F(\{u_j\},w_{\rm R},w_{\rm L},\{\sigma_j\},\tau_{\rm{R}},\tau_{\rm{L}}),
\label{eq: th-LM2}
\end{equation}
with $F(\{u_j\},w_{\rm R},w_{\rm L},\{\sigma_j\},\tau_{\rm{R}},\tau_{\rm{L}})$ defined in (\ref{eq: def-F}). Further, from (\ref{eq: G1}), (\ref{eq: K1}), (\ref{eq: omega-th}), (\ref{eq: P-th}), (\ref{eq: sum-th}), and (\ref{eq: th-LM2}), the expression (\ref{eq: th-G1}) for $\lim_{\rm t.d.l.} G^{(1)}(x,t)$ follows.

\section{Discussion}\label{sec: discussion}
Here we make two remarks on technical points in the derivation presented in section~\ref{sec: finite-size} and \ref{sec: th}; physical implication of main results was discussed in section~\ref{sec: main result}.

One is the expression for the particle propagator in terms of the rapidities $\tilde{\rho}_{\rm L}$,$\tilde{\rho}_{\rm R}$,$\check{\zeta}_j$ and the spins $\tau_{\rm L}$,  $\tau_{\rm R}$, $\sigma_j$ of quasi-particles and quasi-holes. We showed in \cite{hp} that the hole propagator in a finite-size system is expressed in a concise form in terms of the rapidities and the spins of the quasi-holes. The hole propagator and dynamical density correlation functions of the finite-sized scalar CS model have simpler expressions in terms of the rapidities of elementary excitations than those in terms of the momenta. 
It is important to make the results as compact as possible in finite-sized systems, especially in the calculation of the particle propagator; the procedure of taking the thermodynamic limit of the particle propagator is much more involved, compared to the procedure for other dynamical correlation functions. The rapidity-spin description will be useful to calculate dynamical correlation functions in the CS model for particles with SU($K$) internal symmetry with $K\ge 3$. 

The other point of importance is evaluation of the contributions from multiple excitations of the quasi-particles and the quasi-holes in the thermodynamic limit. Most of dynamical correlation functions (the hole propagator, the density correlation function and the spin correlation function) have been evaluated in the thermodynamic limit in a common way; first the form factor is written as a product of Gamma functions, variables of which are the rapidities (the scalar CS model) or the rapidities and the spins (the spin CS model) and then use the formula (\ref{eq: GammaGamma}) for Gamma functions. Evaluation of the particle propagator, on the other hand, contains an involved process both in the scalar and the spin CS models. Furthermore, the presence of the spin degrees of freedom in elementary excitations makes calculation of the particle propagator more involved and nontrivial in the spin CS model, compared to the scalar model. This is the reason why we present details of derivation in section~\ref{subsec: th-G1} and Appendix B. 

\section{Conclusion}\label{sec: conclusion}
In the present paper, we derive the exact explicit expression for the particle propagator of the spin 1/2 Calogero-Sutherland model and discuss physical properties and interpretation of the spectral functions in the full range of the energy and momentum space. Combining this result with the result on the hole propagator obtained in our previous paper\cite{hp}, we obtained a full knowledge of the single-particle Green's function of this model. 

\ack
This work was supported in part by 
Global COE Program  ``the Physical Sciences Frontier'', MEXT, Japan.

\appendix

\section{Derivation of scalar products and the matrix elements related to $G^{(1)}$ in finite systems}\label{FiniteDetail}

In this section, we derive the finite-size representation of the scalar product $\{P^{(\lambda')}_\nu,P^{(\lambda')}_\nu\}_{N+1,\lambda}$ and $\psi_{\nu\mu}^{(\lambda')}$ for $\nu$ satisfying (\ref{eq: 0L0Rlambdahole}).
\begin{figure}[t]
 \begin{center}
\unitlength 0.1in
\begin{picture}( 38.0000, 28.8000)(  4.0000,-30.1000)
\put(34.0000,-5.5500){\makebox(0,0)[lb]{III}}%
%
\special{pn 8}%
\special{pa 1800 406}%
\special{pa 2800 406}%
\special{pa 2800 606}%
\special{pa 1800 606}%
\special{pa 1800 406}%
\special{fp}%
%
\special{pn 8}%
\special{pa 2800 606}%
\special{pa 3200 606}%
\special{fp}%
\special{pa 2800 406}%
\special{pa 4200 406}%
\special{fp}%
\special{pa 4200 406}%
\special{pa 4200 606}%
\special{fp}%
\special{pa 4200 606}%
\special{pa 3800 606}%
\special{fp}%
%
\special{pn 8}%
\special{pa 1800 606}%
\special{pa 1400 606}%
\special{fp}%
\special{pa 1800 406}%
\special{pa 400 406}%
\special{fp}%
\special{pa 400 406}%
\special{pa 400 606}%
\special{fp}%
\special{pa 400 606}%
\special{pa 800 606}%
\special{fp}%
%
\special{pn 8}%
\special{pa 1800 406}%
\special{pa 1800 3006}%
\special{fp}%
\special{pa 1800 3006}%
\special{pa 400 3006}%
\special{fp}%
\special{pa 400 3006}%
\special{pa 400 406}%
\special{fp}%
%
\special{pn 8}%
\special{pa 1800 2606}%
\special{pa 2000 2606}%
\special{fp}%
\special{pa 2000 2606}%
\special{pa 2000 2206}%
\special{fp}%
%
\special{pn 8}%
\special{pa 800 606}%
\special{pa 1400 606}%
\special{fp}%
\special{pa 3200 606}%
\special{pa 3800 606}%
\special{fp}%
\put(22.0000,-5.5500){\makebox(0,0)[lb]{II}}%
\put(22.0000,-11.5500){\makebox(0,0)[lb]{V}}%
\put(10.0000,-11.5500){\makebox(0,0)[lb]{IV}}%
\put(10.0000,-5.5500){\makebox(0,0)[lb]{I}}%
%
\special{pn 8}%
\special{pa 2800 606}%
\special{pa 2800 1006}%
\special{fp}%
%
\special{pn 8}%
\special{sh 1}%
\special{ar 2100 1810 10 10 0  6.28318530717959E+0000}%
\special{sh 1}%
\special{ar 2300 1610 10 10 0  6.28318530717959E+0000}%
\special{sh 1}%
\special{ar 2500 1410 10 10 0  6.28318530717959E+0000}%
%
\special{pn 8}%
\special{pa 3200 350}%
\special{pa 2800 350}%
\special{fp}%
\special{sh 1}%
\special{pa 2800 350}%
\special{pa 2868 370}%
\special{pa 2854 350}%
\special{pa 2868 330}%
\special{pa 2800 350}%
\special{fp}%
\special{pa 3200 350}%
\special{pa 4200 350}%
\special{fp}%
\special{sh 1}%
\special{pa 4200 350}%
\special{pa 4134 330}%
\special{pa 4148 350}%
\special{pa 4134 370}%
\special{pa 4200 350}%
\special{fp}%
%
\special{pn 8}%
\special{pa 2200 350}%
\special{pa 1800 350}%
\special{fp}%
\special{sh 1}%
\special{pa 1800 350}%
\special{pa 1868 370}%
\special{pa 1854 350}%
\special{pa 1868 330}%
\special{pa 1800 350}%
\special{fp}%
\special{pa 2200 350}%
\special{pa 2800 350}%
\special{fp}%
\special{sh 1}%
\special{pa 2800 350}%
\special{pa 2734 330}%
\special{pa 2748 350}%
\special{pa 2734 370}%
\special{pa 2800 350}%
\special{fp}%
%
\special{pn 8}%
\special{pa 800 350}%
\special{pa 400 350}%
\special{fp}%
\special{sh 1}%
\special{pa 400 350}%
\special{pa 468 370}%
\special{pa 454 350}%
\special{pa 468 330}%
\special{pa 400 350}%
\special{fp}%
\special{pa 800 350}%
\special{pa 1800 350}%
\special{fp}%
\special{sh 1}%
\special{pa 1800 350}%
\special{pa 1734 330}%
\special{pa 1748 350}%
\special{pa 1734 370}%
\special{pa 1800 350}%
\special{fp}%
%
\special{pn 8}%
\special{pa 1730 810}%
\special{pa 1730 410}%
\special{fp}%
\special{sh 1}%
\special{pa 1730 410}%
\special{pa 1710 478}%
\special{pa 1730 464}%
\special{pa 1750 478}%
\special{pa 1730 410}%
\special{fp}%
\special{pa 1730 810}%
\special{pa 1730 3010}%
\special{fp}%
\special{sh 1}%
\special{pa 1730 3010}%
\special{pa 1750 2944}%
\special{pa 1730 2958}%
\special{pa 1710 2944}%
\special{pa 1730 3010}%
\special{fp}%
\put(15.2000,-29.1000){\makebox(0,0)[lb]{$N$}}%
%
\put(20.9000,-2.9000){\makebox(0,0)[lb]{}}%
\put(18.5000,-3.3000){\makebox(0,0)[lb]{$\lambda'(=2\lambda+1)$}}%
\put(34.0000,-3.0000){\makebox(0,0)[lb]{$\rho_{{\rm L}}$}}%
\put(10.0000,-3.0000){\makebox(0,0)[lb]{$\rho_{{\rm R}}$}}%
%
\special{pn 8}%
\special{pa 2800 1000}%
\special{pa 2800 1200}%
\special{fp}%
\special{pa 2800 1200}%
\special{pa 2600 1200}%
\special{fp}%
%
\special{pn 8}%
\special{pa 400 410}%
\special{pa 400 310}%
\special{fp}%
%
\special{pn 8}%
\special{pa 1800 410}%
\special{pa 1800 310}%
\special{fp}%
%
\special{pn 8}%
\special{pa 2800 410}%
\special{pa 2800 310}%
\special{fp}%
%
\special{pn 8}%
\special{pa 4200 410}%
\special{pa 4200 310}%
\special{fp}%
\end{picture}%
 \end{center}
 \caption{Decomposition of the Young diagram in $D(\nu_+)\cup D(\tilde{\nu}_+)$.}
 \label{part}
\end{figure}
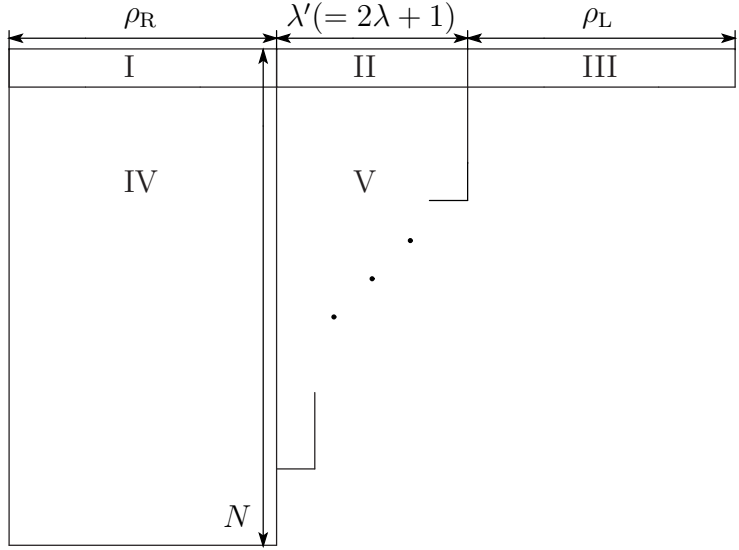
\subsection{The scalar product $\{P^{(\lambda')}_\nu,P^{(\lambda')}_\nu\}_{N+1,\lambda}$}
\label{subsec: normofnu}
For later convenience, we set 
\begin{eqnarray*}
\tilde{\nu}&=&(\lambda',\nu_2,\cdots,\nu_N,0)=(\lambda',\zeta_1,\cdots,\zeta_{N-1},0)\in \Lambda_{N+1}
\end{eqnarray*}
and decompose the scalar product $\{P^{(\lambda')}_\nu,P^{(\lambda')}_\nu\}_{N+1,\lambda}$ as
$$
\{P^{(\lambda')}_\nu,P^{(\lambda')}_\nu\}_{N+1,\lambda}
= \{P^{(\lambda')}_{\tilde{\nu}},P^{(\lambda')}_{\tilde{\nu}}\}_{N+1,\lambda} 
\times \mathcal{N}_{\nu\tilde{\nu}}
$$
with
$$
\mathcal{N}_{\nu\tilde{\nu}}=
\frac{\{P^{(\lambda')}_\nu,P^{(\lambda')}_\nu \}_{N+1,\lambda}}
     {\{P^{(\lambda')}_{\tilde{\nu}},P^{(\lambda')}_{\tilde{\nu}}\}_{N+1,\lambda}}. 
$$
\subsubsection{Expression for $\{P^{(\lambda')}_{\tilde{\nu}},P^{(\lambda')}_{\tilde{\nu}}\}_{N+1,\lambda}$}
First we consider $\{P^{(\lambda')}_{\tilde{\nu}},P^{(\lambda')}_{\tilde{\nu}}\}_{N+1,\lambda}$, which is given from (\ref{norm-nu-def}) by 
\begin{equation}
\frac{c^{(\lambda',2)}_{N+1} Y'_{\tilde{\nu}}(1)Z_{\tilde{\nu}}(1/(2\lambda'))}{Y'_{\tilde{\nu}}((\lambda'+1)/(2\lambda'))Z_{\tilde{\nu}}(1/2)}.\label{norm-tildenu-def}
\end{equation}
$Z_{\tilde{\nu}}(r)$ is decoupled into the contribution from the first row in $D(\tilde{\nu})$ and other rows. In the first row, $a(s)=\lambda'-j$ and $l(s)=\zeta'_j$ and hence $s\in H_2(\tilde{\nu})$ in the first row is given by $s=(1,j)$ with $\sigma_j=-1/2$ . We thus obtain 
\begin{eqnarray}
\prod_{s=(1,j)\in H_2(\tilde{\nu})}\left(\frac{a(s)}{2\lambda'}+\frac{l(s)}{2}+r\right)&=&\prod_{j=1}^{\lambda'}\left(\frac{\zeta'_j+1}{2}-\frac{j}{2\lambda'}+r\right)^{\delta_{\sigma_j,\downarrow}}\nonumber\\
&=&\prod_{j=1}^{\lambda'}\left(\frac{\check{\zeta}_j-\check{\zeta}_{\lambda'+1}-1/\lambda'}{2}+r\right)^{\delta_{\sigma_j,\downarrow}}.
\end{eqnarray}
Noting that the contribution from the other rows is given by $Z_\zeta(r)$ with $\zeta=(\zeta_1,\cdots,\zeta_N)\in \Lambda_N$, we obtain
\begin{equation}
 Z_{\tilde{\nu}}(r)=Z_\zeta(r)\prod_{j=1}^{\lambda'}\left(\frac{\check{\zeta}_j-\check{\zeta}_{\lambda'+1}-1/\lambda'}{2}+r\right)^{\delta_{\sigma_j,\downarrow}}.\end{equation}
The expression for $Z_\zeta(r)$ is available in (96) in \cite{hp} with replacement of $\tilde{\mu'}_j$ by $\tilde{\zeta'}_j$  and it is given by
\begin{eqnarray}
&Z_{\tilde{\nu}}(r)=\Gamma\left[r+1/2\right]^{-\lambda'}
  \prod_{j=1}^{\lambda'}
  \Gamma[(\check{\zeta}_j-\check{\zeta}_{\lambda'+1}+1-1/\lambda'+\delta_{\sigma_j,\sigma_{\lambda'+1}})/2+r]\nonumber\\
 &\qquad\times \prod_{1\le j<k\le \lambda'}
  \frac{\Gamma\left[(\check{\zeta}_j-\check{\zeta}_k+1-1/\lambda'-\delta_{\sigma_j,\sigma_k})/2+r\right]}
  {\Gamma\left[(\check{\zeta}_j-\check{\zeta}_k+\delta_{\sigma_j,\sigma_k})/2+r\right]}.\label{ztildenufinal}
\end{eqnarray}
On the other hand, $Y'_{\tilde{\nu}}(r)$ is decoupled into the product of the contribution from the squares $s=(\zeta'_j+1,j)$ at the bottom of each column and the remaining squares. When $s=(\zeta'_j+1,j)\in D(\tilde{\nu})\setminus C_2(\tilde{\nu})$, $\sigma_j=-1/2$ and we obtain 
\begin{eqnarray}
&&\prod_{s=(\zeta'_j+1,j)\in D(\tilde{\nu})\setminus C_2(\tilde{\nu})}\left(\frac{a'(s)}{2\lambda'}+\frac{N-1-l'(s)}{2}+r\right)\nonumber\\
&&\quad\quad=
\prod_{j=1}^{\lambda'}\left(\frac{\check{\zeta}_0 -\check{\zeta}_j-1/\lambda'}{2}+r\right)^{\delta_{\sigma_j,\downarrow}}.
\end{eqnarray}
The remaining contribution to $Y'_{\tilde{\nu}}(r)$ is given by $Y'_{\zeta}(r)$ and as a result, $Y'_{\tilde{\nu}}(r)$ is written as
\begin{equation}
 Y'_{\tilde{\nu}}(r)=Y'_\zeta(r)\prod_{j=1}^{\lambda'}
\left(\frac{\check{\zeta}_0 -\check{\zeta}_j-1/\lambda'}{2}+r\right)^{\delta_{\sigma_j,\downarrow}}.
\label{Yprimenu}
\end{equation}
The expression for $Y'_\zeta(r)$ has been derived in \cite{hp2} as
\begin{equation}
Y'_\zeta(r)=\prod_{j=1}^{\lambda'}\frac{\Gamma\left[N/2 +(j-1)/\lambda'+r\right]}{\Gamma\left[(\check{\zeta}_0-\check{\zeta}_j+2-1/\lambda'-\delta_{\sigma_0,\sigma_j})/2+r\right]}.
\label{Yprimezeta}
\end{equation}
From (\ref{Yprimenu}) and (\ref{Yprimezeta}), we obtain 
\begin{eqnarray}
 &Y'_{\tilde{\nu}}(r)=\prod_{j=1}^{\lambda'}\frac{\Gamma\left[N/2 +(j-1)/\lambda'+r\right]}{\Gamma\left[(\check{\zeta}_0-\check{\zeta}_j-1/\lambda'+\delta_{\sigma_0,\sigma_j})/2+r\right]}\label{Yprimenu-final}.
\end{eqnarray}

The scalar product $\{P^{(\lambda')}_{\tilde{\nu}},P^{(\lambda')}_{\tilde{\nu}}\}_{N+1,\lambda}$ is explicitly given by
\begin{eqnarray}
&&\{P^{(\lambda')}_{\tilde{\nu}},P^{(\lambda')}_{\tilde{\nu}}\}_{N+1,\lambda}\nonumber\\
&&=\frac{c^{(\lambda',2)}_{N+1}}{(\lambda')^\lambda\left(\Gamma\left[\frac{\lambda'+1}{2\lambda'}\right]\right)^{\lambda'}}
\frac{\Gamma\left[\left(1+\frac{N}{2}\right)\lambda'\right]}
     {\Gamma\left[\frac{1+(N+1)\lambda'}{2}\right]}
\nonumber\\
&&\times\prod_j
\frac{\Gamma\left[(\check{\zeta}_0-\check{\zeta}_j +1+\delta_{\sigma_j,\sigma_0})/2\right]}
{\Gamma\left[(\check{\zeta}_0-\check{\zeta}_j +2-1/\lambda'+\delta_{\sigma_j,\sigma_0})/2\right]}
\frac{\Gamma\left[(\check{\zeta}_j-\check{\zeta}_{\lambda'+1} +1+\delta_{\sigma_j,\sigma_{\lambda'+1}})/2\right]}
{\Gamma\left[(\check{\zeta}_j-\check{\zeta}_{\lambda'+1} +2-1/\lambda'+\delta_{\sigma_j,\sigma_{\lambda'+1}})/2\right]}
\nonumber\\
&&\times\prod_{j<k}\frac{\Gamma\left[(\check{\zeta}_j-\check{\zeta}_k +1-\delta_{\sigma_j,\sigma_k})/2\right]\Gamma\left[(\check{\zeta}_j-\check{\zeta}_k +1+\delta_{\sigma_j,\sigma_k})/2\right]}
{\Gamma\left[(\check{\zeta}_j-\check{\zeta}_k +2-1/\lambda'-\delta_{\sigma_j,\sigma_k})/2\right]\Gamma\left[(\check{\zeta}_j-\check{\zeta}_k +1/\lambda'+\delta_{\sigma_j,\sigma_k})/2\right]}.\label{eq: normofnutilde}
\end{eqnarray}
\subsubsection{Expression for $\mathcal{N}_{\nu,\tilde{\nu}}$}
The expression $\mathcal{N}_{\nu,\tilde{\nu}}$ is given by
$$
\mathcal{N}_{\nu\tilde{\nu}}=
\frac{\{P^{(\lambda')}_{\nu_+},P^{(\lambda')}_{\nu_+} \}_{N+1,\lambda}}
     {\{P^{(\lambda')}_{\tilde{\nu}_+},P^{(\lambda')}_{\tilde{\nu}_+}\}_{N+1,\lambda}},
$$
where
\begin{eqnarray}
\nu_+&=&(\nu_1 +\rho_{\rm R},\cdots,\nu_{N+1}+\rho_{\rm R})\nonumber\\
     &=&(\lambda'+\rho_{\rm L}+\rho_{\rm R},\zeta_1+\rho_{\rm R},\cdots,\zeta_{N-1}+\rho_{\rm R},0)\in \Lambda_{N+1}
\end{eqnarray}
and
\begin{eqnarray}
\tilde{\nu}_+&=&(\tilde{\nu}_1+\rho_{\rm R},\cdots,\tilde{\nu}_{N+1}+\rho_{\rm R})\nonumber\\
             &=&(\lambda'+\rho_{\rm R},\zeta_1+\rho_{\rm R},\cdots,\zeta_{N-1}+\rho_{\rm R}
,\rho_{\rm R})\in \Lambda_{N+1}.
\end{eqnarray}
Let $\mathcal{N}_{\nu,\tilde{\nu}}$ be decoupled into following five parts in $D(\nu_+)\cup D(\tilde{\nu}_+)$:
\begin{enumerate}
\item[(I)] 
the contribution $\mathcal{N}^{(\rm I)}$ from $s=(1,j)$ with $j\in [1,\rho_{\rm R}]$. 
\item[(II)]
the contribution $\mathcal{N}^{(\rm II)}$ from $s=(1,j)$ with $j\in [\rho_{\rm R} +1,\rho_{\rm R}+\lambda']$.
\item[(III)]
the contribution $\mathcal{N}^{(\rm III)}$ 
from $s=(1,j)$ with $j\in [\rho_{\rm R}+\lambda'+1,\rho_{\rm L}+\rho_{\rm R}+\lambda']$.

\item[(IV)]
 the contribution $\mathcal{N}^{(\rm IV)}$ from $s=(i,j)$ with $i\in [2,N+1]$ and $j\in [1,\rho_{\rm R}]$.
\item[(V)]
the contribution $\mathcal{N}^{(\rm V)}$ 
from $s=(i,j)$ with $i\in [2,\zeta'_1 +1]$ and $j\in [\rho_{\rm R} +1,\rho_{\rm R}+\zeta_{i-1}]$.
\end{enumerate}
See also Fig.~\ref{part}. 

The contribution $\mathcal{N}^{(\rm I)}$ results from $Z_{\nu}(1/(2\lambda'))/Z_{\nu}(1/2)$ divided by $Z_{\tilde{\nu}}(1/(2\lambda'))/Z_{\tilde{\nu}}(1/2)$ and is explicitly expressed as
\begin{eqnarray}
\mathcal{N}^{(\rm I)}
=&\frac{\Gamma\left[(1+\lambda'(N+1))/2\right]}{\Gamma\left[\lambda'(N+2)/2\right]}
 \frac
     {\Gamma\left[(\tilde{\rho}_{\rm L}-\tilde{\rho}_{{\rm R},0}+\lambda'+\delta_{\tau_{\rm L},\uparrow})/2\right]}
     {\Gamma\left[
(\tilde{\rho}_{\rm L}-\tilde{\rho}_{{\rm R},0}+1+\delta_{\tau_{\rm L},\uparrow})/2
            \right]}
\nonumber\\
&\times \frac{\Gamma\left[(\tilde{\rho}_{{\rm L},0}-\tilde{\rho}_{\rm R}+2\lambda'+\delta_{\tau_{\rm R},\uparrow})/2\right]}{\Gamma\left[
(\tilde{\rho}_{{\rm L},0}-\tilde{\rho}_{\rm R}+\lambda'+1+\delta_{\tau_{\rm R},\uparrow})/2
            \right]}
     \frac{\Gamma\left[(\tilde{\rho}_{\rm L}-\tilde{\rho}_{\rm R}+2-\lambda'-\delta_{\tau_{\rm L},\tau_{\rm R}})/2\right]}{\Gamma\left[(\tilde{\rho}_{\rm L}-\tilde{\rho}_{\rm R}+1-\delta_{\tau_{\rm L},\tau_{\rm R}})/2\right]}\label{mathcalNI}.  \nonumber\\
\end{eqnarray}
Similarly, $\mathcal{N}^{(\rm II)}$ and $\mathcal{N}^{(\rm III)}$ are expressed as 
\begin{eqnarray}
\mathcal{N}^{(\rm II)}=
 \prod_{j}
 \left(\frac{\check{\zeta}_j -\check{\zeta}_{\lambda'+1}+1-1/\lambda'}
            {\check{\zeta}_j -\check{\zeta}_{\lambda'+1}}
 \right)^{1-\delta_{\sigma_j\uparrow}}
 \left(\frac{\tilde{\rho}_{\rm L}+\lambda'\check{\zeta}_j-(\lambda'-1)/2}
            {\tilde{\rho}_{\rm L}+\lambda'\check{\zeta}_j+(\lambda'-1)/2}
 \right)^{1-\delta_{\sigma_j\tau_{\rm L}}}\label{mathcalNII}
\end{eqnarray}
and 
\begin{eqnarray}
&\mathcal{N}^{(\rm III)}= 
 \frac{\Gamma\left[(1+\lambda')/2\right]
  \Gamma\left((\tilde{\rho}_{\rm L}-\tilde{\rho}_{{\rm L},0}-\lambda'+1+\delta_{\tau_{\rm L},\uparrow})/2)\right)}{\Gamma\left[
	      (\tilde{\rho}_{\rm L}-\tilde{\rho}_{{\rm L},0}+\delta_{\tau_{\rm L},\uparrow})/2)\right]}\nonumber\\
 &\times\frac
 {\Gamma\left[(\tilde{\rho}_{{\rm L},0}-\tilde{\rho}_{\rm R}+\lambda'+1+\delta_{\tau_{\rm R},\uparrow})/2
	\right]}
 {\Gamma\left[(\tilde{\rho}_{{\rm L},0}-\tilde{\rho}_{\rm R}+
	     2\lambda'+\delta_{\tau_{\rm R},\uparrow})/2\right]}
 \frac
 {\Gamma\left[(\tilde{\rho}_{\rm L}-\tilde{\rho}_{\rm R}+\lambda'+\delta_{\tau_{\rm L},\tau_{\rm R}})/2\right]}
 {\Gamma\left[(\tilde{\rho}_{\rm L}-\tilde{\rho}_{\rm R}+1+
	      \delta_{\tau_{\rm L},\tau_{\rm R}})/2
	    \right]}\label{mathcalNIII}
\end{eqnarray}
The contribution to $\mathcal{N}^{(\rm IV)}$ consists of  $Z_{\nu}(1/(2\lambda'))/Z_{\nu}(1/2)$ divided by $Z_{\tilde{\nu}}(1/(2\lambda'))/Z_{\tilde{\nu}}(1/2)$ from the $i$th row with $i\in [2,N]$ and  $Y'_{\tilde{\nu}}(1/2+1/(2\lambda'))/Y'_{\tilde{\nu}}(1)$ from $N+1$th row in $D(\tilde{\nu}_+)$;
\begin{eqnarray}
\mathcal{N}^{(\rm IV)}= 
&\prod_{j}
 \left(\frac{\check{\zeta}_0-\check{\zeta}_j +1-1/\lambda'}
            {\check{\zeta}_0-\check{\zeta}_j }
 \right)
^{\delta_{\sigma_{j}\uparrow}}
 \left(\frac{-\tilde{\rho}_{\rm R}-\lambda'\check{\zeta}_j-(\lambda'-1)/2}
            {-\tilde{\rho}_{\rm R}-\lambda'\check{\zeta}_j+(\lambda'-1)/2}
 \right)
^{1-\delta_{\sigma_j,\tau_{\rm R}}} \nonumber\\
&\times \frac{\Gamma\left[\left(\tilde{\rho}_{{\rm L},0}-\tilde{\rho}_{\rm R}+\lambda'+\delta_{\tau_{\rm R},\downarrow}\right))/2\right]}
{\Gamma\left[\left(\tilde{\rho}_{{\rm L},0}-\tilde{\rho}_{\rm R}+1+\delta_{\tau_{\rm R},\downarrow}\right))/2\right]}
\frac{\Gamma\left[(2+\lambda'N)/2\right]}{\Gamma\left[(1+\lambda'(N+1))/2\right]}\nonumber\\
&\times
\frac
 {\Gamma\left[(1+\lambda')/2\right] \Gamma\left[(\tilde{\rho}_{{\rm R},0}-\tilde{\rho}_{\rm R}+1-\lambda'+\delta_{\tau_{\rm R},\downarrow})/2\right]
}{\Gamma\left[(\tilde{\rho}_{{\rm R},0}-\tilde{\rho}_{\rm R}+\delta_{\tau_{\rm R},\downarrow})/2
	     \right]}.\label{mathcalNIV}
\end{eqnarray}
It is easily seen that $\mathcal{N}^{\rm V}=1$ because the contribution from $D(\nu_+)$ exactly cancels with that from $D(\tilde{\nu}_+)$. 

\subsubsection{Final expression for $\{P^{(\lambda')}_{\nu},P^{(\lambda')}_{\nu}\}_{N+1,\lambda}$}
From (\ref{eq: normofnutilde}), (\ref{mathcalNI}), (\ref{mathcalNII}), (\ref{mathcalNIII}) and (\ref{mathcalNIV}), the final expression for the scalar product $\{P^{(\lambda')}_{\nu},P^{(\lambda')}_{\nu}\}_{N+1,\lambda}$ is given by
\begin{equation}
\{P^{(\lambda')}_{\nu},P^{(\lambda')}_{\nu}\}_{N+1,\lambda}=\mathcal{K}\mathcal{A}_\nu\mathcal{B}_\nu,
\label{Pnunormfinal}
\end{equation}
with
\begin{eqnarray}
\mathcal{K}&=&\frac{c_{N+1}^{(\lambda',2)}\left(\Gamma\left[\frac{1+\lambda'}{2}\right]\right)^2\Gamma\left[\frac{2+N\lambda'}{2}\right]}
{(\lambda')^{\lambda}\left(\Gamma\left[\frac{\lambda'+1}{2\lambda'}\right]\right)^{\lambda'}\Gamma\left[\frac{1+(N+1)\lambda'}{2}\right]},
\label{mathcalK}
\end{eqnarray}
\begin{eqnarray}
\mathcal{A}_\nu&=&\prod_{j<k}
\frac{\Gamma\left[(\check{\zeta}_j-\check{\zeta}_k+1-\delta_{\sigma_j,\sigma_k})/2\right]\Gamma\left[(\check{\zeta}_j-\check{\zeta}_k+1+\delta_{\sigma_j,\sigma_k})/2\right]}
     {\Gamma\left[(\check{\zeta}_j-\check{\zeta}_k+2-1/\lambda'-\delta_{\sigma_j,\sigma_k})/2\right]\Gamma\left[(\check{\zeta}_j-\check{\zeta}_k+1/\lambda'+\delta_{\sigma_j,\sigma_k})/2\right]}\nonumber\\
&\times&
\prod_{j}\left(\frac{\tilde{\rho}_{\rm L}+\lambda'\check{\zeta}_j-(\lambda'-1)/2}{\tilde{\rho}_{\rm L}+\lambda'\check{\zeta}_j+(\lambda'-1)/2}\right)^{1-\delta_{\sigma_j\tau_{\rm L}}}
\left(\frac{\tilde{\rho}_{\rm R}+\lambda'\check{\zeta}_j+(\lambda'-1)/2}{\tilde{\rho}_{\rm R}+\lambda'\check{\zeta}_j-(\lambda'-1)/2}\right)^{1-\delta_{\sigma_j\tau_{\rm R}}}\nonumber\\
&\times&
\frac{\Gamma\left[(\tilde{\rho}_{\rm L}-\tilde{\rho}_{\rm R}+2-\lambda'-\delta_{\tau_{\rm L},\tau_{\rm R}})/2\right]\Gamma\left[(\tilde{\rho}_{\rm L}-\tilde{\rho}_{\rm R}+\lambda'+\delta_{\tau_{\rm L},\tau_{\rm R}})/2\right]}{\Gamma\left[(\tilde{\rho}_{\rm L}-\tilde{\rho}_{\rm R}+1-\delta_{\tau_{\rm L},\tau_{\rm R}})/2\right]\Gamma\left[(\tilde{\rho}_{\rm L}-\tilde{\rho}_{\rm R}+1+\delta_{\tau_{\rm L},\tau_{\rm R}})/2\right]},
\label{mathcalA}
\end{eqnarray}
and
\begin{eqnarray}
\mathcal{B}_\nu&=&\prod_{i=0,\lambda'+1}\prod_{j=1}^{\lambda'}
\frac{\Gamma\left[(|\check{\zeta}_i-\check{\zeta}_{j}|+1-\delta_{\sigma_i,\sigma_{j}})/2\right]
      }{
      \Gamma\left[(|\check{\zeta}_i-\check{\zeta}_j|+2-1/\lambda'-\delta_{\sigma_i,\sigma_j})/2\right]}\nonumber\\
&\times&
\frac
     {\Gamma\left[(\tilde{\rho}_{\rm L}-\tilde{\rho}_{{\rm R},0}+\lambda'+\delta_{\tau_{\rm L},\uparrow})/2\right]\Gamma\left((\tilde{\rho}_{\rm L}-\tilde{\rho}_{{\rm L},0}-\lambda'+1+\delta_{\tau_{\rm L},\uparrow})/2)\right)}
     {\Gamma\left[
(\tilde{\rho}_{\rm L}-\tilde{\rho}_{{\rm R},0}+1+\delta_{\tau_{\rm L},\uparrow})/2
            \right]\Gamma\left[
	      (\tilde{\rho}_{\rm L}-\tilde{\rho}_{{\rm L},0}+\delta_{\tau_{\rm L},\uparrow})/2)\right]}\nonumber\\
&\times&
\frac{\Gamma\left[\left(\tilde{\rho}_{{\rm L},0}-\tilde{\rho}_{\rm R}+\lambda'+\delta_{\tau_{\rm R},\downarrow}\right))/2\right]\Gamma\left[(\tilde{\rho}_{{\rm R},0}-\tilde{\rho}_{\rm R}+1-\lambda'+\delta_{\tau_{\rm R},\downarrow})/2\right]}
{\Gamma\left[\left(\tilde{\rho}_{{\rm L},0}-\tilde{\rho}_{\rm R}+1+\delta_{\tau_{\rm R},\downarrow}\right))/2\right]\Gamma\left[(\tilde{\rho}_{{\rm R},0}-\tilde{\rho}_{\rm R}+\delta_{\tau_{\rm R},\downarrow})/2
	     \right]}.\nonumber\\
\label{mathcalB}
\end{eqnarray}

\subsection{$\psi_{\nu,\mu}^{(\lambda')}$}\label{psinumu}
\begin{figure}
\input{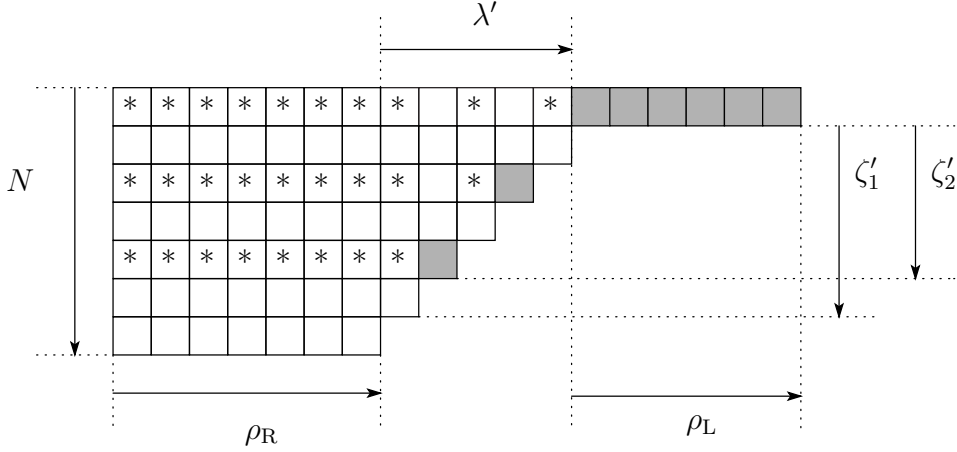}
\caption{Diagram of $\nu_+$. Shaded squares represent those in $D(\nu_+)\setminus D(\mu_+)$. The squares marked by $*$ represent those squares 
belonging to ${\rm C}_{\nu_+/\mu_+}\setminus {\rm R}_{\nu_+/\mu_+}$.}
\label{fig: horizontal-strip3}
\end{figure}

Let $\psi_{\nu,\mu}^{(\lambda')}$ be decoupled into four parts 
$$
\psi_{\nu,\mu}^{(\lambda')}=V_{\nu,\mu}^{({\rm I})}V_{\nu,\mu}^{({\rm II})}
V_{\nu,\mu}^{({\rm IV})}
V_{\nu,\mu}^{({\rm  V})}.
$$
Each part denotes, respectively,  
\begin{itemize}
\item 
the contributions $V_{\nu,\mu}^{({\rm I})}$ from $s=(1,j)\in H_2(\nu_+)$ and $H_2(\mu_+)$ with $j\in [1,\rho_{\rm R}]\cap {\rm C}_{\nu_+/\mu_+}$. 
\item
the contributions $V_{\nu,\mu}^{({\rm II})}$ from $s=(1,j)\in H_2(\nu_+)$ and $H_2(\mu_+)$ with $j\in [\rho_{\rm R} +1,\rho_{\rm R}+\lambda']\cap {\rm C}_{\nu_+/\mu_+}$.
\item
the contributions $V_{\nu,\mu}^{({\rm IV})} $ from $s=(i,j)\in H_2(\nu_+)$ and $H_2(\mu_+)$ with $i\in [2,N+1]$ and $j\in [1,\rho_{\rm R}]\cap{\rm C}_{\nu_+/\mu_+}$.
\item
the contributions $V_{\nu,\mu}^{({\rm V})} $ 
from $s=(i,j)\in H_2(\nu_+)$ and $H_2(\mu_+)$ with $i\in [2,\zeta'_1 +1]$ and $j\in [\rho_{\rm R} +1,\rho_{\rm R}+\zeta_{i-1}]\cap{\rm C}_{\nu_+/\mu_+}$.
\end{itemize}
The sets of squares which belong to (I)$\sim$(V) in $D(\nu_+)$ are shown in Fig.~\ref{part}. In Fig.~\ref{fig: horizontal-strip3}, the squares marked by $*$ represent squares belonging to ${\rm C}_{\nu_+/\mu_+}\setminus {\rm R}_{\nu_+/\mu_+}$ and contributing to $\psi_{\nu,\mu}^{(\lambda')}$. Obviously, there are no squares in (III) contributing to $\psi_{\nu,\mu}^{(\lambda')}$.
Each part is given, respectively, by
\begin{eqnarray}
V_{\nu,\mu}^{({\rm I})}=
 &\frac{\Gamma\left(1+\lambda'N/2\right)
\Gamma\left[(\tilde{\rho}_{{\rm L},0}-\tilde{\rho}_{\rm R}+\lambda'+\delta_{\tau_{\rm R},\downarrow})/2
	     \right]}
 {\Gamma\left((1+\lambda'(N+1))/2\right)\Gamma
\left[(\tilde{\rho}_{{\rm L},0}-\tilde{\rho}_{\rm R}+1+\delta_{\tau_{\rm R},\downarrow})/2
	     \right]}
\nonumber\\
 &\times\frac{\Gamma\left[(\tilde{\rho}_{\rm L}-\tilde{\rho}_{{\rm R},0}+\lambda'+\delta_{\tau_{\rm L},\uparrow})/2\right]
	      \Gamma\left[(\tilde{\rho}_{\rm L}-\tilde{\rho}_{\rm R}+2-\lambda'-\delta_{\tau_{\rm L},\tau_{\rm R}})/2\right]}
 {\Gamma\left[\tilde{\rho}_{\rm L}-\tilde{\rho}_{{\rm R},0}+1+\delta_{\tau_{\rm L},\uparrow})/2\right]
  \Gamma\left[(\tilde{\rho}_{\rm L}-\tilde{\rho}_{\rm R}+1-\delta_{\tau_{\rm L},\tau_{\rm R}})/2\right]}
\nonumber\\
\end{eqnarray}
\begin{eqnarray}
V_{\nu,\mu}^{({\rm II})}=
 \prod_{j\in J}
  \left(\frac{\check{\zeta}_j-\check{\zeta}_{\lambda'+1}+1-1/\lambda'}
  {\check{\zeta}_j-\check{\zeta}_{\lambda'+1}}\right)^{\delta_{\sigma_j,\sigma_{\lambda'+1}}}
  \left(\frac{\tilde{\rho}_{\rm L}+\lambda'\check{\zeta}_j+(1-\lambda')/2}
             {\tilde{\rho}_{\rm L}+\lambda'\check{\zeta}_j-(1-\lambda')/2}
\right)
  ^{1-\delta_{\sigma_j\tau_{\rm L}}}
\end{eqnarray}
\begin{eqnarray}
V_{\nu,\mu}^{({\rm IV})}=
 &\prod_{j\in I}
\left(\frac{\check{\zeta}_{0}-\check{\zeta}_j+1-1/\lambda'}
  {\check{\zeta}_0-\check{\zeta}_j}\right)^{\delta_{\sigma_j,\sigma_{0} }}
\left(\frac{\tilde{\rho}_{\rm R}+\lambda'\check{\zeta}_j+(\lambda'-1)/2}
             {\tilde{\rho}_{\rm R}+\lambda'\check{\zeta}_j-(\lambda'-1)/2}
\right)
  ^{1-\delta_{\sigma_j\tau_{\rm R}}}
\end{eqnarray}
\begin{eqnarray}
V_{\nu,\mu}^{({\rm V})}=
 \underset{{\rm s.t.}j\in J,k\in I}{\prod_{j<k}}
  &\left(\frac{\check{\zeta}_j-\check{\zeta}_k+1-1/\lambda'}
  {\check{\zeta}_j-\check{\zeta}_k}\right)^{\delta_{\sigma_j\sigma_k}}
\left(\frac{\check{\zeta}_j-\check{\zeta}_k+1/\lambda'}
  {\check{\zeta}_j-\check{\zeta}_k+1}\right)^{1-\delta_{\sigma_j\sigma_k}}.
\nonumber\\
\end{eqnarray}

\section{Proof of (i), (ii) and (iii) in Section~\ref{sec: th}}\label{proof}
\subsection{Outline of proof}
Let $F(u_1,\cdots,u_{\lambda'};S)$ be the following expression:
\begin{eqnarray}
F(u_1,\cdots,u_{\lambda'};S)=\underset{{\rm s.t.}\,|{\cal I}|=\lambda,\lambda+1}{\sum_{{\cal I}\subset[1,\lambda']}}(-1)^{|S\cap {\cal I}|}
 \,\,\underset{(j,k)\in ({\cal I},{\cal I})\cup(\bar{{\cal I}},\bar{\cal I})}{\prod_{j<k}}(u_k-u_j).\label{ap_non0}
\end{eqnarray}
The symbol $S$ denotes a subset of $\{1,2,\cdots,\lambda'\}$. 
In \ref{sec: App-B2}, we show that
\begin{eqnarray}
&&\underset{n_{\mu}=\lambda,\lambda+1,\,\,\mu_1\le \lambda'}
  {\underset{{\rm s.t.}\nu/\mu:{\rm h.s.}}
  {\sum_{\mu \in\Lambda_N}}}
  \prod_{j<k}
  (u_k-u_j)^{\delta_{\sigma_j^{\mu}\sigma_k^{\mu}}}
  \prod_{j\in {\rm odd}}e_j\prod_{j\in Q}e_j\nonumber\\
&&=(-1)^{|S(Q)\cap \bar{Q}^{(-)}_\sigma|}F(u_1,u_2,\cdots,u_{\lambda'};S(Q)), 
\label{eq: appB2}
\end{eqnarray}
with $S(Q)=({\rm odd}\cap \bar{Q})\cup({\rm even}\cap Q)$,
where $Q$ is a subset of $\{1,2,\cdots,\lambda'\}$,
$Q^{(-)}_\sigma$ is the set of columns $j\in[1,\lambda']$ such that $\zeta_j'-j$ is even,
and $n_{\mu}$ is the number of $j\in[1,\lambda']$ for which $\mu'_j-j$ is even.

In \ref{sec: App-B3}, we show that 
\begin{eqnarray}
&&F(u_1,\cdots,u_{\lambda'};S)\nonumber\\
&&=\left\{
\begin{array}{cc}
2^{\lambda+1}(-1)^{|S\cap {\rm even}|} \underset{j,k\in {\cal I}_S {\rm or} \bar{{\cal I}}_S}{\prod_{j<k}}(u_k-u_j), &\mbox{ if }|S\cap{\rm odd}|=|S\cap{\rm even}|\label{eq: goalofappB}\\
0,&\mbox{ otherwise}\\
\end{array}
\right.
\end{eqnarray}
with ${\cal I}_S=({\rm even}\cap S)\cup({\rm odd}\cap \bar{S})$.

We then arrive at the relation
\begin{eqnarray}
 &\underset{n_{\mu}=\lambda,\lambda+1,\,\,\mu_1\le \lambda'}
  {\underset{{\rm s.t.}\nu/\mu:{\rm h.s.}}
  {\sum_{\mu \in\Lambda_N}}}
  \prod_{j<k}
  (u_k-u_j)^{\delta_{\sigma_j^{\mu}\sigma_k^{\mu}}}
  \prod_{j\in {\rm odd}}e_j\prod_{j\in Q}e_j \nonumber\\
  &\qquad=
\left\{
\begin{array}{cc}
2^{\lambda+1}\varepsilon(Q,\{\sigma_j\})
  {\underset{j,k\in (Q,Q)\cup(\bar{Q},\bar{Q})}
  {\prod_{j<k}}}(u_k-u_j)&\mbox{ if }|\bar{Q}\cap{\rm odd}|=|Q\cap{\rm even}|\\
0,&\mbox{otherwise}\\
\end{array}
\right.
\label{eq: appBfinal}
\end{eqnarray}
with $\varepsilon(Q,\{\sigma_j\})=(-1)^{|S(Q)\cap \bar{Q}^{(-)}_\sigma|+|S(Q)\cap{\rm even}|}$, which coincides with (\ref{eq: signfactor}).
The relation (\ref{eq: appBfinal}) can be obtained using (\ref{eq: appB2}) and (\ref{eq: goalofappB}) and the relation ${\cal I}_{S(Q)}=Q$. The relation $|Q\cap \mbox{even}|=|\bar{Q}\cap \mbox{odd}|$ and $|Q|=\lambda+1$ follows from $|S(Q)\cap \mbox{even}|=|S(Q)\cap \mbox{odd}|$. 

The relations (ii) and (iii) in section~\ref{sec: th} immediately follow from (\ref{eq: appBfinal}). The relation (i) also follows from (\ref{eq: appBfinal}) when the powers of $e_j$ in monomials in the polynomial $f$ are less than two for all $j$. When the powers of $e_j$ for some $j$s are larger than one in $f$, the expression (\ref{Msumex}) reduces to that for a polynomial $g$ that contains monomials in which the powers of $e_j$ are less than two for all $j$. The order of $g$ is not larger than that of $f$. Thus (i) in section~\ref{sec: th} holds even when the powers of $e_j$ for some $j$s are larger than one in $f$.

\subsection{Proof of (\ref{eq: appB2})}\label{sec: App-B2}
In the left-hand side of (\ref{eq: appB2}), we first note that
\begin{equation}
\left[\prod_{j\in \mbox{{\footnotesize odd}}}e_j\right] 
\left[\prod_{j\in Q}e_j\right]=\prod_{j\in \underbrace{[\mbox{{\footnotesize odd}}\cap \bar{Q}]\cup[\mbox{{\footnotesize even}}\cap Q]}_{S(Q)}}e_j=(-1)^{|S(Q)\cap I|} 
\label{eq: appB-lhs1}
\end{equation}
We also note that the set $I$, which is defined as the set of columns so that $\zeta'_j=\mu'_j$, is expressed as
\begin{equation}
I=[Q_\mu \cap Q^{(-)}_\sigma]\cup[\bar{Q}_\mu \cap \bar{Q}^{(-)}_\sigma]
\label{eq: appB-lhs2}
\end{equation}
in terms of $Q_\mu$ ($Q^{(-)}_\sigma$) defined as the set of columns where $\mu'_j-j$ ($\zeta'_j-j$) is even. From (\ref{eq: appB-lhs1}) and (\ref{eq: appB-lhs2}), left-hand side of (\ref{eq: appB2}) becomes
\begin{eqnarray}
&&\mbox{LHS of (\ref{eq: appB2})}=\underset{|Q_{\mu}|=\lambda,\lambda+1}{\sum_{Q_\mu\in [1,\lambda']}}
(-1)^{|S(Q)\cap [[Q_\mu \cap Q^{(-)}_\sigma]\cup[\bar{Q}_\mu \cap \bar{Q}^{(-)}_\sigma]]|}
\underset{j,k\in (Q_\mu,Q_\mu)\cup(\bar{Q}_\mu,\bar{Q}_\mu)}{\prod_{j<k}(u_k-u_j)}.\nonumber\\
\label{eq: LHS-2}
\end{eqnarray}
We see that the relation
\begin{eqnarray}
&&(-1)^{|S(Q)\cap [[Q_\mu \cap Q^{(-)}_\sigma]\cup[\bar{Q}_\mu \cap \bar{Q}^{(-)}_\sigma]]|}
(-1)^{|S(Q)\cap \bar{Q}^{(-)}_\sigma|}(-1)^{|S(Q)\cap Q_\mu|}\nonumber\\
&&=(-1)^{|S(Q)\cap Q_\mu \cap Q^{(-)}_\nu|}
\underbrace{(-1)^{|S(Q)\cap\bar{Q}_\mu \cap \bar{Q}^{(-)}_\sigma|}(-1)^{|S(Q)\cap \bar{Q}^{(-)}_\sigma|}}_{(-1)^{|S(Q)\cap Q_\mu \cap \bar{Q}^{(-)}_\sigma|}}(-1)^{|S(Q)\cap Q_\mu|}
=1\nonumber\\
\end{eqnarray}
holds and hence we obtain
\begin{eqnarray}
&&\mbox{LHS of (\ref{eq: appB2})}=(-1)^{|S(Q)\cap \bar{Q}^{(-)}_\sigma|}\underset{|Q_{\mu}|=\lambda,\lambda+1}{\sum_{Q_\mu\in [1,\lambda']}}
(-1)^{|S(Q)\cap Q_\mu|}\underset{j,k\in (Q_\mu,Q_\mu)\cup(\bar{Q}_\mu,\bar{Q}_\mu)}{\prod_{j<k}(u_k-u_j)}.\nonumber\\
\label{eq: RHS-2}
\end{eqnarray}
Replacing the dummy index $Q_\mu$ in (\ref{eq: RHS-2}) by ${\cal I}$, (\ref{eq: RHS-2}) becomes RHS of (\ref{eq: appB2}). 
\subsection{Proof of (\ref{eq: goalofappB})}\label{sec: App-B3}
First, we show that $F(u_1,\cdots,u_{\lambda'};S)$ is zero when $|S|$ is odd. 
In the right-hand side of (\ref{ap_non0}), the summands for ${\cal I}={\cal I}'$ with $|{\cal I}'|=\lambda$ and its counterpart $\bar{{\cal I}'}\equiv[1,\lambda']\setminus {\cal I}'$ 
with $|\bar{{\cal I}'}|=\lambda+1$ differ only in their signs $(-1)^{|S\cap {\cal I}|}$ in (\ref{ap_non0}).
Noting
\begin{eqnarray}
  (-1)^{|S\cap {\cal I}'|} (-1)^{|S\cap \bar{{\cal I}}'|}=(-1)^{|S|},
\end{eqnarray}
we see that the summand for ${\cal I}'$ and $\bar{{\cal I}}'$ cancel with each other and consequently $F(u_1,\cdots,u_{\lambda'};S)$ vanishes when $|S|$ is odd.
In the following, we consider $|S|$ to be even. 
Since the sums over ${\cal I}$ satisfying $|{\cal I}|$ being $\lambda$ and $|{\cal I}|$ being $\lambda+1$ give the same contribution, we only consider the sum over ${\cal I}$ with $|{\cal I}|$ being $\lambda+1$. Twice the result gives $F(u_1,\cdots,u_{\lambda'};S)$.

Now we expand (\ref{ap_non0}) with respect to monomials of $u_1, u_2,\cdots $ and $u_{\lambda'}$,  
\begin{eqnarray}
&&F(u_1,\cdots,u_{\lambda'};S)\nonumber\\
&&=2\sum_{(A,B)\in S_{\lambda'}}{\rm sgn}(A){\rm sgn}(B)
(-1)^{|A\cap S|} u_{a_1}^0u_{a_2}^1\cdots u_{a_{\lambda}}^{\lambda-1}
  u_{a_{\lambda+1}}^{\lambda}u_{b_1}^0
  u_{b_2}^1\cdots u_{b_{\lambda}}^{\lambda-1}.
\label{eq: expandFu}
\nonumber\\
\end{eqnarray} 
Here the sum with respect to $(A,B)$ with $A=(a_1,\cdots,a_{\lambda+1})$ and $B=(b_1,\cdots,b_{\lambda})$ runs over all permutations of $(1,2,\cdots,\lambda')$. 
We denote by ${\rm sgn}(A)$ the signature of the permutation $\sigma\in S_{\lambda+1}$ such that $a_{\sigma(1)}>a_{\sigma(2)}>\cdots a_{\sigma(\lambda+1)}$. The symbol ${\rm sgn}(B)$ is defined in the same way. 
The symbol $(-1)^{|A\cap S|}$ is defined as $(-1)^{\sharp\{a_i\in S|i\in [1,\lambda+1]\}}$. 
The polynomial 
\begin{eqnarray}
 u_{a_1}^0u_{a_2}^1\cdots u_{a_{\lambda}}^{\lambda-1}
  u_{a_{\lambda+1}}^{\lambda}u_{b_1}^0
  u_{b_2}^1\cdots u_{b_{\lambda}}^{\lambda-1}, \label{ap_term}
\end{eqnarray}
is obviously invariant with 
respect to interchange $a_i$ and $b_i$ for $i\in[1,\lambda]$.
Thus there appear $2^{\lambda}$ terms of the form (\ref{ap_term}) in the right-hand side of (\ref{eq: expandFu}). 

We seek for the condition that the polynomial (\ref{ap_term}) survives after the summation in (\ref{eq: expandFu}). First we consider the relative sign of ${\rm sgn}(A){\rm sgn}(B)(-1)^{|A\cap S|}$ and 
${\rm sgn}(A'){\rm sgn}(B')(-1)^{|A'\cap S|}$, where
$$
 \begin{array}{l}
  A'=\{a_1,\cdots,a_{i-1},b_i,a_{i+1},\cdots,a_{\lambda+1}\}\\
  B'=\{b_1,\cdots,b_{i-1},a_i,b_{i+1},\cdots,b_{\lambda}\}
  \end{array}.
$$
We assume $a_i<b_i$ for simplicity. The opposite case $a_i>b_i$ can be discussed in a similar way.  
The relative sign ${\rm sgn}(A')/{\rm sgn}(A)$ is given by
$(-1)^{\sharp\{a_j (j\neq i)|a_j \in [a_i,b_i]\}}$,
and ${\rm sgn}(B')/{\rm sgn}(B)$ by $(-1)^{\sharp\{b_j (j\neq i)|b_j \in [a_i,b_i]\}}$. The other factor $(-1)^{|A'\cap S|-|A\cap S|}$ is given by 
$(-1)^{\sharp\{a_i,b_i \in S\}}$.
We thus obtain 
\begin{eqnarray}
\frac{{\rm sgn}(A'){\rm sgn}(B')(-1)^{|A'\cap S|}}
     {{\rm sgn}(A){\rm sgn}(B)(-1)^{|A\cap S|}}=(-1)^{a_i-b_i-1}(-1)^{\sharp\{a_i,b_i \in S\}}. \label{ap_cond}
\end{eqnarray}
This result depends only on $a_i-b_i$ and whether $a_i$ or $b_i$
belongs to $S$, not on $a_j$ or $b_j$ with $j(\ne i)$.
Taking account of (\ref{ap_cond}), we rewrite (\ref{eq: expandFu}) as we explain below.
First we define an equivalent relation as
\begin{eqnarray}
&&(A',B')=(a'_1,\cdots,a'_{\lambda+1},b'_1,\cdots,b'_{\lambda})\sim (A,B),\nonumber\\
&&\mbox{ if}\quad a'_i=\left\{
\begin{array}{cc}
a_i,\quad & i\in[1,\lambda+1]\setminus J\\
b_i,\quad & i\in J\\
\end{array}
\right.
\quad
b'_i=\left\{
\begin{array}{cc}
b_i,\quad & i\in[1,\lambda]\setminus J\\
a_i,\quad & i\in J\\
\end{array}
\right.,\nonumber\\ 
\end{eqnarray}
with a set $J\subset[1,\lambda]$. We denote the equivalent class with the representative $(A,B)$ by $[(A,B)]=\{(A',B')\in S_{\lambda'}|(A',B')\sim (A,B)\}$. Further we denote the quotient set of $S_{\lambda'}$ with respect to $\sim $ by 
$S_{\lambda'}/\sim$. With these notations, (\ref{eq: expandFu}) is rewritten as
  \begin{eqnarray}
F(u_1,\cdots,u_{\lambda'};S)&=&2\sum_{[(A,B)]\in S_{\lambda'}/\sim}
 u_{a_1}^0u_{a_2}^1\cdots u_{a_{\lambda}}^{\lambda-1}
  u_{a_{\lambda+1}}^{\lambda}u_{b_1}^0
  u_{b_2}^1\cdots u_{b_{\lambda}}^{\lambda-1}\nonumber\\
&&\times\sum_{(A',B')\sim (A,B)}{\rm sgn}(A'){\rm sgn}(B')
(-1)^{|A'\cap S|}. \label{eq: expandFu-rewrite}
\end{eqnarray} 
Considering the result (\ref{ap_cond}), we see that the second line of the right-hand side of (\ref{eq: expandFu-rewrite}) becomes
\begin{eqnarray}
&&\sum_{(A',B')\sim (A,B)}{\rm sgn}(A'){\rm sgn}(B')
(-1)^{|A'\cap S|}\nonumber\\
&&={\rm sgn}(A){\rm sgn}(B)
(-1)^{|A\cap S|}\prod_{i=1}^\lambda\left(
1+
(-1)^{a_i-b_i-1}(-1)^{\sharp\{a_i,b_i \in S\}}\right).\label{eq: sgmAsgmB}
\end{eqnarray}
Thus $[(A,B)]$ contributes to the sum in (\ref{eq: expandFu-rewrite}) only when the condition
\begin{equation}
(-1)^{a_i-b_i-1}(-1)^{\sharp\{a_i,b_i \in S\}}=1$\quad\mbox{ for all }$i\in [1,\lambda] \label{eq: abcond}
\end{equation}
is satisfied.

Next we prove that 
\begin{itemize}
\item a sequence $(A,B)\in S_{\lambda'}$ satisfying
(\ref{eq: abcond}) exists only when the relation 
\begin{equation}
|S\cap{\rm even}|=|S\cap{\rm odd}| 
\label{eq: Seven_Sodd}
\end{equation}
holds.
\item
As a representative of the equivalent class $[(A,B)]$ with $(A,B)$ satisfying
(\ref{eq: abcond}), we can choose $(A,B)$ satisfying
\begin{eqnarray}
&&\{a_1,\cdots,a_{\lambda+1}\}={\cal I}_S\equiv({\rm odd}\cap \bar{S})\cup({\rm even}\cap S) \label{ap_i}\\
&&\{b_1,\cdots,b_{\lambda}\}=\bar{{\cal I}}_S=[1,\lambda']\setminus {\cal I}_S.\label{ap_i2}
\end{eqnarray}
\end{itemize}

\begin{enumerate}
 \item \,$|S|=0$\\
The condition (\ref{eq: abcond}) is equivalent to $a_i-b_i$ being odd for $i\in [1,\lambda]$.
Only the pairs 
$(a_i,b_i)$ being $({\rm even},{\rm odd})$ or $({\rm odd},{\rm even})$
for $i\in [1,\lambda]$ can be a solution. As a result, $a_{\lambda+1}$ is odd, since $\{a_i,b_i\}$ for $i\in [1,\lambda]$ exhausts $\lambda$ even numbers and $\lambda$ odd numbers in $[1,\lambda']$. As a representative of $[(A,B)]$, we can take $a_i$ to be odd for $[1,\lambda+1]$. 
The resultant $A$ is $(1,3,\cdots,\lambda')$ or its permutation, which agrees with (\ref{ap_i}).

\item \,$|S|=2$\\
For a given $S$, there are six cases, which we examine below. 
\begin{enumerate}
\item
For the case where $a_j\in S$ and $b_j\in S$ for an index $j\in [1,\lambda]$, the condition (\ref{eq: abcond}) holds only when $a_i-b_i$ is odd for all $i\in [1,\lambda]$. It then follows that $a_{\lambda+1}$ is odd and $|S\cap {\rm even}|=|S\cap {\rm odd}|=1$. As a representative of $[(A,B)]$, we can take $a_i$ to be odd for all $i\in [1,\lambda+1]$ but $a_j$ even. The set $A$ then satisfies
 (\ref{ap_i}). 
\item
For the case where $a_j\in S$ and $a_k\in S$ for a pair $(j,k)$ satisfying $1\le j\ne k\le \lambda$, it follows from (\ref{eq: abcond}) that $a_i-b_i$ is odd for $i\in [1,\lambda]\setminus\{j,k\}$ and both $a_j-b_j$ and $a_k-b_k$ are even. For $i\in[1,\lambda]\setminus\{j,k\}$, $\{a_i,b_i\}$ exhausts $\lambda-2$ even and $\lambda-2$ odd numbers. The remaining two even and three odd integers are available for $\{a_j,a_k,a_{\lambda+1},b_j,b_k\}$. if $a_j$ is taken to be even, $b_j$ then becomes even and the even numbers are exhausted. Consequently $a_k$, $b_k$ and $a_{\lambda+1}$ are forced to be odd. Similarly, when $a_j$ is odd, it follows that $a_k$ is even and $a_{\lambda+1}$ is odd. We assume that $a_j$ is even and $a_k$ is odd.  As a representative of $[(A,B)]$, we can take $a_i$ to be odd for $i\in [1,\lambda+1]\setminus\{j\}$ but even for $i=j$. Alternatively, we can take $(A',B')$ with
\begin{eqnarray*}
&&A'=(a_1,\cdots,a_{k-1},b_k,a_{k+1},\cdots,a_{\lambda+1})\\
&&B'=(b_1,\cdots,b_{k-1},a_k,b_{k+1},\cdots,b_{\lambda})
\end{eqnarray*}
as a representative of $[(A,B)]$. The set $A'$ then satisfies (\ref{ap_i}).  
\item
For the case where $b_j\in S$ and $b_k\in S$ for a pair $(j,k)$ satisfying $1\le j<k\le \lambda$, we can prove in the same way as (b). 
\item
For the case where $a_j\in S$ and $b_k\in S$ for a pair $(j,k)$ satisfying $1\le j\ne k\le \lambda$, we can prove in the same way as (b). 
\item
For the case where $a_j\in S$ for an index $j\in [1,\lambda]$ and $a_{\lambda+1}\in S$, we see from (\ref{eq: abcond}) that $a_i -b_i$ is odd for $i\in [1,\lambda]\setminus\{j\}$ and $a_j-b_j$ is even. The set $\{a_i,b_i\}$ for $i\in [1,\lambda]\setminus\{j\}$ thus exhausts $\lambda-1$ even numbers and $\lambda-1$ odd numbers. For $\{a_j,b_j,a_{\lambda+1}\}$, an even and two odd numbers are available. We see that $a_j$ and $b_j$ are odd and $a_{\lambda+1}$ is even. The relation $|S\cap {\rm even}|=|S\cap {\rm odd}|(=1)$ is thus proven. 
 As a representative of $[(A,B)]$, we can take $a_i$ to be odd for $i\in [1,\lambda]$ and $a_{\lambda+1}$ to be even. Alternatively, we can take $(A',B')$ with
\begin{eqnarray*}
&&A'=(a_1,\cdots,a_{j-1},b_j,a_{j+1},\cdots,a_{\lambda+1})\\
&&B'=(b_1,\cdots,b_{j-1},a_j,b_{j+1},\cdots,b_{\lambda})
\end{eqnarray*}
as another choice of the representative of $[(A,B)]$. The set $A'$ then satisfies (\ref{ap_i}).  
 
\item
For the case where $b_j\in S$ for an index $j\in [1,\lambda]$ and $a_{\lambda+1}\in S$, we can prove in the same way as (e). 

\end{enumerate}

\item $|S|\ge 2$\\
First we decompose the set $[1,\lambda]$ as
\begin{equation}
[1,\lambda]={\cal I}^{(++)}\oplus {\cal I}^{(+-)}\oplus {\cal I}^{(-+)}\oplus 
{\cal I}^{(--)},
\end{equation} 
with 
\begin{equation}
\begin{array}{l}
{\cal I}^{(++)}=\{i\in [1,\lambda]|a_i\in S, \quad b_i\in S\}\\
{\cal I}^{(+-)}=\{i\in [1,\lambda]|a_i\in S, \quad b_i\notin S\}\\
{\cal I}^{(-+)}=\{i\in [1,\lambda]|a_i\notin S, \quad b_i\in S\}\\ 
{\cal I}^{(--)}=\{i\in [1,\lambda]|a_i\notin S, \quad b_i\notin S\}.\\
\end{array}
\label{eq: Ipm}
\end{equation}
The condition (\ref{eq: abcond}) is satisfied when
\begin{equation}
a_i-b_i\in \left\{
\begin{array}{cc}
{\rm odd},\quad &i\in {\cal I}^{(++)}\cup {\cal I}^{(--)}\\
{\rm even},\quad &i\in {\cal I}^{(+-)}\cup {\cal I}^{(-+)},\\
\end{array}
\right.
\label{eq: conds2n}
\end{equation}
for $i\in [1,\lambda']$. 
In terms of (\ref{eq: Ipm}), $|S|$ is written as
\begin{equation}
|S|=2|{\cal I}^{(++)}|+|{\cal I}^{(+-)}|+|{\cal I}^{(-+)}|+\left\{
\begin{array}{cc}
1,\quad& a_{\lambda+1}\in S\\
0,\quad & a_{\lambda+1}\notin S.\\
\end{array}
\right.
\label{eq: modulusS}
\end{equation}
We consider the following cases (a)$\sim$(e).
\begin{enumerate}
\item
When $|{\cal I}^{(+-)}|=|{\cal I}^{(-+)}|=0$, $|S|$ is given by
$$
2|{\cal I}^{(++)}|+\left\{
\begin{array}{cc}
1,\quad& a_{\lambda+1}\in S\\
0,\quad & a_{\lambda+1}\notin S.\\
\end{array}
\right.
$$
Since $|S|$ is even, $a_{\lambda+1}\notin S$ and the set $S$ is written as
$$
S=\{a_i,b_i\}_{i\in {\cal I}^{(++)}},
$$
from which and (\ref{eq: conds2n}), the relation $|S\cap {\rm even}|=|S\cap {\rm odd}|$ follows. The set $\{a_i,b_i\}_{i\in {\cal I}^{(++)}\cap {\cal I}^{(--)}}$ exhausts $\lambda$ even and $\lambda$ odd integers in $[1,\lambda']$ and thus $a_{\lambda+1}$ is odd. As a representative of $[(A,B)]$, we can take $a_i$ to be even for $i\in {\cal I}^{(++)}$ and odd for $i\in {\cal I}^{(--)}$. The set $A$ then satisfies (\ref{ap_i}). 
\item
When $a_i$ and $b_i$ are even for $i\in {\cal I}^{(+-)}\cup {\cal I}^{(-+)}$,
the number of odd integers in $\{a_{1},\cdots,a_{\lambda},b_{1},\cdots,b_{\lambda}\}$ is given by
\begin{equation}
|{\cal I}^{(++)}|+|{\cal I}^{(--)}|\label{eq: numofodd}
\end{equation}
and that of even integers is given by
\begin{equation}
|{\cal I}^{(++)}|+|{\cal I}^{(--)}|+2|{\cal I}^{(+-)}|+2|{\cal I}^{(-+)}|.\label{eq: numofeve}
\end{equation}
(\ref{eq: numofodd}) is either equal to or larger (by one) than (\ref{eq: numofeve}), depending on whether $a_{\lambda+1}$ is odd or even. Only the case with $|{\cal I}^{(+-)}|=|{\cal I}^{(-+)}|=0$ is possible, which we have examined in (a). 
\item
When $a_i$ and $b_i$ are odd for $i\in {\cal I}^{(+-)}\cup {\cal I}^{(-+)}$,
the number of odd integers in $\{a_{1},\cdots,a_{\lambda},b_{1},\cdots,b_{\lambda}\}$ is given by
\begin{equation}
|{\cal I}^{(++)}|+|{\cal I}^{(--)}|+2|{\cal I}^{(+-)}|+2|{\cal I}^{(-+)}|\label{eq: numofodd2}
\end{equation}
and that of even integers is given by
\begin{equation}
|{\cal I}^{(++)}|+|{\cal I}^{(--)}|.\label{eq: numofeven2}
\end{equation}
The difference between (\ref{eq: numofodd2}) and (\ref{eq: numofeven2}) should be equal to 0 (2) when $a_{\lambda+1}$ is odd (even). Namely, we see that
\begin{eqnarray}
&&|{\cal I}^{(+-)}|=|{\cal I}^{(-+)}|=0,\quad \mbox{ when }a_{\lambda+1}\mbox{ is odd}\label{eq: I00}\\
&&(|{\cal I}^{(+-)}|,|{\cal I}^{(-+)}|)=(1,0)\mbox{ or }(0,1),\quad \mbox{ when }a_{\lambda+1}\mbox{ is even}.\label{eq: I1001} 
 \end{eqnarray} 
The case (\ref{eq: I00}) has been already considered in (a). The other case (\ref{eq: I1001}) is compatible with $|S|$ being an even integer only when $a_{\lambda+1}\in S$. We then see that
\begin{eqnarray*}
&&|S\cap \mbox{ odd }|=|{\cal I}^{(++)}|+\underbrace{|{\cal I}^{(+-)}|+|{\cal I}^{(-+)}|}_{1}\\
&&|S\cap \mbox{ even }|=|{\cal I}^{(++)}|+1,
\end{eqnarray*}
which leads to $|S\cap \mbox{ odd }|=|S\cap \mbox{ even }|$. 
As a representative of $[(A,B)]$, we can take $a_i$ to be even for $i\in {\cal I}^{(++)}$ and odd for $i\in {\cal I}^{(--)}$.    
Instead, we can take another choice of the representative as
\begin{equation}
A'=(a'_1,\cdots,a'_{\lambda+1})\quad B'=(b'_1,\cdots,b'_{\lambda}), 
\end{equation}
with $a'_i=a_i $ for $i\in {\cal I}^{(++)}\cup {\cal I}^{(--)}\cup {\cal I}^{(-+)}\cup \{\lambda+1\}$ 
and $a'_i=b_i $ for $i \in {\cal I}^{(+-)}$. 
The set $A'$ agrees with (\ref{ap_i}). 
\item
When $a_i$ and $b_i$ are even (odd) for $i\in {\cal I}^{(+-)}$ (${\cal I}^{(-+)}$). The number of odd integers in  
$\{a_{1},\cdots,a_{\lambda},b_{1},\cdots,b_{\lambda}\}$ is given by
\begin{equation}
|{\cal I}^{(++)}|+|{\cal I}^{(--)}|+2|{\cal I}^{(-+)}|\label{eq: numofodd-d}
\end{equation}
and that of even integers is given by
\begin{equation}
|{\cal I}^{(++)}|+|{\cal I}^{(--)}|+2|{\cal I}^{(+-)}|.\label{eq: numofeve-d}
\end{equation}
The difference between (\ref{eq: numofodd-d}) and (\ref{eq: numofeve-d}) 
is 0 (2) when $a_{\lambda+1}$ is odd (even). Namely, 
\begin{eqnarray}
&&|{\cal I}^{(+-)}|=|{\cal I}^{(-+)}|,\quad \mbox{ when }a_{\lambda+1}\mbox{ is odd}\label{eq: I00-d}\\
&&|{\cal I}^{(-+)}|=|{\cal I}^{(+-)}|+1,\quad \mbox{ when }a_{\lambda+1}\mbox{ is even}.\label{eq: 1001-d} 
 \end{eqnarray} 
The case (\ref{eq: I00-d}) is compatible with the condition that $|S|$ is an even integer only when $a_{\lambda+1}\notin S$. We then obtain
\begin{eqnarray*}
&&|S\cap \mbox{ odd }|=|{\cal I}^{(++)}|+|{\cal I}^{(-+)}|\\
&&|S\cap \mbox{ even }|=|{\cal I}^{(++)}|+|{\cal I}^{(+-)}|.
\end{eqnarray*}
We thus obtain $|S\cap \mbox{ odd }|=|S\cap \mbox{ even }|$. 
As a representative of $[(A,B)]$, we can take $a_i$ to be even for $i\in {\cal I}^{(++)}$ and odd for $i\in {\cal I}^{(--)}$. The set $A$ agrees with (\ref{ap_i}). 

The case (\ref{eq: 1001-d}) is compatible with $|S|$ being an even integer only when $a_{\lambda+1}\in S$. We then obtain
\begin{eqnarray}
&&|S\cap \mbox{ odd }|=|{\cal I}^{(++)}|+|{\cal I}^{(-+)}|\nonumber\\
&&|S\cap \mbox{ even }|=|{\cal I}^{(++)}|+|{\cal I}^{(+-)}|+1.\label{eq: sevenplus1}
\end{eqnarray}
The last term (=1) of the right-hand side in (\ref{eq: sevenplus1}) comes from $a_{\lambda+1}$. We see that $|S\cap \mbox{ odd }|=|S\cap \mbox{ even }|$. 
We can take a representative of $[(A,B)]$ as $A$ with $a_i$ to be even for $i\in {\cal I}^{(++)}$ and odd for $i\in {\cal I}^{(--)}$. The set $A$ agrees with (\ref{ap_i}). 

\item

Other cases reduce to (d), with an appropriate choice of a representative $(A,B)$.

\end{enumerate}

\end{enumerate}

Now we derive (\ref{eq: goalofappB}) from (\ref{eq: expandFu-rewrite}). 
Let ${\mathcal S}({\cal I}_S)$ be the set of $A$ satisfying (\ref{ap_i}) and
let ${\mathcal S}(\bar{{\cal I}}_S)$ be the set of $B$ satisfying (\ref{ap_i2}).  
As shown above, it suffices to consider the contributions from $[(A,B)]$ with $A\in {\mathcal S}({\cal I}_S)$ and $B\in {\mathcal S}(\bar{{\cal I}}_S)$ in (\ref{eq: expandFu-rewrite}). For $(A',B')\in [(A,B)]$ with $A\in {\mathcal S}({\cal I}_S)$ and $B\in {\mathcal S}(\bar{{\cal I}}_S)$, (\ref{eq: sgmAsgmB}) becomes $2^{\lambda}{\rm sgn}(A){\rm sgn}(B) (-1)^{|S\cap {\rm even}|}$. The expression (\ref{eq: expandFu-rewrite}) then reduces to 
\begin{eqnarray}
F(u_1,\cdots,u_{\lambda'};S)&=&2^{\lambda+1}(-1)^{|S\cap {\rm even}|}\sum_{A\in \mathcal{S}({\cal I}_S)}{\rm sgn}(A)u_{a_1}^0u_{a_2}^1\cdots u_{a_{\lambda}}^{\lambda-1}
  u_{a_{\lambda+1}}^{\lambda}\nonumber\\
&&\times\sum_{B\in \mathcal{S}(\bar{{\cal I}}_S)}{\rm sgn}(B)
 u_{b_1}^0
  u_{b_2}^1\cdots u_{b_{\lambda}}^{\lambda-1}\nonumber\\
&=&2^{\lambda+1}(-1)^{|S\cap {\rm even}|}
\prod_{(j<k)\in {\cal I}_S}(u_k -u_j)\prod_{(j<k)\in \bar{{\cal I}}_S}(u_k -u_j),
\nonumber\\
\end{eqnarray}
which coincides with (\ref{eq: goalofappB}).

\end{document}